\documentclass[11pt,letterpaper]{article}
\usepackage{jheppub}
\usepackage{style}

\begin{document}

\title{Ultralight Axial Dark Matter}

\author[1]{Anson~Hook\,\orcidlink{0009-0005-7319-1439},}
\affiliation[1]{Maryland Center for Fundamental Physics, University of Maryland, College Park, MD 20742, USA}
\author[2]{Junwu~Huang\,\orcidlink{0000-0001-6007-7315},}
\affiliation[2]{Perimeter Institute for Theoretical Physics, 31 Caroline St. N., Waterloo, Ontario N2L 2Y5, Canada}
\author[3,4]{and~Kevin~Zhou\,\orcidlink{0000-0002-9810-3977}}
\affiliation[3]{Theory Group, Lawrence Berkeley National Laboratory, Berkeley, CA 94720, USA}
\affiliation[4]{Leinweber Institute for Theoretical Physics, University of California, Berkeley, CA 94720, USA}

\emailAdd{hook@umd.edu}
\emailAdd{jhuang@perimeterinstitute.ca}
\emailAdd{kzhou7@berkeley.edu}

\newcommand{\kz}[1]{{\color{blue}{\sf[KZ: #1]}} }
\newcommand{\ah}[1]{{\color{red}{\sf[AH: #1]}} }
\newcommand{\jh}[1]{{\color{orange}{\sf[JH: #1]}} }

\abstract{Ultralight dark matter could be an axial vector field, a simple possibility which motivates new experiments. Couplings to fermions, through an axial vector current or a dark electric dipole moment operator, lead to enhanced spin torques compared to axion dark matter. An axial vector also has an analogue of the axion-photon coupling, though its consistent realization requires a photon mass. Under this ``axial-photon'' coupling, the effect of a background magnetic field is suppressed, and the strongest experimental probes involve polarimetry, electric fields in superconducting cavities, and the cosmic microwave background. Since a massive axial vector is dual to a massive two-form, our results also apply to ``Kalb--Ramond'' dark matter motivated by string theory.}

\maketitle
\setcounter{page}{2}

\vspace{-8mm}
\paragraph{Conventions and Notation.} We use natural units, $\hbar = c = 1$, with rationalized/Heaviside--Lorentz electromagnetic units (i.e., SI units with $\epsilon_0 = \mu_0 = 1$). We use a mostly-negative spacetime metric and define $\epsilon^{0123} = 1$, $\gamma^5 = i \gamma^0 \gamma^1 \gamma^2 \gamma^3$, and $\sigma^{\mu\nu} = (i/2) [\gamma^\mu, \gamma^\nu]$. The axial vector field $V^\mu = (V^0, \v{V})$ has mass $m$ and field strength $F^V_{\mu\nu}$, the photon $A^\mu = (A^0, \v{A})$ has mass $m_\gamma$ and field strength $F_{\mu\nu}$, and a fermion $\psi$ has mass $m_\psi$ and electric charge $q_\psi$. The dual field strength is $\tilde{F}^{\mu\nu} = \epsilon^{\mu\nu\rho\sigma} F_{\rho\sigma}/2$. The local dark matter density is $\rho_\DM = 0.4 \, \mathrm{GeV}/\mathrm{cm}^3$. 

%!TEX root = main.tex

\section{Introduction}
\label{sec:intro}

In the past decade, there has been much interest in the possibility that dark matter (DM) is a single ultralight bosonic field, $m \lesssim 10 \, \mathrm{eV}$, with high typical mode occupancy in the galaxy~\cite{Irastorza:2018dyq,Adams:2022pbo,Antypas:2022asj,Berlin:2024pzi}. This is appealing on theoretical grounds because such fields arise ubiquitously in extensions of the Standard Model, and because they furnish one of the most minimal explanations of DM. Furthermore, modern technology makes it possible for new small-scale experiments to search for weakly coupled fields across orders of magnitude in mass with exquisite precision. 

Another key factor, often left implicit, is that the minimality of ultralight DM means there are only a few ways it can interact with photons, electrons, and nucleons, consistent with effective field theory. This prevents the combinatorial explosion that often plagues searches for physics beyond the Standard Model, and sharpens the focus of new experiments. 

In Table~\ref{tab:dm_fields} we categorize ultralight DM candidates by their spin $S$ and intrinsic parity $\eta_P$. The case $S = 0$ and $\eta_P = -1$ is the pseudoscalar ``axion'' field $a$, whose leading derivative couplings are
\begin{equation} \label{eq:axion_couplings}
\mathcal{L}_{\mathrm{int}} = g_{a\psi} (\del_\mu a) \bar{\psi} \gamma^\mu \gamma^5 \psi - \frac{g_{a\gamma\gamma}}{4} (\del_\mu a) K^\mu \, ,
\end{equation}
where $\psi$ is an electron or nucleon, and $K^\mu = \epsilon^{\mu\nu\rho\sigma} A_\nu F_{\rho\sigma}$ is the Chern--Simons current. These are the axion-fermion and axion-photon couplings, respectively. We can also consider nonderivative couplings, such as the axion-gluon coupling or the couplings of a scalar (dilaton) $\phi$, though these can substantially affect the DM mass.

For nonzero spin, nonderivative interactions often still allow for an ultralight mass, since the number of degrees of freedom changes discontinuously at $m = 0$. However, theories with $S = 2$ generically suffer from very low cutoffs~\cite{Hinterbichler:2011tt,Bellazzini:2023nqj}, and the higher spin case $S > 2$ faces even more difficulties~\cite{Bekaert:2010hw}. Focusing on $S = 1$, the case $\eta_P = 1$ corresponds to a vector field $A'_\mu$, often called the dark photon. Its leading interactions are kinetic mixing with the photon, $\epsilon F'_{\mu\nu} F^{\mu\nu}/2$, and coupling to a vector current $A'_\mu J^\mu$, such as the $B-L$ current.

There is one case remaining: a $S = 1$ particle with $\eta_P = -1$, which can be created or annihilated by an axial vector field $V^\mu$. This simple possibility has been almost completely unexplored, possibly because of an expectation that it behaves like an axion. It has also been claimed~\cite{PhysRevD.37.1237} that it cannot couple to on-shell photons because of the Landau--Yang theorem. On the contrary, in this work we show that axial vector couplings can arise in viable models, and lead to new experimental signatures that differ dramatically from axions and dark photons. 

In Sec.~\ref{sec:fermion_couplings} we consider the simplest fermion couplings,
\begin{equation} \label{eq:axial_fermion_couplings}
\mathcal{L}_{\mathrm{int}}^\psi = \frac{m}{f_\psi} \, V_\mu \bar{\psi} \gamma^\mu \gamma^5 \psi + \frac{1}{2f_\psi^E} \, F_{\mu\nu}^V \bar{\psi} i \sigma^{\mu\nu} \gamma^5 \psi \, ,
\end{equation}
which we call the axial-fermion and dark electric dipole moment (EDM) couplings, respectively. When relativistic vectors are produced through the axial-fermion coupling, the longitudinal mode $V_\mu \supset \del_\mu \alpha_V / m$ dominates, and for this mode the axial-fermion coupling reduces to the axion-fermion coupling with $1/f_\psi \lra g_{a\psi}$. However, in general axial vector DM behaves differently from axion DM, since DM is highly nonrelativistic, with typical speed $v_\DM \sim 10^{-3}$.

Explicitly, for a randomly polarized vector we expect 
\begin{equation} \label{eq:random_pol}
\langle V_i V_j \rangle = \frac{\rho_\DM}{3 m^2} \, \delta_{ij}
\end{equation}
while the vector's free equation of motion 
\begin{equation} \label{eq:vector_eom}
\del_\mu F_V^{\mu\nu} + m^2 V^\nu = 0
\end{equation}
implies $\del_\mu V^\mu = 0$. Since time and space derivatives scale as $m$ and $m v_\DM$ respectively, with $v_\DM \sim 10^{-3}$, this implies $V^0 \sim v_\DM V^i$ for a vector, in contrast to $\del^i a \sim v_\DM \del^0 a$ for a scalar. We will see that this leads to greatly enhanced spin torques for both couplings in Eq.~\eqref{eq:axial_fermion_couplings}, so that small-scale experiments already probe well beyond astrophysical bounds.

\begin{table}[t]
\begin{center}
\begin{tabular}{r|c|c}
& spin $S = 0$ & spin $S = 1$ \\ \hline
\makecell{parity \\ $\eta_P = -1$} & \makecell{\underline{pseudoscalar/axion $a$} \\ axion-fermion \\ axion-photon \\ axion-gluon \\ (equivalent to 3-form)} & \makecell{\underline{axial vector $V_\mu$} (this work) \\ axial-fermion (Sec.~\ref{sec:axial_fermion}) \\ dark electric dipole (Sec.~\ref{sec:dark_dipole}) \\ axial-photon (Sec.~\ref{sec:axial_photon}, Sec.~\ref{sec:axial_photon_pheno}, App.~\ref{app:axial_photon}) \\ equivalent to 2-form (Sec.~\ref{sec:two_form}, App.~\ref{app:dualities})} \\ \hline
\makecell{parity \\ $\eta_P = 1$} & \makecell{\underline{scalar/dilaton $\phi$} \\ dilaton-fermion \\ dilaton-photon \\ dilaton-gluon \\ (requires mass tuning)} & \makecell{\underline{vector/dark photon $A'_\mu$} \\ kinetic mixing \\ vector current \\ dark magnetic dipole (Sec.~\ref{sec:dark_dipole})} 
\end{tabular}
\end{center}
\caption{Candidates for an ultralight bosonic DM field, and their leading couplings, visible in low-energy laboratory experiments. This work investigates the single remaining candidate.}
\label{tab:dm_fields}
\end{table}

The simplest coupling of an axial vector to photons is the axial-photon coupling
\begin{equation} \label{eq:axial_photon_coupling}
\mathcal{L}_{\mathrm{int}}^A = - \frac{m}{4f_\gamma} \, V_\mu K^\mu
\end{equation}
where the prefactor is defined so that, for relativistic longitudinal vectors, it reduces to the axion-photon coupling with $1/f_\gamma \lra g_{a\gamma\gamma}$. As we discuss in Sec.~\ref{sec:axial_photon}, this ``generalized Chern--Simons'' term can arise when a would-be axion-like particle is eaten by a gauge boson, in the presence of fermions with nontrivial $U(1)_D U(1)_{\mathrm{EM}}^2$ anomaly cancellation. 

The axial-photon coupling violates electromagnetic gauge invariance; accordingly, in App.~\ref{app:axial_photon} we present a viable model in which the photon obtains a mass, while remaining consistent with photon mass bounds. This in turn induces a wide variety of new effects, associated with background vector potentials and longitudinal photons. 

In Sec.~\ref{sec:axial_photon_pheno} we investigate the experimental signatures of the axial-photon coupling. Here the effects of background magnetic fields, almost universally used to probe axion DM, become velocity suppressed; instead background electric fields can provide better sensitivity. In addition, polarization signals in both laboratory experiments and the CMB are dramatically enhanced, with the latter due to $1/m_\gamma$-enhanced processes involving longitudinal photons. To our knowledge, the axial-photon coupling has only been considered in Refs.~\cite{Antoniadis:2006wp,Antoniadis:2007sp}, but these works incorrectly set $m_\gamma = 0$, and did not find any of the novel signatures discussed here. 

A massive spin $1$ particle can also be created or annihilated by a massive antisymmetric rank $2$ tensor field $B_{\mu\nu}$, and these two-form (Kalb--Ramond) fields arise naturally in higher-dimensional theories. In Sec.~\ref{sec:two_form} we review well-known dualities which relate massive axial vectors and two-forms in $d = 4$. We ultimately find that all physical effects are equivalent, but that the duality modifies the EFT power counting. In particular, the dark EDM coupling in Eq.~\eqref{eq:axial_fermion_couplings} is dimension $5$ on the vector side, but becomes the leading coupling to fermions on the two-form side, motivating further study of its effects. Finally, we conclude in Sec.~\ref{sec:discussion} by discussing other axial vector couplings, production mechanisms, and open questions. 

%!TEX root = main.tex

\section{Couplings to Fermions}
\label{sec:fermion_couplings}

The couplings of axial vector DM to fermions manifest as spin torques and spin-dependent forces. We find that many results can be recast directly from those for axion-fermion couplings, recently reviewed in Ref.~\cite{Berlin:2023ubt}, but their relative sensitivity shifts because of changes in the pattern of velocity suppression, shown in Table~\ref{tab:velocity_suppressions}. 

In Sec.~\ref{sec:axial_fermion} we consider the axial-fermion coupling, and show that its stronger spin precession effects imply that existing experiments already reach well beyond astrophysical bounds. In Sec.~\ref{sec:dark_dipole} we consider the dark EDM coupling, whose effects scale like those of the axial-fermion coupling. The dark magnetic dipole moment (MDM) appears for an ordinary vector rather than an axial vector, but we also analyze it in passing.

\subsection{The Axial-Fermion Coupling}
\label{sec:axial_fermion}

Following the same logic as in Ref.~\cite{Berlin:2023ubt}, for a fermion $\psi$ of mass $m_\psi$ and charge $q_\psi$, the axial-fermion coupling in Eq.~\eqref{eq:axial_fermion_couplings} corresponds to the nonrelativistic interaction Hamiltonian\footnote{Technically one instead has the symmetric ordering $\{V^0 , (\v{p} - q_\psi \v{A}) \cdot \bm{\sigma}\}/2$, but this only differs by a velocity-suppressed term proportional to $\nabla V^0 \sim m v_\DM^2 |\v{V}|$. We drop such terms here and in Sec.~\ref{sec:dark_dipole}.}
\begin{equation} \label{eq:axial_fermion_ham}
H_{\mathrm{int}} \simeq \frac{m}{f_\psi} \, \v{V} \cdot \bm{\sigma} - \frac{m V^0}{m_\psi f_\psi} \, (\v{p} - q_\psi \v{A}) \cdot \bm{\sigma}.
\end{equation}
For brevity, in the rest of this section we set the vector potential $\v{A}$ to zero, but its effect can be trivially recovered by mapping $\v{p} \to \v{p} - q_\psi \v{A}$. 

In the presence of DM, the first term in Eq.~\eqref{eq:axial_fermion_ham} induces a spin torque scaling as 
\begin{equation} \label{eq:axial_fermion_torque}
\bm{\tau} \simeq \frac{m}{f_\psi} (\v{V} \times \hat{\v{s}}) \sim \frac{\sqrt{\rho_\DM}}{f_\psi}, 
\end{equation}
where we identified $\hat{\v{s}} = \langle \bm{\sigma} \rangle$ as the fermion's spin polarization vector. Both terms in Eq.~\eqref{eq:axial_fermion_ham} induce a spin-dependent force. Defining $\v{F} = m_\psi \, d \langle \v{v} \rangle / dt$, and dropping terms suppressed by the fermion velocity, we have
\begin{equation} \label{eq:axial_fermion_force}
\v{F} \simeq - \frac{m}{f_\psi} \left( \nabla (\v{V} \cdot \hat{\v{s}}) + \frac{d}{dt} (V^0 \hat{\v{s}}) \right) \sim \frac{\sqrt{\rho_\DM}}{f_\psi} \, m v_\DM.
\end{equation}
Parametrically, these results are similar to those of an axion coupled to fermions with $g_{a\psi} \lra 1/f_\psi$, but in that case the ``axion wind'' spin torque is velocity suppressed, and the ``axioelectric'' force is not; here the location of the velocity suppression is reversed. 

\begin{table}
\begin{center}
\begin{tabular}{r|c|c}
& $\tau / \sqrt{\rho_\DM}$ & $F / \sqrt{\rho_\DM}$ \\ \hline
axion-fermion coupling & $v g_{a\psi}$ & $m g_{a\psi}$ \\
axial-fermion coupling & $1/f_\psi$ & $m v / f_\psi$ \\
dark electric dipole moment & $1/f_\psi^E$ & $m v / f_\psi^E$ \\
dark magnetic dipole moment & $v / f_\psi^M$ & $m / f_\psi^M$
\end{tabular}
\end{center}
\caption{Velocity scaling of torques and forces for ultralight DM couplings to fermions. Generically we have $v \sim \max(v_\DM, v_\psi)$, and in the text we focus on $v_\DM$ effects. The second and third rows are relevant for axial vector DM. Stellar cooling bounds for all four dimensionful couplings are comparable.}
\label{tab:velocity_suppressions}
\end{table}

\begin{figure} 
\centering
\includegraphics[scale=0.7]{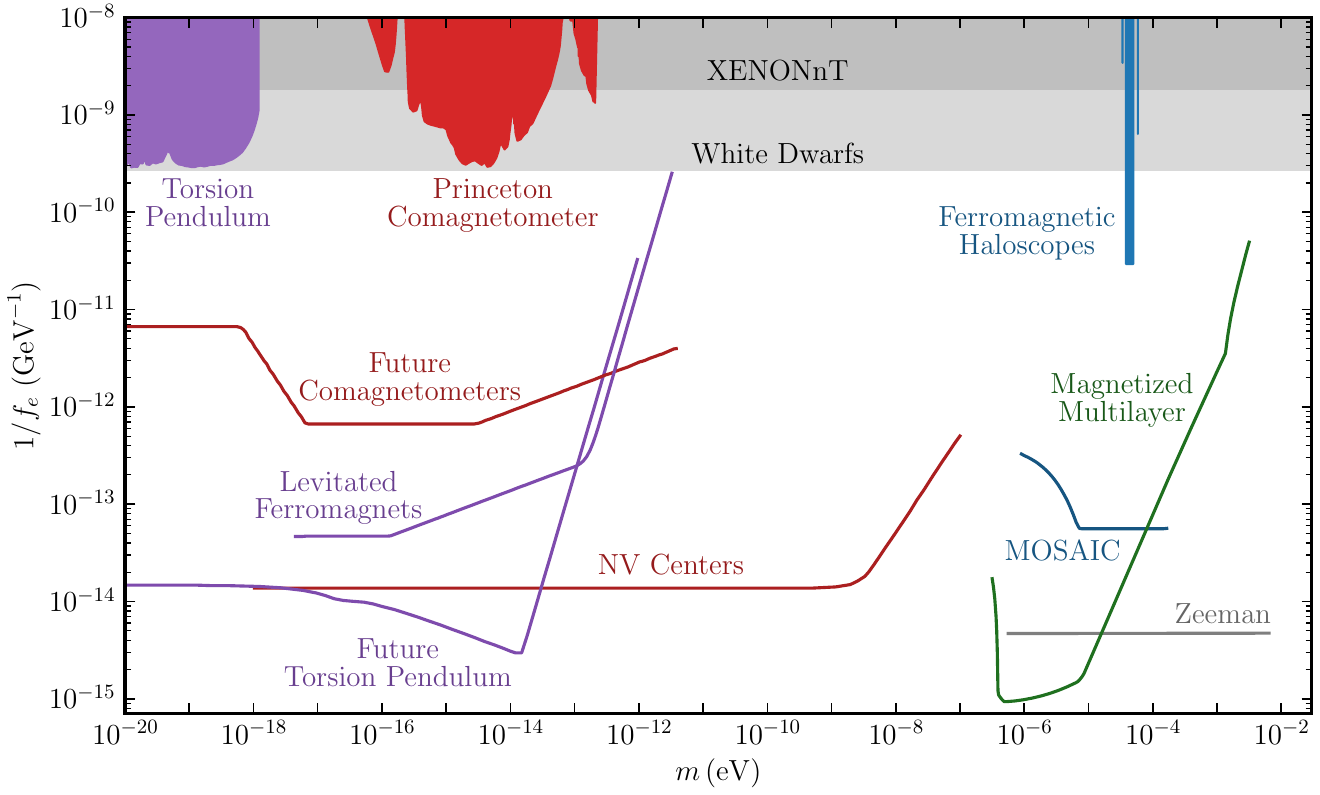}
\caption{Current constraints and projected future sensitivity to the axial-electron coupling. We show astrophysical bounds~\cite{MillerBertolami:2014rka,XENON:2022ltv} and the reach of existing and future comagnetometers~\cite{Bloch:2019lcy}, existing~\cite{Terrano:2015sna} and future~\cite{Graham:2017ivz} torsion pendulums, levitated ferromagnets~\cite{Kalia:2024eml}, NV centers~\cite{Chigusa:2023roq}, existing ferromagnetic haloscopes~\cite{Crescini:2018qrz,QUAX:2020adt,Flower:2018qgb,Ikeda:2021mlv} and the MOSAIC proposal~\cite{Chang:2025sno} involving an array of YIG spheres, magnetized multilayers~\cite{Berlin:2023ubt}, and Zeeman transitions~\cite{Sikivie:2014lha}.}
\label{fig:axial_electron}
\end{figure}

In Fig.~\ref{fig:axial_electron}, we show the reach of existing and future experiments to the axial-electron coupling. As a baseline, we show bounds from energy loss in white dwarfs~\cite{MillerBertolami:2014rka} and solar axion searches in XENONnT~\cite{XENON:2022ltv}. Since these bounds involve emission of highly relativistic axial vectors, the longitudinal mode $V_\mu \supset \del_\mu \alpha_V / m$ dominates, and one can directly infer the limit from the corresponding axion result by recasting $g_{ae} \lra 1/f_e$, which yields $f_e \gtrsim 10^{10} \, \mathrm{GeV}$. 

\paragraph{Dark Matter Searches.} We first discuss the DM searches shown in Fig.~\ref{fig:axial_electron}. Signatures involving nonrelativistic electrons experiencing a DM spin torque are strongly enhanced for axial vector DM, and here one can recast via $g_{ae} v_\DM \lra 1/f_e$.\footnote{There is an $\mathcal{O}(1)$ factor due to the difference in statistics between $\v{V}$, a spatial vector, and $\nabla a$, the gradient of a scalar. For the rough estimates here, we absorb this into the uncertainty on $v_\DM$ and take $v_\DM \sim 10^{-3}$.} This implies that existing comagnetometer limits are already competitive with astrophysical bounds, as was previously noted in Ref.~\cite{Alonso:2018dxy}. By recasting the projections of Ref.~\cite{Bloch:2019lcy}, we find that future comagnetometers can probe well beyond astrophysical constraints. The same is true for searches for mechanical torques in spin-polarized torsion pendulums~\cite{Terrano:2015sna,Graham:2017ivz} or levitated ferromagnets~\cite{Kalia:2024eml}. Experiments searching for spin precession in NV centers~\cite{Chigusa:2023roq} could also have strong reach. 

At higher frequencies, ``ferromagnetic haloscope'' experiments using ferromagnetic resonance with $\mathrm{mm}$-scale YIG spheres~\cite{Crescini:2018qrz,QUAX:2020adt,Flower:2018qgb,Ikeda:2021mlv} already probe well beyond astrophysical bounds. Ferromagnetic resonance in larger-scale experiments could be even stronger; we show the array-based MOSAIC proposal~\cite{Chang:2025sno} and the magnetized multilayer concept of Ref.~\cite{Berlin:2023ubt}. We also show the projection of Ref.~\cite{Sikivie:2014lha} involving atomic Zeeman transitions.

A variety of other proposals are too weak to appear in Fig.~\ref{fig:axial_electron}. First, absorption in semiconductors~\cite{Hochberg:2016sqx} and superconductors~\cite{Hochberg:2016ajh}, and the ``piezoaxionic'' effect of Ref.~\cite{Arvanitaki:2021wjk}, rely on the second term in Eq.~\eqref{eq:axial_fermion_ham}, and thus receive a velocity suppression for axial vector DM, $g_{ae}/v_\DM \lra 1/f_e$. Next, Ref.~\cite{Galon:2020lol} considered the spin precession of relativistic electrons in storage rings, where the relevant velocity is $v_e \sim 1$. This effect is thus neither enhanced nor suppressed by $v_\DM$, but in any case, it only reaches $f_e \sim 10^{-2} \, \mathrm{GeV}$. Finally, Ref.~\cite{Gaul:2020bdq} considered spectroscopy of chiral molecules; its most ambitious projection is $(m/f_e) V^0 \sim 10^{-17} \, \mathrm{GeV}$, which only corresponds to $f_e \sim 10^{-7} \, \mathrm{GeV}$.

\paragraph{Generic Constraints.} There are other constraints on the axial-fermion coupling which are independent of the DM density, but they are all far weaker than astrophysical bounds. 

First, virtual axial vectors mediate a spin-dependent force similar to that of an axion~\cite{Fadeev:2018rfl,PhysRevA.105.022812,Cong:2024qly}. Neglecting the fermion velocity, the potential mediated by the axial vector is 
\begin{equation}
V_A(\v{r}) = V_a(\v{r}) - \frac{m^2}{f_e^2} \frac{e^{-mr}}{4 \pi r} \, \bm{\sigma}_1 \cdot \bm{\sigma}_2, 
\end{equation}
where $V_a(\v{r})$ is the spin-dependent potential mediated by an axion with $g_{ae} \lra 1/f_e$. Fifth force searches are effective for $mr \lesssim 1$, and the additional term in $V_A$ is suppressed, relative to the leading terms in $V_a$, by $(mr)^2$. Thus, fifth force constraints on axial vectors are essentially the same as for axions, up to an order-one difference near $mr \sim 1$. The strongest existing bound comes from the torsion pendulum experiment of Ref.~\cite{Terrano:2015sna}, which reaches $f_e \sim 10^5 \, \mathrm{GeV}$. 

Next, loop corrections with an axial vector shift the electron $g$-factor by order $\sim m_e^2 / (4 \pi f_e)^2$~\cite{Fayet:2007ua,NA64:2021xzo}. The $g$-factor of the electron is measured to $\sim 10^{-13}$ precision~\cite{Fan:2022eto}, but this only implies $f_e \gtrsim 10^2 \, \mathrm{GeV}$. (However, an axial coupling to muons would yield $f_\mu \gtrsim 10^3 \, \mathrm{GeV}$ from muon $g-2$~\cite{Aliberti:2025beg}, which could be competitive with other probes.)

Light axial vectors in the $\mathrm{MeV}$ range have been considered as force carriers or DM mediators~\cite{Kahlhoefer:2015bea,Kahn:2016vjr,Gori:2025jzu}, and they are known to be strongly constrained due to $E/m$ enhanced amplitudes involving the longitudinal mode~\cite{Dror:2017ehi,Dror:2017nsg,Ekhterachian:2021rkx}. These processes also exist for an ultralight axial vector, but they are significantly weaker than astrophysical bounds. For example, constraints on the decay $\pi^+ \to e^+ \nu_e V$ imply $f_e \gtrsim 10^2 \, \mathrm{GeV}$~\cite{Dror:2017nsg}. The strongest collider bound comes from measurements of Bhabha scattering; in the equivalent language of two-forms, to be reviewed in Sec.~\ref{sec:two_form}, Ref.~\cite{Malta:2025ydq} showed these imply $f_e \gtrsim 10^3 \, \mathrm{GeV}$. 

Finally, we can consider the UV cutoff implied from perturbative unitarity. The constraint must depend on $m_e$, as when $m_e = 0$ the axial vector's one-loop self-energy due to the axial coupling is the same as for a vector coupling, where there is no UV cutoff. For nonzero $m_e$, it was shown in Ref.~\cite{Ekhterachian:2021rkx} that the scale at which this coupling becomes strongly coupled is $\Lambda \sim 4 \pi f_e \log^{3/2} (f_e/m_e)$. Imposing $\Lambda \gtrsim \mathrm{TeV}$ gives a constraint far weaker than other bounds. 

\begin{figure} 
\centering
\includegraphics[scale=0.7]{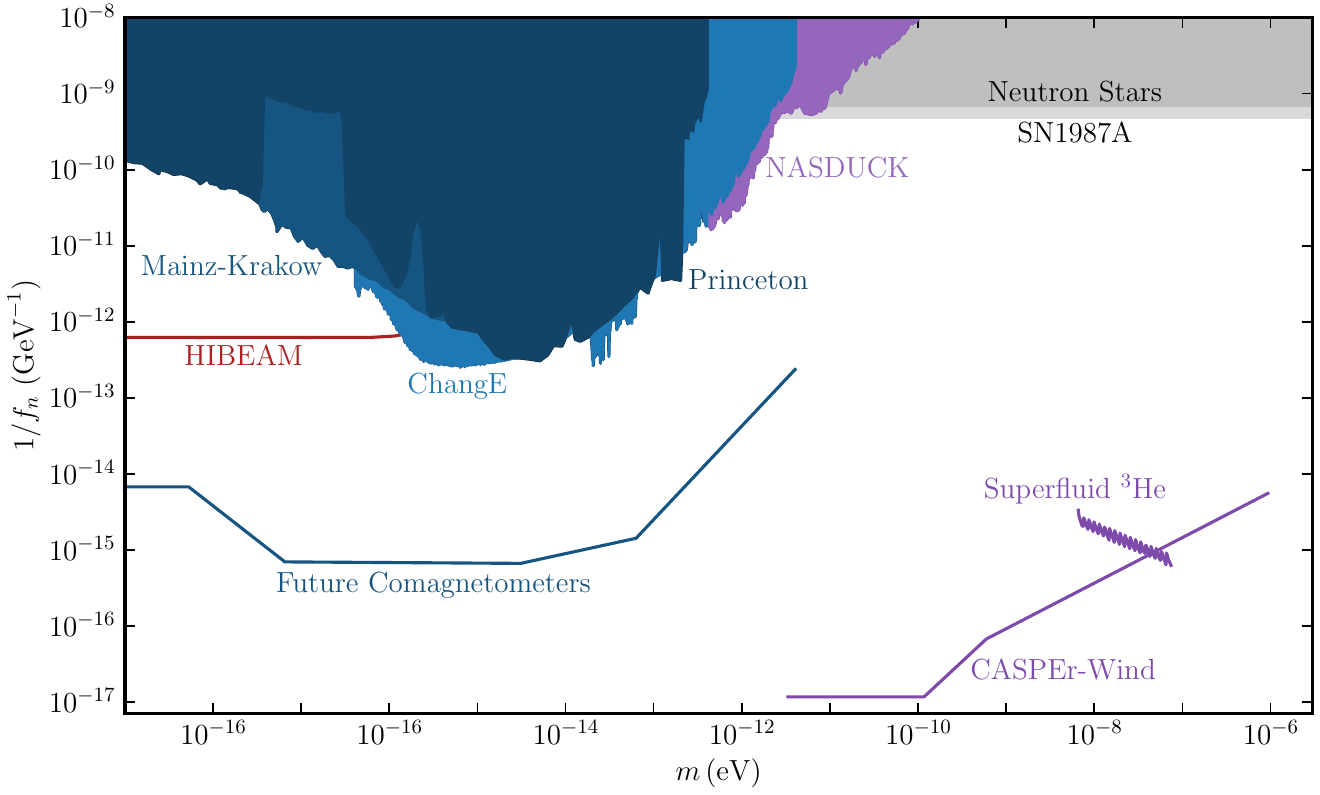}
\caption{Current constraints and projected future sensitivity to the axial-neutron coupling. We show existing bounds from the cooling of neutron stars~\cite{Buschmann:2021juv} and SN1987A~\cite{Carenza:2019pxu}, and from the Princeton~\cite{Bloch:2019lcy} and Mainz-Krakow~\cite{Gavilan-Martin:2024nlo} comagnetometers and the NASDUCK~\cite{Bloch:2021vnn,Bloch:2022kjm} and ChangE~\cite{Wei:2023rzs,Xu:2023vfn} experiments. (See Ref.~\cite{Zhang:2025zcq} for another bound, only applicable at $m \lesssim 10^{-20} \, \mathrm{eV}$.) We also show projections for future comagnetometers~\cite{Bloch:2019lcy}, the HIBEAM neutron beamline~\cite{Fierlinger:2024aik}, and NMR experiments, including CASPEr-wind~\cite{JacksonKimball:2017elr} and a proposal using superfluid ${}^3\mathrm{He}$~\cite{Gao:2022nuq}.}
\label{fig:axial_neutron}
\end{figure}

\paragraph{Nucleon Couplings.} Similar conclusions hold for the axial-neutron and axial-proton couplings; here we consider the axial-neutron coupling for concreteness. 

Independent of the DM density, constraints on spin-dependent neutron interactions mediated by an axial vector are relatively weak~\cite{Cong:2024qly}, partly due to the difficulty of polarizing nuclear spins. If the axial-neutron coupling arises from axial-quark couplings, constraints on $B \to KV$ and $K \to \pi V$ imply $f_q \gtrsim 10^5 \, \mathrm{GeV}$ and $f_q \gtrsim 10^6 \, \mathrm{GeV}$ respectively, for a generation-universal quark-level $f_q$~\cite{Dror:2017nsg}. Astrophysical bounds involving cooling via emission of relativistic vectors are by far the strongest DM-independent bounds; they can be recast via $g_{an} \lra 1/f_n$, yielding $f_n \gtrsim 10^9 \, \mathrm{GeV}$.

For the axion-nucleon couplings, competitive DM searches all involve nuclear spin precession, and in some cases benefit from the very high quality factors achievable with nuclear magnetic resonance. (Signals from the force in Eq.~\eqref{eq:axial_fermion_force} are much weaker, since the resulting acceleration is suppressed by the mass of the nucleus.) We thus recast the sensitivities of these searches to the axial-neutron coupling by $g_{an} v_\DM \lra 1/f_n$, and show the results in Fig.~\ref{fig:axial_neutron}. 

\subsection{Dark Dipole Moments}
\label{sec:dark_dipole}

Here we consider the dark EDM and MDM interactions
\begin{equation}
\mathcal{L}_{\mathrm{int}} = \frac{1}{2f_\psi^E} \, F_{\mu\nu}^V \bar{\psi} i \sigma^{\mu\nu} \gamma^5 \psi + \frac{1}{2f_\psi^M} \, F_{\mu\nu}^V \bar{\psi} \sigma^{\mu\nu} \psi,
\end{equation}
which appear for an axial vector and an ordinary vector respectively. These terms have been discussed for massless vectors in Refs.~\cite{Dobrescu:2004wz,Fabbrichesi:2020wbt}, and for vector DM with $\mathrm{eV}$ to $\mathrm{MeV}$ mass~\cite{Krnjaic:2022wor,Krnjaic:2023nxe}. Our new contribution is to demonstrate the pattern of velocity suppression shown in the last two rows in Table~\ref{tab:velocity_suppressions}.

Since the effects depend on the field strength $F_{\mu\nu}^V$, we define the components
\begin{subequations}
\begin{align}
\v{E}^V &= - \dot{\v{V}} - \nabla V^0 \sim \sqrt{\rho_\DM}, \label{eq:EV_scaling} \\
\v{B}^V &= \nabla \times \v{V} \sim v_\DM |\v{E}^V|, \label{eq:BV_scaling}
\end{align}
\end{subequations}
where the first term in $\v{E}^V$ dominates, and we used the scaling in Eq.~\eqref{eq:random_pol}. In components, the equation of motion Eq.~\eqref{eq:vector_eom} reads
\begin{subequations}
\begin{align}
\nabla \cdot \v{E}^V &= - m^2 V^0, \label{eq:massive_gauss} \\
\nabla \times \v{B}^V &= \dot{\v{E}}^V - m^2 \v{V}, \label{eq:massive_ampere}
\end{align}
\end{subequations}
where the quantity $- m^2 V^\mu$ acts as an effective current. 

\paragraph{Dark Electric Dipole Moment.} First, to estimate the stellar cooling bound, note that for a relativistic vector $V$ with energy $\omega \gg m$, the cross section for the process $\gamma e^- \to V e^-$ scales as $\sigma \sim \alpha \omega^2 / (m_e f_e^E)^2$. This is parametrically the same as the cross section for $\gamma e^- \to a e^-$ through an axion-electron coupling $g_{ae} = 1/f_e^E$~\cite{Pospelov:2008jk}. The main difference compared to the axion case is that the dark EDM term involves the field strength $F_{\mu\nu}^V$, so the vector's two transverse modes dominate, while the effect of the vector's longitudinal mode is suppressed. 

Thus, up to a factor of $\sqrt{2}$, we can recast stellar cooling bounds via $g_{ae} \lra 1/f_e^E$, which implies $f_e^E \gtrsim 10^{10} \, \mathrm{GeV}$. This matches the estimate in Ref.~\cite{Dobrescu:2004wz} and the calculation in Ref.~\cite{Fabbrichesi:2020wbt}. (These works set $m = 0$, but for transverse polarizations this has negligible impact on the result.) Recently, Ref.~\cite{Bai:2025nzd} considered couplings to neutrons and found $f_n^E \gtrsim 2 \times 10^9 \, \mathrm{GeV}$. 

As for laboratory probes, the dark EDM term mediates a dipole-dipole interaction~\cite{Fadeev:2018rfl}, like an axion with $g_{a\psi} \sim 1/f_\psi^E$, but constraints on such an interaction are weaker than astrophysical bounds. To compute DM effects, we use the standard nonrelativistic interaction Hamiltonian for an EDM, which is
\begin{equation} \label{eq:dark_edm_int}
H_{\mathrm{int}} \simeq - \frac{1}{f_\psi^E} \, \bm{\sigma} \cdot \v{E}^V - \frac{1}{m_\psi f_\psi^E} \, \bm{\sigma} \cdot (\v{p} \times \v{B}^V).
\end{equation}
This corresponds to a canonical momentum $\v{p} = m_\psi \v{v} - (\hat{\v{s}} \times \v{B}^V) / f_\psi^E$. Dropping terms higher-order in the DM or fermion velocities, the leading spin torque and force are then
\begin{equation}
\bm{\tau} \simeq \frac{1}{f_\psi^E} \, \hat{\v{s}} \times \v{E}^V \sim \frac{\sqrt{\rho_\DM}}{f_\psi^E}, \qquad \v{F} \simeq \frac{1}{f_\psi^E} \, (\hat{\v{s}} \cdot \nabla) \v{E}^V \sim \frac{\sqrt{\rho_\DM}}{f_\psi^E} \, m v_\DM,
\end{equation}
which are simply the familiar classical results. 

The magnitudes of the torque and force are parametrically the same as for the axial-fermion coupling, given in Eqs.~\eqref{eq:axial_fermion_torque} and~\eqref{eq:axial_fermion_force}, with $f_{\psi} \lra f_{\psi}^E$. Thus, the phenomenology of the dark EDM term is similar to the axial-fermion coupling: laboratory probes due to spin torques become stronger relative to astrophysical bounds by a factor of $1/v_\DM$. However, the two terms are different in general. For example, the relativistic corrections to the torque and force are different; moreover, the axial-fermion coupling yields the same spin torque for a fermion and antifermion with the same spin, while the dark EDM term yields opposite torques.

\paragraph{Dark Magnetic Dipole Moment.} Like the dark EDM term, the dark MDM term mediates a dipole-dipole interaction~\cite{Fadeev:2018rfl}, like an axion with $g_{a\psi} \sim 1/f_\psi^M$. Since $m \ll \omega$, stellar cooling bounds are identical to the dark EDM case~\cite{Bai:2025nzd}, so $f_e^M \gtrsim 10^{10} \, \mathrm{GeV}$ and $f_n^M \gtrsim 2 \times 10^9 \, \mathrm{GeV}$.

The nonrelativistic interaction Hamiltonian of the dark MDM term is dual to that of the dark EDM term, so that 
\begin{equation} \label{eq:dark_mdm_int}
H_{\mathrm{int}} \simeq - \frac{1}{f_\psi^M} \, \bm{\sigma} \cdot \v{B}^V + \frac{1}{m_\psi f_\psi^M} \, \bm{\sigma} \cdot (\v{p} \times \v{E}^V).
\end{equation}
However, the derivation of torques and forces is more subtle than in the dark EDM case. Even though the second term is still suppressed by the fermion velocity $v_\psi$, the first term is now suppressed by $v_\DM$, so we must keep both terms to extract the most important contributions to the spin torque and force. The spin torque is
\begin{equation} \label{eq:dark_mdm_torque}
\bm{\tau} \simeq \frac{1}{f_\psi^M} \, \hat{\v{s}} \times (\v{B}^V - \v{v} \times \v{E}^V) \sim \frac{\sqrt{\rho_\DM}}{f_\psi^M} \, \max(v_\DM, v_\psi),
\end{equation}
where either term could be larger, depending on the situation. This is the same scaling as the torque due to the axion-fermion coupling, but their directional dependence is order-one different; for each momentum mode $\v{k}$, the gradient $\nabla a$ is parallel to $\v{k}$ while $\v{B}^V = \nabla \times \v{V}$ is transverse to it. To our knowledge, the detailed statistics of $\v{B}^V$ have not been worked out. 

To find the force, we must account for the fact that the canonical momentum is 
\begin{equation}
\v{p} = m_\psi \v{v} + \frac{1}{f_\psi^M} \, \hat{\v{s}} \times \v{E}^V,
\end{equation}
where the second term is the famously subtle ``hidden'' momentum of a magnetic dipole in an electric field~\cite{PhysRevLett.18.876,Coleman:1968zza}. Dropping terms suppressed by $v_\psi$, the force $\v{F} = m_\psi \, d \langle \v{v} \rangle / dt$ is
\begin{align} \label{eq:mdm_force}
\v{F} \simeq \frac{1}{f_\psi^M} \left( \nabla (\hat{\v{s}} \cdot \v{B}^V) - \del_t (\hat{\v{s}} \times \v{E}^V) \right),
\end{align}
where the second term is a direct result of the hidden momentum. This force was derived in Ref.~\cite{10.1119/1.16260} in the context of ordinary electromagnetism. That work showed that the second force term appears for Ampere dipoles, but not Gilbert dipoles; since an EDM is the dual of a Gilbert dipole, this term appears for the dark MDM term, but not the dark EDM term. 

To estimate the force, we drop the contribution of $\del_t \hat{\v{s}}$ because it is higher-order in $1/f_\psi^M$, assuming the spin torque is given by Eq.~\eqref{eq:dark_mdm_torque}. Further simplifying using Eq.~\eqref{eq:massive_ampere}, we have 
\begin{equation}
\v{F} \simeq \frac{1}{f_\psi^M} \, (\hat{\v{s}} \cdot \nabla) \v{B}^V - \frac{1}{f_\psi^M} \, \hat{\v{s}} \times (m^2 \v{V}) \sim \frac{\sqrt{\rho_\DM}}{f_\psi^M} \, m,
\end{equation}
where the first term is suppressed by $v_\DM^2$, while the more subtle second term is not velocity suppressed at all. (The $v_\psi$ terms we dropped earlier turn out to be suppressed by $v_\DM v_\psi$.) The leading force has the same scaling as for the axion-fermion coupling. 

If a dark sector generates a dark MDM term, it also generically generates a kinetic mixing with the photon, whose effect can dominate over the dark MDM. This can be mitigated in appropriate models. For example, Ref.~\cite{Dobrescu:2004wz} considers a dark sector where kinetic mixing cannot be generated at one loop, while Ref.~\cite{Barducci:2021egn} considers one where the one-loop contributions to kinetic mixing cancel. However, Ref.~\cite{Krnjaic:2022wor} argued that even if one begins with no kinetic mixing, loops of the charged fermion $\psi$ generate 
\begin{equation} \label{eq:induced_kinetic_mixing}
\epsilon \sim \frac{q_\psi}{4 \pi^2} \frac{m_\psi}{f_\psi^M} \sim 10^{-15} \, \frac{10^{10} \, \mathrm{GeV}}{f_e^M}
\end{equation}
up to a logarithmic factor. In the second step, we let $\psi$ be the electron and normalized to the astrophysical limit. This induced $\epsilon$ is comparable to the reach of haloscopes searching for dark photons~\cite{Berlin:2024pzi}, which suggests that one should generically consider the effects of the dark MDM and kinetic mixing terms together. (When they occur together, an ordinary MDM is also generated, but this effect is negligible.) One could further map out the parameter space, but since our main focus here is axial vector DM, we now proceed to the axial-photon coupling. 

%!TEX root = main.tex

\section{Overview of the Axial-Photon Coupling}
\label{sec:axial_photon}

We now turn to the axial-photon coupling, defined in Eq.~\eqref{eq:axial_photon_coupling}. If we restrict to a relativistic longitudinal vector, and map it to an axion via $V^\mu \to \del^\mu a / m$, the axial-photon coupling becomes the axion-photon coupling with $1/f_\gamma \to g_{a\gamma\gamma}$. Since axion DM yields an effective current $\v{J}_{\mathrm{eff}} = g_{a\gamma\gamma} (\v{E} \times \nabla a - \dot{a} \v{B})$, we naively expect that for axial vector DM, 
\begin{equation} \label{eq:axial_photon_current}
\v{J}_{\mathrm{eff}} = - \frac{m}{f_\gamma} \, (\v{E} \times \v{V} + V^0 \v{B}).
\end{equation}
Most laboratory haloscopes use static magnetic background fields, but for axial vector DM this effect is velocity-suppressed. Instead, as we will see in Sec.~\ref{sec:axial_photon_pheno}, the strongest probes involve radiation fields, as these have $E_0 \sim B_0$. 

However, the phenomenology of the axial-photon coupling is far richer than this single change. For an axion, one can convert $\epsilon^{\mu\nu\rho\sigma} (\del_\mu a) A_\nu F_{\rho\sigma}$ into the manifestly gauge invariant $(-1/2) a \epsilon^{\mu\nu\rho\sigma} F_{\mu\nu} F_{\rho\sigma}$ by integrating by parts, but this is not possible for an axial vector. Instead, the axial-photon coupling is \textit{not} gauge invariant, so for consistency one must include a photon mass $m_\gamma$. This introduces effects associated with background vector potentials and longitudinal photons, along with constraints on the photon mass.

In App.~\ref{app:axial_photon}, we present a simple UV completion which generates the axial-photon coupling by integrating out chiral fermions with nontrivial $U(1)_D U(1)_{\mathrm{EM}}^2$ anomaly cancellation. (If $U(1)_D$ were a global $U(1)_{\mathrm{PQ}}$ symmetry, instead of a gauge symmetry, these triangle diagrams would generate the usual axion-photon coupling.) We exhibit model parameters which permit photon vortex formation, implying that $m_\gamma = 10^{-15} \, \mathrm{eV}$ is allowed by existing bounds.

The effective theory with the axial-photon coupling and the fermions integrated out can be perturbative up to high scales, and the fermions can be taken heavier than laboratory and astrophysical scales. This renders the experimental signatures of the axial-photon coupling largely independent of the UV completion. In this section, we will compute the equations of motion and the full effective current, then categorize the new phenomena that can arise.

\paragraph{Equations of Motion.} The Lagrangian for a massive axial vector and massive photon is
\begin{equation} \label{eq:axial_vector_two_photon}
\mathcal{L}[A, V] = - \frac14 F_{\mu\nu}^2 - \frac14 (F_{\mu\nu}^V)^2 + \frac12 m^2 V_\mu V^\mu - \frac{g}{4} \, V_\mu K^\mu + \frac12 m_\gamma^2 A_\mu A^\mu
\end{equation}
where we defined the dimensionless coupling $g = m / f_\gamma$. We will first derive the equations of motion for the axial vector and photon, then show how their longitudinal modes are sourced.

First, consider the equation of motion for the axial vector $V^\mu$, 
\begin{equation}
\del_\mu F_V^{\mu\nu} + m^2 V^\nu = \frac{g}{4} \, K^\nu. \label{eq:V_vec_eom} 
\end{equation}
Taking the divergence of Eq.~\eqref{eq:V_vec_eom} and plugging it back into its left-hand side gives
\begin{equation} \label{eq:vector_eom_2}
(\del^2 + m^2) V^\nu = \frac{g}{4} \, K^\nu + \frac{g}{4m^2} \, \del^\nu (\del_\mu K^\mu).
\end{equation}
In components, the Chern--Simons current $K^\mu = \epsilon^{\mu\nu\rho\sigma} A_\nu F_{\rho\sigma}$ obeys
\begin{subequations}
\begin{align}
K^\mu &= 2 (\v{B} \cdot \v{A}, \v{E} \times \v{A} + A^0 \v{B}), \\
\del_\mu K^\mu &= - 4 \v{E} \cdot \v{B}.
\end{align}
\end{subequations}
To interpret these results, we note the second term in Eq.~\eqref{eq:vector_eom_2} describes sourcing of longitudinal vectors, and if we identify $V^\mu \to \del^\mu a / m$ and $g/m \to g_{a\gamma\gamma}$, we simply recover the source term for an axion. More generally, longitudinal vectors are sourced by the divergence $\del_\mu K^\mu$, and the longitudinal amplitude is enhanced by powers of $1/m$.

To compute the photon's equation of motion, note that $K^\mu$ depends on the photon field in two places, yielding two source terms, 
\begin{equation} \label{eq:em_modified_eom}
\del_\mu F^{\mu\nu} + m_\gamma^2 A^\nu = \frac{g}{4} \, \epsilon^{\mu\nu\rho\sigma} (2 V_\mu F_{\rho\sigma} + A_\sigma F^V_{\rho \mu}) \equiv J^\nu_{\mathrm{eff}}. 
\end{equation}
The components of the effective current $J^\mu_{\mathrm{eff}} = (\rho_{\mathrm{eff}}, \v{J}_{\mathrm{eff}})$ are
\begin{subequations}
\begin{align}
\rho_{\mathrm{eff}} &= - g \left( \v{B} \cdot \v{V} - \frac12 (\nabla \times \v{V}) \cdot \v{A} \right), \\
\v{J}_{\mathrm{eff}} &= - g \left( \v{E} \times \v{V} + V^0 \v{B} + \frac{1}{2} (\dot{\v{V}} + \nabla V^0) \times \v{A} - \frac12 A^0 (\nabla \times \v{V}) \right).
\end{align}
\end{subequations}
This differs from the naive Eq.~\eqref{eq:axial_photon_current} by terms which depend explicitly on the scalar and vector potential, though these terms vanish for the vector's longitudinal mode $V^\mu \to \del^\mu \alpha_V / m$. 

Taking the divergence of Eq.~\eqref{eq:em_modified_eom} gives $m_\gamma^2 \del_\mu A^\mu = \del_\mu J_{\mathrm{eff}}^\mu$, and plugging it back in gives
\begin{align} \label{eq:divergent_source_new}
(\del^2 + m_\gamma^2) A^\nu = J_{\mathrm{eff}}^\nu + \frac{1}{m_\gamma^2} \, \del^\nu (\del_\mu J^\mu_{\mathrm{eff}}). 
\end{align}
The divergence $\del_\mu J_{\mathrm{eff}}^\mu$ sources the longitudinal photon mode, enhanced by powers of $1/m_\gamma$. Explicitly, we have
\begin{equation} \label{eq:div_jeff}
\del_\mu J_{\mathrm{eff}}^\mu = - \frac{g}{2} \, \v{B} \cdot (\dot{\v{V}} + \nabla V^0) + \frac{g}{2} \, \v{E} \cdot (\nabla \times \v{V}).
\end{equation}
The sourcing of longitudinal modes is a manifestation of the underlying nontrivial $U(1)_D U(1)_{\mathrm{EM}}^2$ anomaly structure, discussed in App.~\ref{app:photon_model}. Since these processes are enhanced by $1/m$ and $1/m_\gamma$, and both $m$ and $m_\gamma$ can be extremely small, we will have to carefully track them.

\paragraph{Longitudinal Photons.} To build intuition, we now briefly review the properties of longitudinal photons. The free solutions $A^\mu = \epsilon^\mu e^{i k \cdot x}$ satisfy $k \cdot \epsilon = 0$. Choosing $k^\mu = (\omega, 0, 0, k)$ for concreteness, the unit-normalized polarizations are 
\begin{equation} \label{eq:epsilon_solutions}
\epsilon^\mu_x = (0, 1, 0, 0), \qquad \epsilon^\mu_y = (0, 0, 1, 0), \qquad \epsilon^\mu_L = (k/m_\gamma, 0, 0, \omega/m_\gamma).
\end{equation}
In the relativistic limit $\omega \simeq k \gg m_\gamma$, these solutions all have the same mean energy density 
\begin{equation} \label{eq:vacuum_energy_density}
T^{00} = \frac12 (|\v{E}|^2 + |\v{B}|^2) + \frac12 m_\gamma^2 (A_0^2 + |\v{A}|^2).
\end{equation}
For the transverse modes, the energy is mostly in the transverse electric and magnetic field. By contrast, the longitudinal polarization $\epsilon^\mu_L$ has entries enhanced by $\omega/m_\gamma$, so that its sourcing can be enhanced, but it corresponds to a solution with zero magnetic field, and a suppressed longitudinal electric field $E^z \propto \omega \epsilon^z_L - k \epsilon^0_L = (\omega^2 - k^2) / m_\gamma = m_\gamma$. Instead, most of the energy is due to the potentials. In general, the electric field scales with energy density as 
\begin{equation} \label{eq:basic_scalings}
|\v{E}| \sim \begin{cases} \sqrt{T^{00}} & \text{transverse photons,} \\ \sqrt{T^{00}} \, (m_\gamma/\omega) & \text{longitudinal photons}. \end{cases}
\end{equation}
These results show that, compared to transverse photons, relativistic longitudinal photons have a potential enhanced by $\omega/m_\gamma$ but a field strength suppressed by $m_\gamma/\omega$. Since the axial-photon coupling directly depends on the potential, longitudinal photons are often dominantly produced. However, transverse photons can still yield the dominant signals in laboratory experiments, since ordinary detectors couple to the field strength. Thus, in general we must track both the longitudinal and transverse photon modes. 

Above we assumed the photons were in vacuum; in App.~\ref{app:photon_dof} we show that in a plasma there is a fourth degree of freedom, the longitudinal plasmon, for which the polarization is not enhanced and the electric field is not suppressed. However, even in plasma, a longitudinal photon mode continues to exist, with the same properties discussed above.  

For a complementary perspective, we can expand Eq.~\eqref{eq:em_modified_eom} in components, giving
\begin{subequations}
\begin{align}
\nabla \cdot \v{E} + m_\gamma^2 A^0 &= \rho_{\mathrm{eff}} \label{eq:modified_gauss} \\
\nabla \times \v{B} - \dot{\v{E}} + m_\gamma^2 \v{A} &= \v{J}_{\mathrm{eff}} \label{eq:modified_ampere}, \\
m_\gamma^2 (\dot{A}^0 + \nabla \cdot \v{A}) &= \dot{\rho}_{\mathrm{eff}} + \nabla \cdot \v{J}_{\mathrm{eff}}.
\end{align}
\end{subequations}
The final equation indicates that when the effective current is not conserved, the potential $A^\mu$ is enhanced by $1/m_\gamma^2$ due to the longitudinal photon. However, since the potential enters with a factor of $m_\gamma^2$ in the first two equations, the fields $\v{E}$ and $\v{B}$ are not enhanced as $m_\gamma \to 0$, consistent with the discussion above. 

\paragraph{Relativistic Longitudinal Modes.} The axial-photon coupling simplifies when any one of the particles involved is a relativistic longitudinal vector. First, if the axial vector $V^\mu$ is relativistic and longitudinal, $V_\mu \to \del_\mu \alpha_V / m$, the axial-photon coupling becomes 
\begin{equation} \label{eq:alpha_V_simp}
\mathcal{L}_{\mathrm{int}} \simeq \frac{\alpha_V}{8 f_\gamma} \, \epsilon^{\mu\nu\rho\sigma} F_{\mu\nu} F_{\rho\sigma},
\end{equation}
which, as already mentioned, is simply the axion-photon interaction with $g_{a\gamma\gamma} \lra 1/f_\gamma$. Since the field strength of a longitudinal photon is suppressed by $m_\gamma / \omega$, this term primarily couples $\alpha_V$ to two transverse photons. Thus, many results involving relativistic axion production, e.g.~in astrophysical systems, can be directly translated to longitudinal vectors. 

Second, if a photon is relativistic and longitudinal, $A_\nu \to \del_\nu \alpha / m_\gamma$, we have
\begin{equation} \label{eq:alpha_simp}
\mathcal{L}_{\mathrm{int}} \simeq - \frac{1}{f_\gamma} \frac{m}{m_\gamma} \frac{\alpha}{8} \, \epsilon^{\mu\nu\rho\sigma} F^V_{\mu\nu} F_{\rho\sigma}.
\end{equation}
This is the ``dark axion portal'' introduced in Ref.~\cite{Kaneta:2016wvf}, where the longitudinal photon and axial vector play the role of the axion and dark photon, respectively. Compared to Eq.~\eqref{eq:alpha_V_simp}, matrix elements involving this interaction are enhanced by $m/m_\gamma$, and thus dominate for $m \gg m_\gamma$. But again, since the field strength of a longitudinal photon is suppressed by $m_\gamma/\omega$, the signal in a laboratory detector is still often dominated by transverse photons.

As we will see below, processes with both a longitudinal vector and photon are naively enhanced by $1/(m_\gamma m)$, but are actually highly suppressed. Heuristically, this is because the axial-photon coupling vanishes if one substitutes both $V_\mu \to \del_\mu \alpha_V / m$ and $A_\nu \to \del_\nu \alpha / m_\gamma$.

\paragraph{Vector Decay.} The Landau--Yang theorem~\cite{PhysRev.77.242} states that a massive spin $1$ particle cannot decay into two massless transverse photons. Since it is derived from kinematics and symmetry, it holds regardless of the decay interaction. The axial-photon coupling obeys this result; moreover, it forbids decay to two transverse photons even when the photons are massive.

To see this, note that the amplitude for a vector, with momentum $q$ and polarization $\epsilon^{(q)}$, to decay to two photons with momenta $k^{(i)}$ and polarizations $\epsilon^{(i)}$ is 
\begin{equation} \label{eq:matrix_elt_vector}
\mathcal{M}_V \sim g \, \epsilon^{\mu\nu\rho\sigma} \epsilon^{(q)}_\mu (k^{(1)} - k^{(2)})_\nu \epsilon^{(1)}_\rho \epsilon^{(2)}_\sigma.
\end{equation}
If the photons are both transverse, this amplitude vanishes because it is the wedge product of four vectors which are all purely spatial in the vector's frame. 

However, there is a subtlety: in Eq.~\eqref{eq:alpha_V_simp} we saw that for a relativistic longitudinal vector, the axial-photon coupling is equivalent to the axion-photon coupling with $g_{a\gamma\gamma} \lra 1/f_\gamma$. But the axion-photon coupling \textit{does} allow axions to decay to two photons, with amplitude
\begin{equation} \label{eq:matrix_elt_axion}
\mathcal{M}_a \sim g_{a\gamma\gamma} \, \epsilon^{\mu\nu\rho\sigma} q_\mu (k^{(1)} - k^{(2)})_\nu \epsilon^{(1)}_\rho \epsilon^{(2)}_\sigma \sim m g.
\end{equation}
This is an apparent failure of the Goldstone boson equivalence theorem. 

We can resolve the puzzle by tracking error terms more carefully. For a vector with $q^\mu = (\omega, 0, 0, k)$, the longitudinal polarization is $\epsilon^{(q)} = (k, 0, 0, \omega) / m = q^\mu/m + \mathcal{O}(m/\omega)$, which approaches $q^\mu/m$ in the relativistic limit. Defining $\mathcal{M}_V = \epsilon_\mu^{(q)} \mathcal{M}_V^\mu$, the corresponding axion decay amplitude is $\mathcal{M}_a = (q_\mu/m) \mathcal{M}^\mu_V$. Therefore, in general we can conclude 
\begin{equation}
\mathcal{M}_V = \mathcal{M}_a + (\epsilon^{(q)}_\mu - q_\mu/m) \mathcal{M}^\mu_V.
\end{equation}
In most applications of the equivalence theorem, the error term is negligible, so that $\mathcal{M}_V \simeq \mathcal{M}_a$. However, for the decay process above we have $\mathcal{M}^\mu_V \sim \omega g$, which implies that the error term is the same order as $\mathcal{M}_a$ itself. This is why it is consistent to have $\mathcal{M}_V = 0$. 

However, nothing forbids a vector from decaying into a longitudinal photon and a transverse photon. For $m \gg m_\gamma$, the amplitude in Eq.~\eqref{eq:matrix_elt_vector} scales as $\mathcal{M}_V \sim g m^2 / m_\gamma$. This scaling can also be seen directly from Eq.~\eqref{eq:alpha_simp}, and it is enhanced by $m/m_\gamma$ relative to the axion decay amplitude; we consider this decay in Sec.~\ref{sec:cooling}. More generally, the Landau--Yang theorem does not prevent the axial-photon coupling from having physical effects. 

%!TEX root = main.tex

\section{Phenomenology of the Axial-Photon Coupling}
\label{sec:axial_photon_pheno}

\begin{figure} 
\centering
\includegraphics[width=\textwidth]{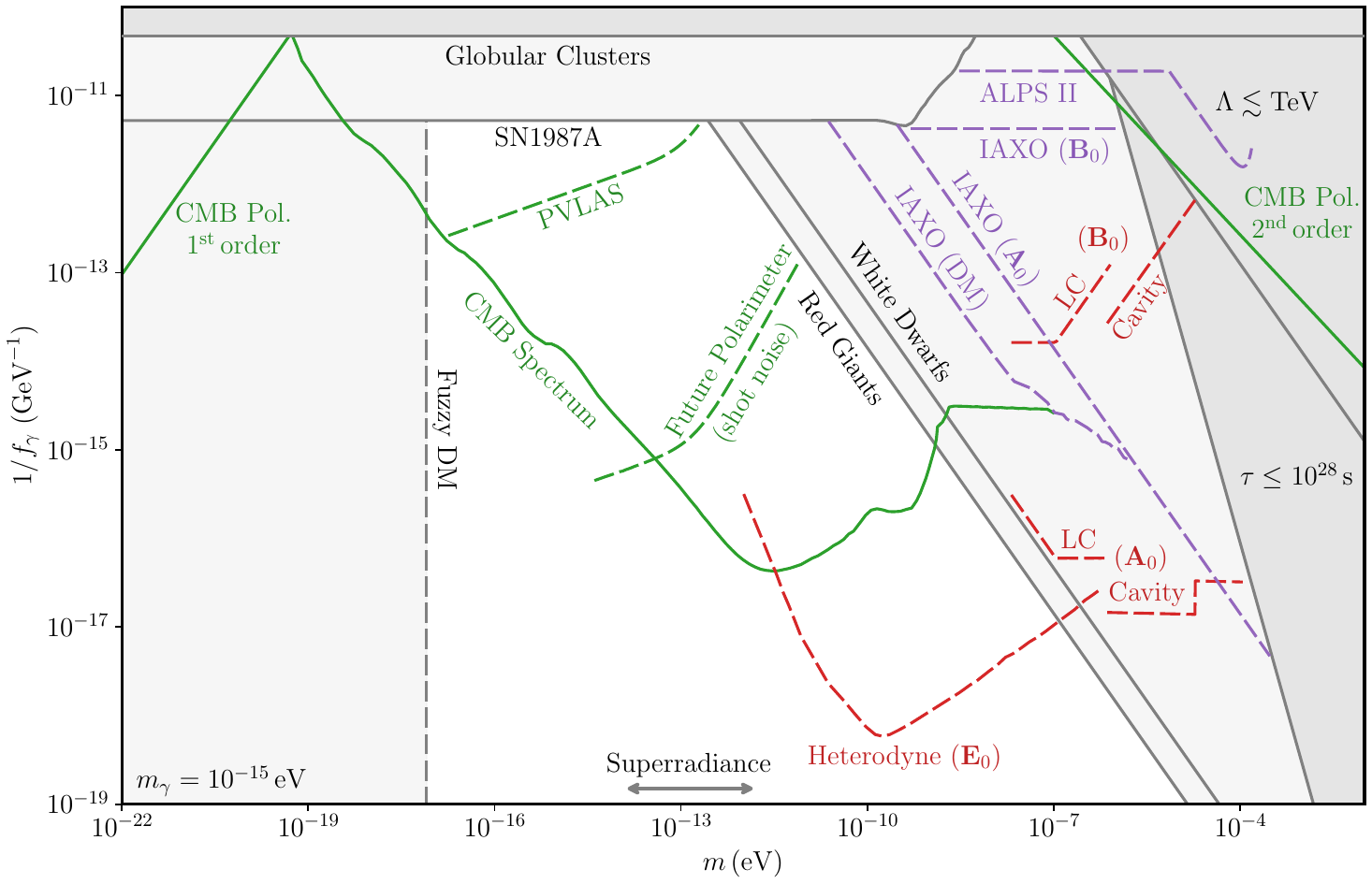}
\caption{Current constraints (solid) and projected future sensitivity (dashed) to the axial-photon coupling, for a benchmark photon mass $m_\gamma = 10^{-15} \, \mathrm{eV}$. DM effects on photon propagation (Sec.~\ref{sec:propagation}) are green, laboratory haloscopes which convert DM to photons (Sec.~\ref{sec:haloscopes}) are red, astrophysical bounds (Sec.~\ref{sec:cooling}) are gray, helioscopes and light shining through walls experiments (Sec.~\ref{sec:rel_mix}) are purple, and the $\Lambda \lesssim \mathrm{TeV}$ region is derived from perturbativity arguments in Eq.~\eqref{eq:axial_photon_lambda}. For clarity, we have cut off some curves when they become weaker than astrophysical bounds; the purple curves all extend straight to arbitrarily low $m$.}
\label{fig:axial_photon}
\end{figure}

We now investigate the observable effects of the axial-photon coupling. In Sec.~\ref{sec:propagation} we consider the effect of DM on photon propagation. We find two strong signatures: the polarization rotation in an ordinary Fabry--Perot cavity is enhanced over many reflections, and CMB spectral distortion is enhanced by $1/m_\gamma$. In Sec.~\ref{sec:haloscopes} we consider conversion of DM to photons in a laboratory haloscope. Here the best sensitivity comes from ``heterodyne'' experiments with superconducting cavities, as they contain strong electric background fields. 

We then consider effects that are largely independent of the DM density. In Sec.~\ref{sec:cooling} we consider astrophysical sources of relativistic vectors and longitudinal photons, the associated cooling bounds, and bounds from DM decay. In Sec.~\ref{sec:rel_mix} we consider the mixing of relativistic vectors and photons, in helioscopes and light shining through walls experiments. 

All main results are shown in Fig.~\ref{fig:axial_photon}, for a benchmark photon mass $m_\gamma = 10^{-15} \, \mathrm{eV}$. For lower photon masses, many of these curves remain the same, while others are enhanced by powers of $1/m_\gamma$. In Sec.~\ref{sec:discussion}, we show an analogous plot for the extreme case $m_\gamma = 10^{-26} \, \mathrm{eV}$.

\subsection{Polarization Rotation Effects}
\label{sec:propagation}

In App.~\ref{app:photon_prop}, we show that for a photon of angular frequency $\omega \gg m, m_\gamma, \omega_p$ propagating along the $z$-direction, the amplitudes of the transverse and longitudinal photon modes evolve as 
\begin{equation} \label{eq:amplitude_evolution_main}
\del_z \begin{pmatrix} \bar{A}^x \\ \bar{A}^y \\ \bar{\alpha} \end{pmatrix} \simeq \frac{g}{2} \begin{pmatrix} 0 & V^z - V^0 & - \dot{V}^y / 2 m_\gamma \\ V^0 - V^z & 0 & \dot{V}^x / 2 m_\gamma \\ \dot{V}^y / 2 m_\gamma & - \dot{V}^x / 2 m_\gamma & 0 \end{pmatrix} \begin{pmatrix} \bar{A}^x \\ \bar{A}^y \\ \bar{\alpha} \end{pmatrix} + \frac{i \omega_p^2}{2 \omega} \begin{pmatrix} 0 \\ 0 \\ \bar{\alpha} \end{pmatrix}
\end{equation}
where $\omega_p$ is the plasma frequency; the final term reflects the fact that the longitudinal photon is not affected by plasma. From here, we can read off the physical effects to be explored below. 

The main effect at $\mathcal{O}(g)$ is transverse polarization rotation,
\begin{equation} \label{eq:vector_pol_rot}
\frac{d\theta}{d z} = \frac{g}{2} (V^0 - V^z).
\end{equation}
This is precisely as expected from known results involving the Carroll--Field--Jackiw term~\cite{Carroll:1989vb}, or axion DM induced polarization rotation~\cite{Harari:1992ea}, 
\begin{equation} \label{eq:axion_pol_rot}
\frac{d\theta}{d z} = \frac{g_{a\gamma\gamma}}{2} (\del_t + \del_z) a = \frac{g_{a\gamma\gamma}}{2} \frac{Da}{Dz}.
\end{equation}
which is related to Eq.~\eqref{eq:vector_pol_rot} by mapping $g V^\mu \to g_{a\gamma\gamma} \, \del^\mu a$. The main difference compared to the axion DM case is that the dominant effect is due to $V^z = \v{V} \cdot \hat{\v{k}}$. As we will see, this will dramatically enhance the sensitivity of laboratory experiments. 

Rotation from a transverse photon to a longitudinal photon is enhanced by $m/m_\gamma$, but as discussed in Sec.~\ref{sec:axial_photon}, the longitudinal electric field has a stronger suppression, by $m_\gamma/\omega$. The dominant signatures of transverse-longitudinal rotation instead appear at $\mathcal{O}(g^2)$, as either a depletion of the transverse photons or a rotation of the transverse polarization. Even though the coupling $g$ is extremely small, these second-order effects can dominate due to a combination of enhancements. In addition to the enhancement by $1/m_\gamma$, transverse-longitudinal conversion can become resonant when $m \simeq \omega_p^2 / 2 \omega$, and second-order transverse polarization rotation need not average to zero over each DM oscillation, leading to a random walk enhancement. We discuss both of these effects in the context of CMB observations below. 

\paragraph{Laboratory Polarimetry.} One can search for the effect in Eq.~\eqref{eq:vector_pol_rot} by measuring the polarization rotation of light in an optical cavity. As a simple example, we consider a Fabry--Perot cavity of length $L$ with reflection coefficients $r_1, r_2 \simeq 1$ and finesse $\mathcal{F} \simeq \pi / (1 - r_1 r_2)$.

To see the difference compared to the axion case, consider the net polarization rotation of light that makes one round trip along the cavity. Ordinarily, the sign of $d\theta/dz$ changes upon reflection, but for an axial vector the sign of $\v{V} \cdot \hat{\v{k}}$ changes as well. Letting $V^z(t) = \text{Re}(V^z_0 e^{imt})$, and letting the initial time be $t_0$, one has a rotation of magnitude
\begin{equation}
\Delta \theta = \frac{g}{2} \int_0^{2L} V^z \, dz = \frac{V^z_0}{f_\gamma} \sin(m L) \cos(m (t_0 + L)).
\end{equation}
By contrast, the analogous calculation for axion DM with $a(t) = \text{Re}(a_0 e^{imt})$ gives 
\begin{align}
\Delta \theta &= \frac{g_{a\gamma\gamma}}{2} (- a(t_0) + 2 a(t_0 + L) - a(t_0 + 2L)) \\
&= 2 g_{a\gamma\gamma} a_0 \sin^2(m L/2) \cos(m (t_0 + L)).
\end{align}
where we used Eq.~\eqref{eq:axion_pol_rot}, which implies that the net polarization rotation in each half of the trip, up to a sign, is the difference in endpoint values. For small $m L$, the contributions of the two halves almost cancel, giving a small rotation.

In both cases, summing over all reflections enhances the rotation of the outgoing light by the usual Fabry--Perot factor $r_1 r_2 / |1 - r_1 r_2 e^{-2 i m L}|$. In the axion case one has~\cite{Nagano:2019rbw}
\begin{equation} \label{eq:delta_theta_a}
\Delta \theta_a = 2 g_{a\gamma\gamma} a_0 \, \frac{r_1 r_2 \sin^2(mL/2)}{|1 - r_1 r_2 e^{- 2 i m L}|} \sim g_{a\gamma\gamma} \sqrt{\rho_\DM} \times \begin{cases} m L^2 \mathcal{F} & m L \ll 1/\mathcal{F}, \\ L & 1/\mathcal{F} \ll mL \ll 1, \end{cases}
\end{equation}
while for an axial vector there is a full enhancement by the cavity finesse,  
\begin{equation} \label{eq:delta_theta_V}
\Delta \theta_V = \frac{V^z_0}{f_\gamma} \, \frac{r_1 r_2 \sin(mL)}{|1 - r_1 r_2 e^{- 2 i m L}|} \sim \frac{\sqrt{\rho_\DM}}{f_\gamma}\, L \mathcal{F} \text{ for } m L \ll 1/\mathcal{F}
\end{equation}
down to arbitrarily low $m$.

There are several proposals to search for axion DM using optical cavities~\cite{DeRocco:2018jwe,Obata:2018vvr,Liu:2018icu,Heinze:2024bdc}, and results from experimental prototypes~\cite{Oshima:2023csb,Heinze:2023nfb,Pandey:2024dcd}. These prototypes all used four-mirror geometries, designed to avoid the strong suppression of the axion sensitivity as $m \to 0$. This more complex geometry led to additional difficulties. For instance, since the light reflected far from normal incidence, small cavity imperfections led to a $\sim \mathrm{MHz}$ splitting between the resonant frequencies for the two polarizations, suppressing sensitivity for lower $m$. 

For axial vector DM, the signal strength is already maximized for an ordinary Fabry--Perot cavity, where these issues are largely avoided. We base our discussion on the PVLAS experiment~\cite{DellaValle:2015xxa}, which sought to measure vacuum magnetic birefringence. This experiment achieved $\mathcal{F} \simeq 7 \times 10^5$ in a cavity of length $L \simeq 3.3 \, \mathrm{m}$, and the splitting of the polarization modes due to mirror birefringence was $\sim 10 \, \mathrm{Hz}$, comparable to the width of the cavity modes. 

The run measuring polarization rotation reported a noise spectral density of $S_\theta = 4.8 \times 10^{-7} \, \mathrm{rad} / \sqrt{\mathrm{Hz}}$ at the modulation frequency $10 \, \mathrm{Hz}$, $\sim 10^2$ times above the shot noise level. A later improved measurement~\cite{Zavattini:2018pbw} traced the noise to birefringence fluctuations in the mirror coatings, also discussed in Refs.~\cite{Hartman:2018svc,Michimura:2023ygm}. In particular, Ref.~\cite{Zavattini:2018pbw} measured rotation noise as a function of frequency, finding $S_\theta = 3.2 \times 10^{-8} \, \mathrm{rad} / \sqrt{\mathrm{Hz}}$ for $f \gtrsim 50 \, \mathrm{Hz}$, rising to $10$ times this level at a few Hz. Measurements were not performed at very low frequency.

To roughly estimate the potential reach of PVLAS, we take a flat $S_\theta = 10^{-8} / \sqrt{\mathrm{Hz}}$ and set
\begin{equation}
\mathrm{SNR} \sim \frac{\Delta \theta_V}{S_\theta} \left( \frac{t_{\mathrm{int}}}{\Delta \omega_{\mathrm{sig}}} \right)^{1/4} \sim 1
\end{equation}
assuming an integration time $t_{\mathrm{int}} = 1 \, \mathrm{year}$, and signal width $\Delta \omega_{\mathrm{sig}} = m / Q_\DM$ with $Q_\DM = 10^6$, cutting off the reach when the DM can no longer be resolved, $t_{\mathrm{int}} \, \Delta \omega_{\mathrm{sig}} \sim 1$. The resulting reach is shown in Fig.~\ref{fig:axial_photon}. 

It would be interesting to explore how one can improve the sensitivity. For shot noise limited operation, the SNR is proportional to $L \mathcal{F} \sqrt{P_{\mathrm{in}}} / f_\gamma$ at low $m$, and since no magnet is required, the length $L$ could be significantly increased. As a benchmark, advanced LIGO has a comparable $L \mathcal{F}$ to PVLAS, and a much larger $P_{\mathrm{in}}$. Though it does not currently measure polarization rotation, its potential reach to axion DM was estimated in Ref.~\cite{Nagano:2019rbw} assuming shot noise limited operation. By comparing Eqs.~\eqref{eq:delta_theta_a} and~\eqref{eq:delta_theta_V}, we can recast this result to axial vector DM by $1/f_\gamma \lra g_{a\gamma\gamma} \tan(mL/2)$, yielding the ``Future Polarimeter'' curve in Fig.~\ref{fig:axial_photon}. 

For a LIGO-scale polarimeter it is not clear what the dominant technical noise sources would be at low frequencies; for instance, LIGO's mirrors are made of fused silica to avoid birefringence noise. In lieu of a full analysis, we simply cut off the recast reach below $1 \, \mathrm{Hz}$. 

\paragraph{CMB Polarization.} The axial-photon coupling can also be probed by CMB polarization measurements. Here we recast the results of Ref.~\cite{Fedderke:2019ajk}, which considered axion DM. 

Referring to Eq.~\eqref{eq:axion_pol_rot}, oscillations of $a(t_f, z_f)$ during a CMB experiment yield an ``AC polarization rotation'' of order $g_{a\gamma\gamma} \sqrt{\rho_\DM}/m$ which can be measured with precision $\Delta \theta \sim 10^{-3}$. Oscillations of $a(t_i, z_i)$ during the CMB's formation yield a ``polarization washout''. Planck is sensitive to a percent-level washout, corresponding only to $\Delta \theta \sim 10^{-1}$, but this signal is slightly stronger than AC polarization rotation due to the higher DM density during recombination, 
\begin{equation}
\rho_\DM^\mathrm{rec} \sim \Omega_\DM \rho_{c, 0} (1 + z_{\mathrm{rec}})^3 \sim 2 \times 10^3 \, \mathrm{GeV}/\mathrm{cm}^3,
\end{equation}
since $\rho_{c,0} = 5 \times 10^{-6} \, \mathrm{GeV}/\mathrm{cm}^3$ and $z_{\mathrm{rec}} \sim 10^3$. We thus focus on the polarization washout. 

Referring to Eq.~\eqref{eq:vector_pol_rot}, the first-order polarization rotation is parametrically the same as that of the axion with $1/f_\gamma \lra g_{a\gamma\gamma}$. More precisely, the sensitivity should be $\mathcal{O}(1)$ different because only the component $\hat{\v{k}} \cdot \v{V}$ along the photon's direction of travel matters. This implies an order-one reduction of the magnitude of the effect, but it also means the AC polarization rotation effect will have a dipole pattern on the sky, which may help distinguish it from noise. As a rough estimate, we simply rescale the axion polarization washout result by $1/f_\gamma \lra g_{a\gamma\gamma}$, yielding the ``$1^\mathrm{st}$ order'' curve in Fig.~\ref{fig:axial_photon}. 

There is one subtlety worth noting in the recast. For axion DM, the total polarization rotation only depends on the difference of the initial and final values of the axion field; there is no $\sqrt{N}$ ``random walk'' enhancement as the light travels over $N$ DM coherence lengths. Instead, the rotation simply cancels over multiple axion oscillations, so that the sensitivity of CMB observations decreases rapidly as $m$ increases. 

For axial vector DM, crossing over a DM coherence length can cause both the polarization and magnitude of $\v{V}$ to change by $\mathcal{O}(1)$. However, there is still no random walk enhancement, because we can write the total rotation angle as an integral with a smooth, rapidly oscillating integrand (see appendix C of Ref.~\cite{Hook:2023smg}). By the principle of stationary phase, we can approximately express the integral in terms of endpoint contributions, which are independent of $N$. Thus, there are no random walk enhancements at $\mathcal{O}(g)$ for either axion or axial vector DM, making the recast straightforward, but we will see that they do appear at $\mathcal{O}(g^2)$.

\paragraph{CMB Spectral Distortion.} At $\mathcal{O}(g^2)$, there is a decrease in the transverse photon amplitude due to conversion to longitudinal photons, which would affect measurements of the CMB spectrum. This effect can be inferred by computing the longitudinal polarization amplitude to $\mathcal{O}(g)$, which is resonantly enhanced if $m \simeq \omega_p^2 / 2 \omega$ at some point after CMB production. 

This signal was studied in Ref.~\cite{Hook:2023smg} in the context of the dark axion portal Eq.~\eqref{eq:alpha_simp}, where the dark photon and axion play the role of the axial vector and longitudinal photon, respectively. We thus recast the result in the top-left panel of Fig.~5 of Ref.~\cite{Hook:2023smg}. That figure assumes the axion is the DM; to convert to the case where the dark photon is the DM, we weaken the bounds by a factor of $\sqrt{3}$. The result is labeled as ``CMB Spectrum'' in Fig.~\ref{fig:axial_photon}.\footnote{There is a distinct CMB spectrum effect due to stimulated decay of axial vector DM, into a transverse and longitudinal photon. This effect was studied in Ref.~\cite{Arias:2020tzl}, also in the context of the dark axion portal. It is relevant for $m \gtrsim 10^{-4} \, \mathrm{eV}$, but ultimately weaker than the astrophysical bounds we consider in Sec.~\ref{sec:cooling}.}

\paragraph{Second-Order Polarization Rotation.} The net second-order polarization rotation due to axial vector DM can build up over many DM oscillations. This enhancement requires the axial vector to have nonzero ellipticity, and occurs because three-dimensional rotations do not commute.\footnote{Recently, Ref.~\cite{Berlin:2026sav} showed that at second order, an elliptically polarized $U(1)_{B-L}$ vector has an enhanced amplitude to flip a neutrino's helicity. This enhancement has a similar origin to the one discussed here.} Note that even if axial vector DM is produced with linear polarization, gravitational self-interactions tend to drive the polarizations to equipartition~\cite{Amaral:2024tjg}, so we generically expect DM to have nonzero ellipticity in the galaxy. 

The dominant second-order effect is associated with rotation from a transverse photon to a longitudinal photon and back to a transverse photon, since this process is enhanced by $1/m_\gamma^2$. We show in App.~\ref{app:propagation} that it yields a transverse polarization rotation 
\begin{equation} \label{eq:second_order_rotation}
\Delta \theta \supset \frac{g^2 m}{32 m_\gamma^2} \int_0^\tau dt \, V^x_0(t) V^y_0(t) \sin \phi(t)
\end{equation}
where $\tau$ is the light's travel time, the $V_0^i$ are the oscillation amplitudes of the components $V^i$ along its path, $\phi$ is the phase difference between $V^x$ and $V^y$, and we neglect the effect of $\omega_p$. Within a DM coherence length, the $V_0^i$ and $\phi$ are approximately constant, and the angle grows proportionally to $\tau$. As the light travels across many DM coherence lengths, their contributions add independently as a random walk, so the rotation angle scales with $\sqrt{\tau}$. 

Assuming $\tau \gtrsim \tau_c$, where $\tau_c$ is the propagation time through a DM coherence length, the total rms rotation angle scales as 
\begin{equation}
\Delta \theta_{\mathrm{rms}} \sim \frac{g^2}{32 m_\gamma^2} \sqrt{\langle (V^x_0)^2 (V^y_0)^2 \sin^2 \phi \rangle} \, m \tau_c \, \sqrt{\frac{\tau}{\tau_c}}. \label{eq:general_second_order}
\end{equation}
Because of this enhancement with $\tau$, the second-order polarization rotation can dominate the first-order polarization rotation for CMB photons. 

Assuming the DM is randomly polarized at recombination, we have $\langle (V_0^i)^2 \rangle = 2 \rho_\DM / 3m^2$ and $\langle \sin^2 \phi \rangle = 1/2$. To be conservative, we will only consider one $e$-fold during recombination, so that $\tau \sim 1 / H_{\mathrm{rec}} \sim 1 / (3 \times 10^{-29} \, \mathrm{eV})$. Standard dark photon production mechanisms, such as that of Ref.~\cite{Graham:2015rva}, produce DM with $v \sim 1$ when $H \sim m$. Taking this as the standard, the small-scale velocity dispersion at recombination is $v_{\mathrm{rec}} \sim \sqrt{H_{\mathrm{rec}}/m}$. This corresponds to a spatial coherence length $\tau_c \sim 1/(mv_{\mathrm{rec}}) \sim 1/\sqrt{H_{\mathrm{rec}} \, m}$, which implies
\begin{equation}
\Delta \theta_{\mathrm{rms}} \sim \frac{1}{48 \sqrt{2}} \frac{\rho_{\DM}^{\mathrm{rec}}}{m_\gamma^2 f_\gamma^2} \left( \frac{m}{H_{\mathrm{rec}}} \right)^{3/4}. 
\end{equation}
Remarkably, this effect gets stronger as $m$ increases, so it greatly dominates over the first-order polarization rotation above ``fuzzy'' DM masses. We conservatively assume it can only be observed as a polarization washout, and set a constraint by imposing $\Delta \theta_{\mathrm{rms}} \lesssim 10^{-1}$, to yield the ``$2^\mathrm{nd}$ order'' curve in Fig.~\ref{fig:axial_photon}. 

This curve is visible on the plot for $m \gtrsim 10^{-7} \, \mathrm{eV}$, and in this regime the DM oscillation is fast enough that the effect of $\omega_p$ is negligible. At much lower values of $m$ the second-order polarization rotation effect can be resonantly enhanced, when $m \simeq \omega_p^2 / 2 \omega$. Moreover, due to the complex phases introduced by $\omega_p$ in Eq.~\eqref{eq:amplitude_evolution_main}, the second-order effect can also induce circular polarization (Stokes $V$) for CMB photons. It would be interesting to investigate these signatures further, though the CMB spectral distortion effect may still dominate, as the CMB's spectrum can be measured more precisely than its polarization. 

\subsection{Laboratory Haloscopes}
\label{sec:haloscopes}

To analyze haloscopes, we focus on the effective current $\v{J}_{\mathrm{eff}}$, because a charge density $\bar{\rho}_{\mathrm{eff}}$ cannot ring up resonant cavity modes~\cite{condon1941forced}. In the presence of background fields $(\v{B}_0, \v{E}_0)$ and potentials $(A_0^0, \v{A}_0)$, the leading contributions to the effective current are
\begin{equation} \label{eq:jbar_full_expr}
\v{J}_{\mathrm{eff}} \simeq - g \left( V^0 \v{B}_0 + \v{E}_0 \times \v{V} + \frac{1}{2} (\dot{\v{V}} + \nabla V^0) \times \v{A}_0 - \frac12 A_0^0 (\nabla \times \v{V}) \right).
\end{equation}
The first two terms, already discussed around Eq.~\eqref{eq:axial_photon_current}, depend on the background field strengths, while the next two terms depend on background potentials. Here we focus on laboratory haloscopes, though we briefly discuss DM to photon conversion on astrophysical scales in App.~\ref{app:photon_prop}. We now discuss each contribution to the effective current in turn. 

\paragraph{Background Magnetic Fields.} Axion haloscopes usually use a large background magnetic field $B \sim 10 \, \mathrm{T}$, because realistic background electric fields are smaller, and electric field effects are suppressed by $v_\DM$ for axion DM. But for axial vector DM, the situation is reversed: instead the magnetic field effect $V^0 \v{B}_0 \sim v_\DM |\v{V}| B_0$ is velocity suppressed. The signal power in a cavity haloscope on resonance, with volume $V_c$ and quality factor $Q_c \lesssim Q_\DM$, is then
\begin{equation} \label{eq:haloscope_signal}
P_{\mathrm{sig}} \sim \frac{\rho_\DM Q_c V_c}{f_\gamma^2 m} \, \times \begin{cases} B_0^2 v_\DM^2 C_B^2 & \text{magnetic\ background}, \\ E_0^2 C_E^2 & \text{electric\ background}, \end{cases}
\end{equation}
where the form factors for a cavity signal mode $n$ in a magnetic or electric background are
\begin{subequations}
\begin{align}
C_B &= \bigg| \int_{V_c} \tilde{\v{E}}_n \cdot \tilde{\v{B}}_0 \bigg|, \\
C_E &= \bigg| \int_{V_c} \tilde{\v{E}}_n \cdot (\tilde{\v{E}}_0 \times \hat{\v{V}}) \bigg| = \bigg| \hat{\v{V}} \cdot \int_{V_c} \tilde{\v{E}}_n \times \tilde{\v{E}}_0 \bigg|.
\end{align}
\end{subequations}
We normalized the profiles as $\int_{V_c} |\tilde{\v{E}}_n|^2 = \int_{V_c} |\tilde{\v{E}}_0|^2 = \int_{V_c} |\tilde{\v{B}}_0|^2 = 1$, so that these form factors are at most $\mathcal{O}(1)$. We have neglected DM spatial gradients, as the DM wavelength is much larger than the length scale $L$ of a cavity haloscope. For a more accurate treatment, one should average over the unknown polarization of $\v{V}$ in $C_E$. As discussed in the paragraph above Eq.~\eqref{eq:second_order_rotation}, we generically expect vector DM in the galaxy to be randomly polarized. 

From the first line of Eq.~\eqref{eq:haloscope_signal}, the sensitivity of a cavity haloscope experiment with a magnetic background field can be roughly recast by $v_\DM / f_\gamma \lra g_{a\gamma\gamma}$. As a concrete example, we show the future ADMX projection from Ref.~\cite{Stern:2016bbw} in Fig.~\ref{fig:axial_photon}. Similar logic applies to lumped element (LC circuit) experiments, targeting $m L \ll 1$. In these experiments the DM effective current is always in the magnetoquasistatic regime, so the signal power is given parametrically by Eq.~\eqref{eq:haloscope_signal} with an additional suppression of $(mL)^2$~\cite{DMRadio:2022pkf}. As an example we show the projection of Ref.~\cite{DMRadio:2022pkf} for a cubic-meter scale experiment, also recast by $v_\DM / f_\gamma \lra g_{a\gamma\gamma}$.

\paragraph{Background Electric Fields.} In all previously known examples, background electric fields yield suppressed sensitivity to ultralight DM. (See App.~\ref{app:electric} for discussion of other claims.) However, for the axial-photon coupling, haloscopes with background electric fields could match the performance of a haloscope with $B_0 \sim 10 \, \mathrm{T}$ with only $E_0 \sim v_\DM B_0 \sim 3 \, \mathrm{MV}/\mathrm{m}$, significantly lower than the electric fields routinely realized in superconducting accelerator cavities~\cite{Padamsee2023}.

As a simple example of a geometry which yields $C_E \sim 1$, suppose $\v{E}_0$ is sourced by a long cylindrical capacitor with a small inner radius. Then $\v{E}_0$ points radially outward, and if $\v{V}$ points along the cylinder's axis, then $\v{E}_0 \times \v{V}$ points tangentially. For a cavity haloscope, the effective current would overlap with the $\mathrm{TE}_{011}$ mode of the same cylinder, while for a lumped element experiment, the current makes an oscillating magnetic field along the cylinder axis, which can drive an emf through a pickup loop. 

\paragraph{Heterodyne Experiments.} In the ``heterodyne'' approach to axion DM detection~\cite{Berlin:2019ahk,Lasenby:2019prg}, the background fields are due to a loaded cavity mode $\ell$. Both magnetic and electric background fields are present, and the form factors are
\begin{equation}
C_B = \bigg| \int_{V_c} \tilde{\v{E}}_n \cdot \tilde{\v{B}}_\ell^* \bigg|, \qquad C_E = \bigg| \hat{\v{V}} \cdot \int_{V_c} \tilde{\v{E}}_n \times \tilde{\v{E}}_\ell^* \bigg|.
\end{equation}
The signal mode $n$ is resonantly excited if the mode frequency difference is equal to the DM frequency, which allows the heterodyne approach to probe $m L \ll 1$. Since $E_0 \sim B_0$, the signal due to the electric field generically dominates.\footnote{These form factors also shed light on the results for Fabry--Perot optical cavities in Sec.~\ref{sec:propagation}. When $n$ and $\ell$ are the two degenerate polarizations of the same cavity mode, $C_B$ vanishes (implying low sensitivity to axion DM as $m \to 0$) while $C_E$ can be $\mathcal{O}(1)$ (implying unsuppressed sensitivity to axial vector DM).}

For an effective heterodyne experiment, one must choose the cavity geometry to maximize $C_E$. As a concrete example, the hybrid mode cavity in Ref.~\cite{Li:2025pyi} contained a pair of modes with electric fields linearly polarized along $\tilde{\v{E}}_\ell \parallel \hat{\v{x}}$ and $\tilde{\v{E}}_n \parallel \hat{\v{y}}$, respectively. The cavity was designed with corrugated endplates that shifted $\tilde{\v{E}}_\ell$ and $\tilde{\v{E}}_n$ by a relative quarter wavelength along $\hat{\v{z}}$, so that $\tilde{\v{B}}_\ell$ was in-phase with $\tilde{\v{E}}_n$, achieving $C_B \simeq 1$. However, this shift sets $C_E \simeq 0$. To get a large value of $C_E$, one could simply remove the endplate corrugations, so that $\tilde{\v{E}}_\ell$ is in-phase with $\tilde{\v{E}}_n$. This achieves $C_E \simeq 1$ for $\v{V} \parallel \hat{\v{z}}$. 

As such, the heterodyne approach is automatically sensitive to axial vector DM, with no velocity suppression, while also avoiding the magnetoquasistatic penalty that applies for static field experiments in the limit $mL \ll 1$. Moreover, adjusting the cavity geometry allows one to distinguish between axions and axial vectors. In Fig.~\ref{fig:axial_photon} we show the potential of this approach by recasting the ``cubic meter'' projection in Ref.~\cite{Adams:2022pbo} with $1/f_\gamma \lra g_{a\gamma\gamma}$. 

\paragraph{Background Potentials.} Within the current in Eq.~\eqref{eq:jbar_full_expr}, we neglect scalar potentials, as they are generally smaller than vector potentials (see Eq.~\eqref{eq:vector_benchmarks}). We also neglect $\nabla V^0$ and $\nabla \times \v{V}$ since they are velocity suppressed, leaving the dominant contribution
\begin{equation}
\v{J}_{\mathrm{eff}}^A = - \frac{g}{2} \dot{\v{V}} \times \v{A}_0.
\end{equation}
To determine $\v{A}_0$, we note that $\nabla \times \v{A}_0 = \v{B}_0$ by definition, while $m_\gamma^2 \del_\mu A^\mu = \del_\mu J^\mu_{\text{eff}}$ implies 
\begin{equation}
\nabla \cdot \v{A}_0 \simeq - \frac{g}{2 m_\gamma^2} \v{B}_0 \cdot \dot{\v{V}}
\end{equation}
where we kept only the leading terms of Eq.~\eqref{eq:div_jeff}, and neglected the scalar potential. That is, $\v{A}_0$ contains a piece which is suppressed by $g$ but enhanced by $1/m_\gamma^2$. Like the second-order effects in Sec.~\ref{sec:propagation}, it is associated with DM sourcing a longitudinal photon and converting it back to a transverse photon. This effect is suppressed relative to the first-order effect by
\begin{equation}
\frac{g \sqrt{\rho_\DM}}{m_\gamma^2} \sim 10^{-5} \, \frac{m}{10^{-6} \, \mathrm{eV}} \frac{10^{17} \, \mathrm{GeV}}{f_\gamma}
\end{equation}
at our benchmark photon mass $m_\gamma = 10^{-15} \, \mathrm{eV}$, and hence negligible here.

Discarding the second-order effect, we have $\nabla \cdot \v{A}_0 \simeq 0$ and the vector potential sourced by a laboratory magnet is $|\v{A}_0| \sim B_0 L_{\mathrm{mag}}$. Meissner screening effects are negligible on the scale of a laboratory magnet, since $m_\gamma L_{\mathrm{mag}} \ll 1$. However, vortices cluster around a laboratory magnet in equilibrium, discharging its vector potential. Further study is required to understand the timescale to reach such a state, and whether the vortices are released when the magnet is switched off. The $\v{A}_0$ sourced by the Earth could be moderately larger, $B_\oplus R_\oplus \sim 200 \, \mathrm{T} \, \mathrm{m}$, but since it is likely discharged by vortices (see App.~\ref{app:vortex}), we do not include it.

Thus, taking $|\v{A}_0| \sim B_0 L_{\mathrm{mag}}$ for concreteness, we have the scaling 
\begin{equation}
|\v{J}_{\mathrm{eff}}^A| \sim g \sqrt{\rho_\DM} B_0 L_{\mathrm{mag}}
\end{equation}
which is $\sim\! m L_{\mathrm{mag}}$ times the usual axion effective current. This restores the sensitivity of cavity haloscopes, which operate with $m L_{\mathrm{mag}} \sim m L \sim 1$. However, the optimal geometry would be significantly different: in a solenoidal magnet where $\v{A}_0 \propto \hat{\bm{\theta}}$, the form factor for the usual $\mathrm{TM}_{010}$ mode would vanish for any direction of $\hat{\v{V}}$. Interestingly, the vector potential effect can exist even when the magnetic field vanishes inside the cavity itself, which would allow the use of superconducting cavities to achieve much higher $Q_c$.

Lumped element experiments operate at $mL \ll 1$, and would thus still have suppressed sensitivity. For $m \lesssim v_\DM / L \sim (10^{-10} \, \mathrm{eV}) (1 \, \mathrm{m} / L)$, the vector potential effect would be even weaker than the velocity-suppressed magnetic field effect. Finally, for heterodyne experiments, the vector potential is never important, because its effect is weaker than that of the electric field by $\sim\! mL$. For heterodyne experiments, a large external static $\v{A}_0$ would be irrelevant, because it would not yield an effective current oscillating at the signal mode frequency. 

Thus, for both cavity haloscope and lumped element experiments, we roughly estimate sensitivity due to $\v{J}_{\mathrm{eff}}^A$ by recasting $(m L/2) / f_\gamma \lra g_{a\gamma\gamma}$ with $L = 1 \, \mathrm{m}$, deferring a more detailed study of optimal geometries to future work. 

\subsection{Astrophysical Constraints}
\label{sec:cooling}

Here we briefly outline the astrophysical constraints that bound the parameter space.

\paragraph{Axion-Like Emission.} As shown in Eq.~\eqref{eq:alpha_V_simp}, emission of a relativistic longitudinal axial vector $V_L$ is analogous to emission of an axion with $g_{a\gamma\gamma} \lra 1/f_\gamma$. We can thus directly recast many axion astrophysical bounds; for simplicity, we focus on a few well-established examples. 

We recast the constraint derived in Ref.~\cite{Dolan:2022kul} from the evolution of stars in globular clusters. At lower $m$, stronger constraints can be derived from observations of SN1987A; in particular, supernova axions can convert to gamma rays in the Galactic magnetic field~\cite{Hoof:2022xbe}. We will show in Sec.~\ref{sec:rel_mix} that this process also applies for $V_L$, so that we can directly recast the result of Ref.~\cite{Hoof:2022xbe}. These results are shown in Fig.~\ref{fig:axial_photon}.

\paragraph{Plasmon Decay.} On-shell plasmon decay can lead to energy loss in stars. We analyze all the decay channels in App.~\ref{app:plasmons} and find that the dominant processes involve transverse plasmons decaying into $V_L + \gamma_L^p$ (a longitudinal axial vector and a longitudinal plasmon), or into $V_T + \gamma_L^0$ (a transverse axial vector and a longitudinal photon). The former process is already included in axion astrophysical constraints, which were already discussed. 

The latter process also occurs in the dark axion portal, normalized in Ref.~\cite{Hook:2021ous} as 
\begin{equation}
\mathcal{L}_{\mathrm{int}} = \frac{\phi}{2 f_a} \frac12 \, \epsilon^{\mu\nu\alpha\beta} F^D_{\mu\nu} F_{\alpha \beta}
\end{equation}
which corresponds to Eq.~\eqref{eq:alpha_simp} by identifying the longitudinal photon with the axion ($\alpha \lra \phi$), the dark photon with the axial vector ($F^V \lra F^D$), and the coupling scale $f_\gamma \lra (m / 2 m_\gamma) f_a$. Several works computed constraints on the dark axion portal from this decay process:
\begin{equation} \label{eq:astro_bounds}
\frac{1}{f_a} \lesssim \begin{cases} 2.3 \times 10^{-7} \, \mathrm{GeV}^{-1} & \text{solar\ luminosity~\cite{Arias:2020tzl}} \\ 6 \times 10^{-9} \, \mathrm{GeV}^{-1} & \text{horizontal\ branch\ stars~\cite{Arias:2020tzl,Kalashev:2018bra}} \\ 2.3 \times 10^{-9} \, \mathrm{GeV}^{-1} & \text{white\ dwarf\ cooling~\cite{Hook:2021ous}} \\ 7.1 \times 10^{-10} \, \mathrm{GeV}^{-1} & \text{red\ giants~\cite{Carenza:2023qxh}\ (see\ also~\cite{Choi:2018mvk})} \end{cases}
\end{equation}
We show the strongest two of these bounds in Fig.~\ref{fig:axial_photon}. Due to the scaling $f_\gamma \propto m/m_\gamma$, they are very strong at higher $m$, but weaker than bounds from $V_L$ at lower $m$. 

These works only considered on-shell plasmon decay. Stronger constraints may follow from considering the decay of off-shell plasmons produced, e.g., from Compton scattering.

\paragraph{Spontaneous DM Decay.} There are also constraints on the decay of axial vector DM. As was discussed in Sec.~\ref{sec:axial_photon}, decay to two transverse photons is forbidden, while decay to a transverse photon and a longitudinal photon is not. In the limit $m \gg m_\gamma$ the outgoing longitudinal photon is relativistic, so the decay process can be described by the dark axion portal. The corresponding lifetime is $\tau \simeq 96 \pi f_a^2/m^3$~\cite{Kaneta:2016wvf}. 

The DM lifetime must be much longer than the age of the universe, due to constraints from cosmological energy injection and direct detection of the decay products. As a benchmark, Ref.~\cite{Janish:2023kvi} recently used JWST data to exclude $\tau \lesssim 10^{27} \, \mathrm{s}$ for eV-scale axion DM decaying to two photons. The decay of axial vector DM is qualitatively similar, though one of the final photons is not visible. In lieu of a full analysis, we show in Fig.~\ref{fig:axial_photon} that even demanding $\tau < 10^{28} \, \mathrm{s}$ yields a constraint weaker than other astrophysical bounds. (Ref.~\cite{Arias:2020tzl} also considered a stimulated decay process due to CMB photons, but found a constraint weaker than the strongest astrophysical bound in Eq.~\eqref{eq:astro_bounds}.)

\paragraph{Fuzzy DM Bound.} ``Fuzzy'' DM masses are constrained by small-scale structure. For example, Ref.~\cite{May:2025ppj} showed that if Ursa Major III is indeed an ultra-faint dwarf galaxy, then ultralight scalar DM must have $m \gtrsim 8 \times 10^{-18} \, \mathrm{eV}$. For concreteness we plot this curve in Fig.~\ref{fig:axial_photon}, though the corresponding bound for vector DM could be order-one different.

\paragraph{Superradiance Bounds.} Superradiance bounds on light vectors require only the vector's gravitational interaction, and hence also apply here. Ref.~\cite{Aswathi:2025nxa} excluded the mass range $1.1 \times 10^{-14} \, \mathrm{eV} \leq m \leq 1.8 \times 10^{-12} \, \mathrm{eV}$ from observations of rapidly spinning primary black holes in binary-merger events such as GW231123, and we show this bound in Fig.~\ref{fig:axial_photon}. 

Limits on light vectors can also come from observations of spinning black holes in X-ray binaries~\cite{Arvanitaki:2014wva,Baryakhtar:2017ngi}. However, the robustness of these spin measurements has been called into question in a recent study of the plunging regions around black holes~\cite{Rule:2026tny}. Additional interactions with matter, such as the axial-fermion coupling in Sec.~\ref{sec:fermion_couplings}, or the axial-photon coupling in a large background $\v{A}$, can lead to changes to accretion disk properties that invalidate the X-ray binary limits~\cite{Siemonsen:2022ivj}. A dedicated numerical study is needed for assessing the impact of these interactions.

Other interactions of the vector can saturate the superradiance instability before the spin of the black hole changes by an $\mathcal{O}(1)$ amount. In our case, the most important is the vector's coupling to the dark Higgs, which generates the vector mass. The growth of the vector field amplitude saturates due to vortex formation in a violent ``string bosenova'' event~\cite{East:2022rsi}. This dynamics is similar in nature to that of the photon vortices discussed in App.~\ref{app:vortex}, and its impact depends on the exact UV completion. However, interactions that are strong enough to saturate the superradiance instability generally also lead to depletion of vector DM in the early universe via vortex formation~\cite{East:2022rsi}, as we discuss further in Sec.~\ref{sec:discussion}.

\paragraph{Other Axial Vector Sources.} There are some other potentially interesting astrophysical sources of axial vectors. First, axions can be sourced from the large values of $\v{E} \cdot \v{B}$ around a pulsar~\cite{Prabhu:2021zve,Noordhuis:2022ljw,Noordhuis:2023wid}. The same is true for longitudinal vectors, as can be seen from the equation of motion Eq.~\eqref{eq:vector_eom_2}. Moreover, Eq.~\eqref{eq:vector_eom_2} also contains source terms involving the potentials, such as $\v{E} \times \v{A}$. Computing their effects requires understanding the vector potential around a pulsar, accounting for its suppression by vortices, as discussed in App.~\ref{app:vortex}.

Second, in the presence of a background electromagnetic field strength or potential, vectors and photons mix. For a kinetically mixed dark photon, the stellar production rate can be written in terms of the photon production rate times an in-medium mixing angle~\cite{An:2013yfc,Redondo:2013lna,Hardy:2016kme}. A similar calculation could be carried out for the axial-photon coupling, though the form of the vector-photon mixing is qualitatively different. However, we expect the mixing angle to be small for $m \ll \mathrm{eV}$, and hence not relevant for our parameter space. 

\subsection{Relativistic Axial Vector-Photon Mixing}
\label{sec:rel_mix}

A relativistic axial vector or photon with energy $\omega \gg m, m_\gamma$ can be sourced astrophysically, as described in Sec.~\ref{sec:cooling}, or in the laboratory, and then mix with other polarizations in a background electromagnetic field strength $F_{\mu\nu}^0$ or potential $A_\mu^0$. Here we analyze these processes, then apply them to helioscopes and light shining through walls experiments. The resulting sensitivity is weaker than astrophysical bounds, but provides complementary information. 

\paragraph{Mixing Processes.} By solving the equations of motion as in App.~\ref{app:photon_prop}, we can estimate the conversion probability for a vector or photon propagating a distance $L$ in the presence of a background field strength $F_{\mu\nu}^0$. Assuming $L$ is small enough, so that the probability is small and the momentum transfer $q$ obeys $q L \ll 1$, we have
\begin{subequations}
\begin{align}
p(V_L \to \gamma_x) &= (g L / 2)^2 ((B_x^0 + E_y^0) / m)^2, \label{eq:axion_mixing} \\
p(V_x \to \gamma_L) &= (g L / 2)^2 ((B_x^0 + E_y^0) / m_\gamma)^2, \label{eq:portal_mixing} \\
p(V_x \to \gamma_y) &= (g L / 2)^2 (E_z^0 / \omega)^2, \\
p(V_L \to \gamma_L) &= (g L / 2)^2 (B_z^0 (m^2 - m_\gamma^2) / m m_\gamma \omega)^2, \label{eq:double_long_mixing}
\end{align}
\end{subequations}
with the same scalings for all reverse processes. Here, Eq.~\eqref{eq:axion_mixing} is precisely what one would expect for axion-photon mixing, consistent with Eq.~\eqref{eq:alpha_V_simp}, while Eq.~\eqref{eq:portal_mixing} is as expected from the dark axion portal Eq.~\eqref{eq:alpha_simp}. The other two results are qualitatively new, but subdominant. In particular, the double longitudinal process Eq.~\eqref{eq:double_long_mixing}, which would naively be enhanced by both $1/m$ and $1/m_\gamma$, is instead highly suppressed. 

Axial vectors and photons also mix in a background electromagnetic potential $A^\mu_0$. Typical vector potentials are estimated in Eq.~\eqref{eq:vector_benchmarks}, and are much greater than scalar potentials, which we neglect. In the presence of a vector potential, the most important mixing process is
\begin{equation} \label{eq:A0_mixing}
p(V_x \to \gamma_y) = (g L / 2)^2 (A^z_0 / 2)^2
\end{equation}
while all other mixing processes are suppressed by powers of $m/\omega$ or $m_\gamma/\omega$. 

\paragraph{Light Shining Through Walls.} Here we focus on the ongoing ALPS II experiment~\cite{Ortiz:2020tgs,ALPSII:2025eri}, which features light propagating through cavities of length $L = 106 \, \mathrm{m}$ with regeneration factor $\beta \sim 10^4$, in a magnet of bore diameter $D = 5.5 \, \mathrm{cm}$. The signal arises from transverse photons in the production cavity converting a species $X$, which propagates through a wall and converts back into a transverse photon in the regeneration cavity. 

In a background $\v{B}_0$, $X$ can be a longitudinal $V$, and from Eq.~\eqref{eq:axion_mixing} we can directly recast sensitivity via $1/f_\gamma \lra g_{a\gamma\gamma}$. In the presence of DM, $X$ can also be a longitudinal photon, but this process is subdominant.\footnote{Though ALPS II is generally more sensitive to DM than PVLAS, the strong PVLAS sensitivity discussed in Sec.~\ref{sec:propagation} involves transverse polarization rotation, which cannot yield a signal in ALPS II.} To see this, note that the DM effect is suppressed unless $m \beta L \lesssim 1$ where $\beta \sim 10^4$ is the regeneration factor. Assuming this holds,
\begin{equation}
\frac{p(\gamma_T \to \gamma_L)}{p(\gamma_T \to V_L)} \sim \frac{\rho_\DM / m_\gamma^2}{B_0^2 / m^2} \sim 10^{-12} \, \left( \frac{m}{m_\gamma} \right)^2 \left( \frac{10 \, \mathrm{T}}{B_0} \right)^2.
\end{equation}
This is a suppression for all relevant $m$, so the DM effect is always weaker than the $\v{B}_0$ effect. 

Finally, in a background $\v{A}_0$, $X$ can be a transverse $V$. Assuming the vector potential is sourced by the ALPS II magnet, it is set by the small diameter $D$, so we can roughly recast sensitivity via $(m D/2) / f_\gamma \lra g_{a\gamma\gamma}$. This exceeds the effect due to $\v{B}_0$ for $m \gtrsim 10^{-5} \, \mathrm{eV}$.

\paragraph{Helioscopes.} Helioscopes seek to detect axions produced in the Sun by converting them to X-ray photons in a laboratory $\v{B}_0$. Here we focus on the proposed IAXO experiment~\cite{IAXO:2019mpb}, which features a magnet which applies $B_0 = 2.5 \, \mathrm{T}$ over length $L = 20 \, \mathrm{m}$, over a region of diameter $D = 5.2 \, \mathrm{m}$. The axion signal applies unchanged if we replace the axion with a longitudinal $V$, and the reach can be recast by $1/f_\gamma \lra g_{a\gamma\gamma}$, as shown in Fig.~\ref{fig:axial_photon}.

The Sun can also produce longitudinal photons and transverse $V$, through the dark axion portal Eq.~\eqref{eq:astro_bounds}, and the rate of this process is parametrically enhanced by $\sim (m/m_\gamma)^2$. The longitudinal photons can convert to transverse photons in the presence of DM, and here we recast the result of Ref.~\cite{Arias:2020tzl} for the dark axion portal. For the masses we plot in Fig.~\ref{fig:axial_photon}, it is much stronger than the $\v{B}_0$ effect, albeit suppressed for $m L \gtrsim 1$. 

Alternatively, the transverse $V$ can convert to transverse photons in background $\v{A}_0$. For $m L \lesssim 1$, the ratio of conversion probabilities is
\begin{equation}
R = \frac{p(V_T \to \gamma_T)}{p(\gamma_L \to \gamma_T)} \sim \frac{A_0^2}{\rho_\DM / m_\gamma^2} \sim 5 \times 10^{-5}
\end{equation}
where we estimated $A_0 \sim B_0 D$. This weakens the coupling reach by $R^{1/4}$, but we also show this result in Fig.~\ref{fig:axial_photon}, as it is independent of the DM density; moreover, while the DM signal is suppressed for $mL \gtrsim 1$, the $\v{A}_0$ effect is unsuppressed until $qL \gtrsim 1$. However, this reach is also subject to the caveats regarding background vector potentials mentioned in Sec.~\ref{sec:haloscopes}. 

%!TEX root = main.tex

\section{Two-Form Dark Matter}
\label{sec:two_form}

Two-form fields, also known as Kalb--Ramond fields~\cite{Kalb:1974yc}, have been used as a ``string-inspired'' model-building ingredient (e.g.~see Refs.~\cite{Majumdar:1999jd,deCesare:2014dga,Dashko:2018dsw}), and have been extensively studied as a source of spontaneous Lorentz violation in ``Kalb--Ramond gravity'' (e.g.~see Refs.~\cite{Altschul:2009ae,Kumar:2020hgm,Yang:2023wtu,Junior:2024ety,Manton:2024hyc}). Recently, massive two-form fields have been considered as an ultralight DM candidate~\cite{Capanelli:2023uwv,Plantier:2025hcm}. (Some of the signals studied in Refs.~\cite{Altschul:2009ae,Kumar:2020hgm,Yang:2023wtu,Junior:2024ety,Manton:2024hyc} can also be recast to ultralight two-form DM, but are highly suppressed unless one has an extremely low DM mass, $m \lesssim 10^{-20} \, \mathrm{eV}$.)

Here we show that our earlier results already cover the main effects of two-form DM. In Sec.~\ref{sec:free_theory}, we review the free massive two-form field, and its duality to a vector field in flat spacetime. In Sec.~\ref{sec:interactions} we use the duality to relate two-form interactions and vector interactions. We find that while the duality modifies operator dimension, and hence EFT power counting, the phenomenology is ultimately equivalent. Notably, the dipole couplings discussed in Sec.~\ref{sec:dark_dipole} become particularly simple in two-form language. 

\subsection{Free Solutions and Dualities}
\label{sec:free_theory}

\paragraph{Equations of Motion.} The Lagrangian for a massive two-form field is
\begin{equation} \label{eq:two_form_initial}
\mathcal{L} = \frac{1}{12} H_{\mu\nu\rho} H^{\mu\nu\rho} - \frac{m^2}{4} B_{\mu\nu} B^{\mu\nu}, 
\end{equation}
where $H_{\mu\nu\rho} = \del_\mu B_{\nu\rho} + \del_\nu B_{\rho\mu} + \del_\rho B_{\mu\nu}$ is the three-form field strength. The corresponding equation of motion is
\begin{equation} \label{eq:massive_eom}
\del_\mu H^{\mu\nu\rho} + m^2 B^{\nu\rho} = 0.
\end{equation}
For nonzero $m$, applying $\del_\nu$ yields the constraint $\del_\nu B^{\nu\rho} = 0$, and plugging it back in gives
\begin{equation} 
(\del^2 + m^2) B^{\mu\nu} = 0, 
\end{equation}
so that the solutions indeed correspond to particles of mass $m$. 

In the specific case of $d = 4$, there is another mass term $\bar{m}^2 \epsilon_{\mu\nu\rho\sigma} B^{\mu\nu} B^{\rho\sigma}/8$. If we only had this mass term in the action, the equation of motion would instead be 
\begin{equation}
\del_\mu H^{\mu\nu\rho} - \frac12 \bar{m}^2 \epsilon^{\nu\rho\alpha\beta} B_{\alpha\beta} = 0.
\end{equation}
However, applying $\del_\nu$ to this equation yields the constraint $H_{\mu\nu\rho} = 0$, which indicates the field is not dynamical. More generally, Ref.~\cite{Plantier:2025hcm} showed that when both mass terms are present, the physical mass-squared is $m_{\mathrm{phys}}^2 = m^2 + \bar{m}^4/m^2$, which diverges for fixed nonzero $\bar{m}$ as $m \to 0$. In principle, we could allow both mass terms simultaneously, but for simplicity we will impose parity to forbid the $\bar{m}^2$ term.

\paragraph{Free Solutions.} In the massive case, we have the free plane wave solutions $B^{\mu\nu} = e^{-i k \cdot x} b^{\mu\nu}$ which obey $k_\mu b^{\mu\nu} = 0$. In the rest frame this condition implies $b^{0i} = 0$, giving three independent solutions parametrized by $\{b^{xy}, b^{yz}, b^{zx}\}$. They transform under rotations as a spatial vector, so Eq.~\eqref{eq:massive_eom} describes a massive spin $1$ particle.

By contrast, when $m = 0$ the field describes a massless helicity $0$ particle. To see this, it is convenient to use the null frame vectors $\{k^\mu, q^\mu, \epsilon_+^\mu, \epsilon_-^\mu\}$ which obey $k \cdot q = 1$ and $\epsilon_+ \cdot \epsilon_- = -2$ with all other inner products vanishing (the $\epsilon_\pm$ are effectively circular polarization vectors). We no longer have the constraint $\del_\mu B^{\mu\nu} = 0$, but the equation of motion $\del_\mu H^{\mu\nu\rho} = 0$ yields $k_\mu k^{[\mu} b^{\nu\rho]} = 0$, which eliminates $b^{\mu\nu} \propto q^{[\mu}_{\phantom{\pm}} \epsilon_\pm^{\nu]}$.

In addition, in the massless case the field has a gauge symmetry $\delta B_{\mu\nu} = \del_\mu \Lambda_\nu - \del_\nu \Lambda_\mu$ for an arbitrary vector field $\Lambda^\mu(x)$. Applying a gauge transformation with $\Lambda^\mu = e^{-i k \cdot x} \lambda^\mu$ induces $\delta b^{\mu\nu} \propto k^{[\mu} \lambda^{\nu]}$, so that polarizations involving $k$ can be gauged away. We are thus left with a single physical solution $b^{\mu\nu} = \epsilon_+^{[\mu} \epsilon_-^{\nu]}$. To see that it carries zero helicity, note that under rotation about $\hat{\v{k}}$, the $\epsilon_\pm$ factors transform with opposite phase factors; alternatively, note that the dual field strength is longitudinal, $\epsilon^{\mu\nu\rho\sigma} H_{\nu\rho\sigma} \propto k^\mu$.

\paragraph{St\"uckelberg Mechanisms.} The counting of massless and massive degrees of freedom can be understood in terms of the St\"uckelberg mechanism. In $d = 4$, massless $p$-form fields (including a scalar for $p = 0$ and a vector for $p = 1$) have $C_p$ dynamical degrees of freedom, where 
\begin{equation} \label{eq:massless_dof}
(C_0, C_1, C_2, C_3, C_4) = (1, 2, 1, 0, 0).
\end{equation}
Massive $p$-form fields have $\bar{C}_p$ degrees of freedom, where $\bar{C}_p \geq C_p$. 

For example, a massive vector field $V^\mu$ has $\bar{C}_1 = 3$, and we can understand this by writing the mass term $m^2 V^\mu V_\mu / 2$ as $(m V_\mu - \del_\mu \phi)^2/2$, where $\phi$ is a St\"uckelberg scalar. The extra degree of freedom comes from the vector ``eating'' the scalar. 

Similarly, for a massive two-form $B_{\mu\nu}$, we can rewrite Eq.~\eqref{eq:two_form_initial} as
\begin{equation} \label{eq:stuckelberg_2}
\mathcal{L} = \frac{1}{12} H_{\mu\nu\rho} H^{\mu\nu\rho} - \frac{1}{4} (m B_{\mu\nu} - f_{\mu\nu})(m B^{\mu\nu} - f^{\mu\nu}), 
\end{equation}
where $f_{\mu\nu} = \del_\mu a_\nu - \del_\nu a_\mu$, $a_\mu$ is a St\"uckelberg vector, and the action is invariant under the joint gauge transformation $\delta B_{\mu\nu} = \del_\mu \Lambda_\nu - \del_\nu \Lambda_\mu$ and $\delta a_\mu = m \Lambda_\mu$. Then a massive two-form has $\bar{C}_2 = C_2 + C_1 = 3$ degrees of freedom, where the two extra degrees of freedom come from ``eating'' the vector. More generally, this logic implies $\bar{C}_p = C_p + C_{p-1}$, so that
\begin{equation} \label{eq:massive_dof}
(\bar{C}_0, \bar{C}_1, \bar{C}_2, \bar{C}_3, \bar{C}_4) = (1, 3, 3, 1, 0).
\end{equation}

From Eqs.~\eqref{eq:massless_dof} and~\eqref{eq:massive_dof}, we have $C_0 = C_2$, as expected from the well-known duality between a massless two-form and a massless pseudoscalar~\cite{Hjelmeland:1997eg,Polchinski:1998rr}, which we review in App.~\ref{app:dualities}. In addition, the result $\bar{C}_0 = \bar{C}_3$ fits the duality between a massive pseudoscalar and a massive three-form, which gets its mass from eating a two-form; this perspective has been used to study the axion quality problem~\cite{Dvali:2005an,Sakhelashvili:2021eid,Choi:2023gin,Burgess:2023ifd}. Here we are concerned with the duality between a massive two-form and a massive axial vector, motivated by $\bar{C}_1 = \bar{C}_2$.

\paragraph{Massive Vector Dualities.} These well-known dualities were discussed by Ref.~\cite{Quevedo:1996uu}, and a clear introduction was recently given by Ref.~\cite{Plantier:2025hcm}. Heuristically, one can think of the dual field strength $\epsilon^{\mu\nu\rho\sigma} H_{\nu\rho\sigma}$ as corresponding to $m V^\mu$ (or $\del^\mu a$ in the massless case). More rigorously, one starts with a ``parent Lagrangian'' $\mathcal{L}[B, V]$ in which neither field has a kinetic term, but where each gains one upon integrating the other out; this yields two Lagrangians $\mathcal{L}[B]$ and $\mathcal{L}[V]$ with equivalent dynamics. We carry this out in App.~\ref{app:dualities} and simply state results here.

Depending on how currents are coupled, there are two distinct dualities. First, we have 
\begin{equation} \label{eq:massive_duality_a}
\mathcal{L}[B] = \frac{1}{12} (H_{\mu\nu\rho} + J_{\mu\nu\rho})^2 - \frac14 m^2 B^2 \enskip \lra \enskip \mathcal{L}[V] = - \frac14 (F^V_{\mu\nu})^2 + \frac12 m^2 V^2 + m V_\mu J^\mu
\end{equation}
where $J^\mu$ and $J^{\mu\nu\rho}$ are currents related by $J_\sigma = \epsilon_{\mu\nu\rho\sigma} J^{\mu\nu\rho}/6$ and $J_{\mu\nu\rho} = - \epsilon_{\mu\nu\rho\sigma} J^\sigma$. Second, 
\begin{equation} \label{eq:massive_duality_b}
\mathcal{L}[B] = \frac{1}{12} H^2 - \frac14 m^2 B^2 + \frac{m}{2} B_{\mu\nu} J^{\mu\nu} \enskip \lra \enskip \mathcal{L}[V] = - \frac14 \left( F^V_{\mu\nu} + \frac{1}{2} \epsilon_{\mu\nu\rho\sigma} J^{\rho\sigma} \right)^2 + \frac12 m^2 V^2
\end{equation}
for an arbitrary current $J^{\mu\nu}$. 

Since $[H_{\mu\nu\rho}] = [F^V_{\mu\nu}] = 2$ and $[B_{\mu\nu}] = [V_\mu] = 1$, both of these dualities change the operator dimension of the interaction and thus reorganize the EFT expansion, as emphasized by Refs.~\cite{Plantier:2026rsv,Burgess:2025geh}. They are not inverses of each other; applying one and then the other in reverse yields the original Lagrangian only up to a field redefinition, as we show in App.~\ref{app:dualities}. 

We can understand the difference between the two dualities in terms of the St\"uckelberg mechanisms derived above. For example, on the left-hand side of Eq.~\eqref{eq:massive_duality_a}, we may smoothly take the $m \to 0$ limit to yield a current coupled to a massless two-form, representing a scalar particle. On the right-hand side of Eq.~\eqref{eq:massive_duality_a} with $V_\mu \to V_\mu - \del_\mu \phi / m$, as $m \to 0$ the vector decouples, and the current is only coupled to the St\"uckelberg scalar $\phi$. The situation is reversed in Eq.~\eqref{eq:massive_duality_b}. The $m \to 0$ limit of the right-hand side yields a current coupled to a massless vector, while on the left-hand side with $B_{\mu\nu} \to B_{\mu\nu} - f_{\mu\nu}/m$, the two-form decouples and the current is coupled to the St\"uckelberg vector $a_\mu$. That is, the two dualities can describe interactions with different $m \to 0$ limits. 

\subsection{Interactions}
\label{sec:interactions}

We now use these dualities to relate theories with two-form interactions to theories with vector interactions. For simplicity, we will impose $P$ and $C$ symmetry, and work through the leading interactions that appear for each assignment of $P$ and $C$ values to the two-form field. 

\paragraph{Axial Vector Behavior.} If $B_{\mu\nu}$ is a tensor and even under $C$, the simplest couplings on the two-form side are dimension $5$, and they are related via Eq.~\eqref{eq:massive_duality_a} to the dimension $4$ axial-fermion and axial-photon couplings. On the vector side these correspond to 
\begin{equation}
\mathcal{L}_{\mathrm{int}}[V] = V_\mu \bigg( \frac{m}{f_\psi} \, \bar{\psi} \gamma^\mu \gamma^5 \psi - \frac{m}{4f_\gamma} \, K^\mu \bigg)
\end{equation}
and on the two-form side we have
\begin{equation} \label{eq:axial_vector_two_form}
\mathcal{L}[B] = \frac{1}{12} \bigg(H_{\mu\nu\rho} - \frac{1}{f_\psi} \, \epsilon_{\mu\nu\rho\sigma} \bar{\psi} \gamma^\sigma \gamma^5 \psi + \frac{3}{2 f_\gamma} A_{[\mu} F_{\nu\rho]}\bigg)^2 - \frac14 m^2 B^2
\end{equation}
where the antisymmetrization carries a factor of $1/6$. These couplings appear qualitatively different, but the derivatives and factors of $m$ ensure the physics is equivalent. 

For example, amplitudes involving high-momentum longitudinal vectors scale with the momentum $p$. On the two-form side, there is no such enhancement, but the interaction vertex itself carries a power of $p$ because it is proportional to the field strength $H_{\mu\nu\rho}$. Thus, in both cases the theory is weakly coupled up to some high scale $p \sim f$. The only conceptual difference is that the large scale $f$ on the two-form side is transmuted into the small dimensionless coupling $m/f$ on the vector side.

Mass renormalization also matches on both sides. For the vector it is proportional to the coupling squared, so that generically $\delta m^2 \sim (m \Lambda/4 \pi f)^2$. On the two-form side, one would naively have a quartic divergence, but the couplings respect the two-form's gauge symmetry at zero mass, which implies the mass renormalization must be proportional to $m^2$. We thus again have $\delta m^2 \sim (m \Lambda/4 \pi f)^2$. 

The same applies to the photon mass renormalization $\delta m_\gamma^2$ due to the axial-photon coupling. At nonzero $m$, this coupling also violates electromagnetic gauge symmetry on the two-form side.\footnote{Again, it is possible to formally restore this gauge symmetry by adding an additional scalar $\varphi$, but the details are different. One must rewrite the mass term as $-(m B_{\mu\nu} + \varphi F_{\mu\nu}/m)^2/4$ and take the simultaneous transformation $\delta A_\mu = \del_\mu \alpha$, $\delta B_{\mu\nu} = -\alpha F_{\mu\nu} / 2 f_\gamma$, and $\delta \varphi = \alpha m^2 / 2 f_\gamma$.} Due to the derivatives in the interaction, $\delta m_\gamma^2$ naively appears quartically divergent. However, this is cancelled by the contribution from the four-photon vertex in Eq.~\eqref{eq:axial_vector_two_form}, and we recover the same quadratically divergent $\delta m_\gamma^2$ found previously. 

Finally, the effects of DM are the same in both cases. For example, we saw in Eq.~\eqref{eq:axial_fermion_ham} that the torque due to the axial-fermion coupling depends on $m V^i / f_\psi \sim \sqrt{\rho_\DM}/f_\psi$. On the two-form side it depends on $H^{0ij}/f_\psi \sim \sqrt{\rho_\DM}/f_\psi$, also with no velocity suppression, since $H^{0ij} \sim \del_t B^{ij}$. In summary, the phenomenology of axial vector interactions carries over unchanged to two-form language, though details of intermediate steps may differ. 

\paragraph{Dark Electric Dipole Moment.} Now suppose $B_{\mu\nu}$ is a $C$-odd tensor, and thus dual to a $C$-odd axial vector. Up through dimension $4$, there are two interactions on the two-form side, 
\begin{equation} \label{eq:scenario_2_lag}
\mathcal{L}_{\mathrm{int}}[B] = \frac{\epsilon'}{2} \, m B_{\mu\nu} F^{\mu\nu} + \frac{m}{2 f_\psi^E} \, B_{\mu\nu} \bar{\psi} \sigma^{\mu\nu} \psi
\end{equation}
where $\epsilon'$ is dimensionless. Both terms arise commonly in string-inspired constructions. The second term is the simplest way a two-form can couple to fermions, normalized so the theory is perturbative up to $\Lambda \sim 4 \pi f_\psi^E$.

Upon applying the duality Eq.~\eqref{eq:massive_duality_b}, the equivalent vector Lagrangian is
\begin{equation} \label{eq:scenario_2_dual}
\mathcal{L}[V] = - \frac14 \bigg(F^V_{\mu\nu} + \frac{\epsilon'}{2} \, \epsilon_{\mu\nu\rho\sigma} F^{\rho\sigma} + \frac{1}{2 f_\psi^E} \, \epsilon_{\mu\nu\rho\sigma} \bar{\psi} \sigma^{\rho\sigma} \psi \bigg)^2 + \frac12 m^2 V_\mu V^\mu.
\end{equation}
This contains the dark EDM term discussed in Sec.~\ref{sec:dark_dipole}, 
\begin{equation} \label{eq:dark_EDM_origin}
\mathcal{L}_{\mathrm{int}}[V] \supset -\frac{1}{4f_\psi^E} \, \epsilon^{\mu\nu\rho\sigma} F_{\mu\nu}^V \bar{\psi} \sigma_{\rho\sigma} \psi = \frac{1}{2f_\psi^E} \, F_{\mu\nu}^V \bar{\psi} i \sigma^{\mu\nu} \gamma^5 \psi
\end{equation}
and we now see that it arises as the simplest two-form fermion coupling for the arguably natural case of tensor $B_{\mu\nu}$. This feature of dipole couplings was previously noted in Ref.~\cite{Dashko:2018dsw}. 

The other terms in Eq.~\eqref{eq:scenario_2_dual} are less phenomenologically important. First, consider the $B_{\mu\nu} F^{\mu\nu}$ term in isolation. On the two-form side, it has no nontrivial effect, because it can be eliminated by a field redefinition of $B_{\mu\nu}$. On the vector side, the order $\epsilon'$ term is proportional to $\epsilon^{\mu\nu\rho\sigma} F^V_{\mu\nu} F_{\rho\sigma}$, which is a total derivative\footnote{Assuming no magnetic charges; for the case with magnetic charges, see Refs.~\cite{Brummer:2009oul,Terning:2018lsv}.}, while the order $(\epsilon')^2$ term is just a rescaling of the photon kinetic term. 

If we consider both couplings at once, then Eq.~\eqref{eq:scenario_2_dual} contains a cross-term that yields an ordinary fermion magnetic dipole moment, 
\begin{equation} \label{eq:scenario_2_cross}
\mathcal{L}_{\mathrm{int}}[V] \supset \frac14 \bigg( \epsilon' F^{\mu\nu} + \frac{1}{f_\psi^E} \, \bar{\psi} \sigma^{\mu\nu} \psi \bigg)^2 \supset \frac{\epsilon'}{2 f_\psi^E} F_{\mu\nu} \bar{\psi} \sigma^{\mu\nu} \psi
\end{equation}
which corresponds to a shift of the $g$-factor of order $\Delta g \sim \epsilon' m_\psi / q_\psi f_\psi^E$. In Ref.~\cite{Plantier:2025hcm}, this was highlighted as a potentially observable prediction of the two-form formalism. The $g$-factor of the electron is measured to $\sim 10^{-13}$ precision~\cite{Fan:2022eto}, so if we assume $\epsilon' \sim 1$, then we are naively sensitive to $f_e^E \sim 10^{13} m_e \sim 10^{10} \, \mathrm{GeV}$, competitive with the probes discussed in Sec.~\ref{sec:dark_dipole}.

However, in our view the cross-term in Eq.~\eqref{eq:scenario_2_cross} is higher-order in effective field theory, so its value is a priori unknown on both sides of the duality. More specifically, by analogy with Eq.~\eqref{eq:induced_kinetic_mixing}, we expect $\epsilon' \sim m_\psi q_\psi / (4 \pi^2 f_\psi^E)$ at one loop. This implies a $g$-factor shift $\Delta g \sim (m_\psi/f_\psi^E)^2 / (4 \pi^2)$, weakening the sensitivity to $f_e^E \sim 10^2 \, \mathrm{GeV}$, which is much weaker than other probes.

\paragraph{Dark Photon Behavior.} For completeness, we briefly consider the case where $B_{\mu\nu}$ is a \textit{pseudotensor}, and thus dual to an ordinary vector. On the two-form side, up through dimension $4$, there are no terms linear in the two-form field that allow it to be $C$-even, so we focus on the $C$-odd case.

The simplest term describing two-form mixing with the photon is 
\begin{equation} \label{eq:two_form_mixing}
\mathcal{L}_{\mathrm{int}}[B] = \frac{\epsilon}{4} \, m \epsilon^{\mu\nu\rho\sigma} B_{\mu\nu} F_{\rho\sigma}.
\end{equation}
This term was originally written down in Ref.~\cite{Cremmer:1973mg}, and has been extensively studied in the context of topological quantum field theory~\cite{Blau:1989bq}. For our purposes, by Eq.~\eqref{eq:massive_duality_b} it is equivalent to the familiar kinetic mixing term for a dark photon, 
\begin{equation} \label{eq:kinetic_mixing_vector}
\mathcal{L}[V] = - \frac14 (F_{\mu\nu}^V - \epsilon F_{\mu\nu})^2 + \frac12 m^2 V^2 \supset \frac{\epsilon}{2} \, F^V_{\mu\nu} F^{\mu\nu}.
\end{equation}
One can also integrate the original interaction by parts, so that one can apply the other duality Eq.~\eqref{eq:massive_duality_a}. The result is the equivalent mass mixing $\mathcal{L}[V] \supset m^2 (V^\mu + \epsilon A^\mu)^2/2$, which can also be reached from Eq.~\eqref{eq:kinetic_mixing_vector} by a field redefinition. Thus, expressing photon mixing with a two-form is essentially equivalent to using a vector. 

As for couplings to fermions, the ordinary coupling to the vector current $V_\mu \bar{\psi} \gamma^\mu \psi$ maps under Eq.~\eqref{eq:massive_duality_a} to a dimension $5$ two-form coupling $\epsilon_{\mu\nu\rho\sigma} H^{\nu\rho\sigma} \bar{\psi} \gamma^\mu \psi/ m$. However, the simplest coupling on the two-form side is instead given by
\begin{equation} \label{eq:scenario_3_lag}
\mathcal{L}_{\mathrm{int}}[B] = -\frac{m}{2 f_\psi^M} \, B_{\mu\nu} \bar{\psi} i \sigma^{\mu\nu} \gamma^5 \psi
\end{equation}
which after applying Eq.~\eqref{eq:massive_duality_b} gives the dark MDM interaction discussed in Sec.~\ref{sec:dark_dipole}, 
\begin{equation} \label{eq:scenario_3_dual}
\mathcal{L}_{\mathrm{int}}[V] \supset \frac{1}{4f_\psi^M} \, \epsilon^{\mu\nu\rho\sigma} F_{\mu\nu}^V \bar{\psi} i \sigma_{\rho\sigma} \gamma^5 \psi = \frac{1}{2f_\psi^M} \, F_{\mu\nu}^V \bar{\psi} \sigma^{\mu\nu} \psi.
\end{equation}
By the same logic below Eq.~\eqref{eq:scenario_2_cross}, in the presence of mixing terms, the cross-term in the current squared also gives a small shift of the fermion $g$-factor, generically of order $(m_\psi / f_\psi^M)^2$. 

\paragraph{More Two-Form Couplings.} For terms linear in the DM field, our discussion is complete up through dimension $4$ on the two-form side. We have tacitly dropped some terms which do not yield physical DM effects at leading order. For example, consider the interaction $(\del_\mu V^\mu) (\del_\nu A^\nu)$. Since $\del_\mu V^\mu$ vanishes on the vector's free equation of motion, this term does not generate mixing with external vectors. It could play a role within a larger process, but this would be higher-order in the weak DM interaction. We can see the same result on the two-form side after applying the duality Eq.~\eqref{eq:massive_duality_a}. The leading term integrates to zero by the Bianchi identity, $\epsilon^{\mu\nu\rho\sigma} \del_\mu H_{\nu\rho\sigma} = 0$, while the term second-order in the current can have a physical effect. (Similar logic holds in reverse for terms involving $\del_\mu B^{\mu\nu}$ on the two-form side.) 

Of course, one could continue to higher order. At dimension $5$ one could write terms like $\bar{\psi} \gamma^\mu \del^\nu \psi \, B_{\mu\nu}$, dual to dimension $6$ vector couplings. Alternatively, one could consider parity-violating scenarios, where one mixes terms from the different scenarios above, and includes the parity-violating mass term $\epsilon^{\mu\nu\rho\sigma} B_{\mu\nu} B_{\rho\sigma}$. This will allow a number of new effects, such as fermion EDMs, but we expect they can also be understood in terms of vector couplings. 

%!TEX root = main.tex

\section{Discussion}
\label{sec:discussion}

We have shown that axial vector DM leads to many new experimental signatures. Here we place our work in broader context, and conclude by reviewing our results. 

\paragraph{Generalized Chern--Simons Terms.} The axial-photon coupling is a ``generalized Chern--Simons term'', introduced in Refs.~\cite{Anastasopoulos:2006cz,Antoniadis:2009ze} in the context of string phenomenology and collider physics. Such triple vector boson vertices have been used in models of particle DM~\cite{Mambrini:2009ad,Dudas:2009uq,Arcadi:2017kky,Arcadi:2017jqd,Catena:2023use}. Much earlier, Refs.~\cite{DHoker:1984izu,DHoker:1984mif} noted that terms of this form arise when integrating out chiral fermions, from the subleading parts of a triangle diagram, as we have used in App.~\ref{app:photon_model}.

The analogous ``axial-gluon'' coupling is unviable, as while one can use small charges in a UV completion of the axial-photon coupling, gluons necessarily interact in the adjoint representation. By the arguments of Ref.~\cite{Preskill:1990fr}, gluon self-interactions lead to an unacceptably low cutoff $\Lambda \lesssim 4 \pi m_g / g_s$, where $m_g$ is the gluon mass and $g_s$ is the strong coupling. 

However, we could exchange the roles of the photon and axial vector in the axial-photon coupling, leading to the two-vector interaction 
\begin{equation} \label{eq:two_vector_coupling}
\mathcal{L}_{\mathrm{int}} = -\frac{g'}{4} \, \epsilon^{\mu\nu\rho\sigma} A_\mu V_\nu F^V_{\rho\sigma}.
\end{equation}
In analogy with our UV completion in App.~\ref{app:photon_model}, this term can arise from fermions with nontrivial $U(1)_{\mathrm{EM}} U(1)_D^2$ anomaly cancellation. 

We have not considered this term in detail because it violates parity, whether $V^\mu$ is an ordinary vector or an axial vector. Still, we can make a few statements about its effects. First, by the same argument used to derive Eq.~\eqref{eq:axial_photon_lambda}, we have $g' \lesssim (4\pi/\Lambda) \min(m, m_\gamma)$. Second, a variety of astrophysical bounds can still be recast from the dark axion portal, except that now the longitudinal axial vector plays the role of the axion. Finally, the interaction is quadratic in the DM field, so the effective current primarily has support at near-zero frequency and double the DM frequency, and it scales with an extra power of $\sqrt{\rho_\DM}$. It would be interesting to consider this term further. 

One can write down even more dimension $4$ terms involving $A_\mu$ and $V_\mu$, such as $(\del_\mu V^\mu) A_\nu A^\nu$, but they are effectively higher-dimensional, in the sense that their amplitudes grow with energy faster than the axial-photon coupling. The only terms that are equally relevant are the electric and magnetic dipole moments $V_\mu^\dagger V_\nu F^{\mu\nu}$ and $V_\mu^\dagger V_\nu \tilde{F}^{\mu\nu}$~\cite{Hisano:2020qkq,Chu:2023zbo,Bertuzzo:2024bwy}, but these are only nonzero for a complex vector, which would require doubling the DM degrees of freedom. 

\paragraph{DM Production and UV Completions.} In the early universe, axial vectors can be produced by the same mechanisms as ordinary vectors. Inflationary fluctuations generate the observed vector dark matter abundance for $m \gtrsim 10^{-5} \, \mathrm{eV}$~\cite{Graham:2015rva}, and Ref.~\cite{Capanelli:2024rlk} argued that nonminimal couplings to gravity extend this mechanism to lower masses. All these results carry over directly to the axial case, since they are independent of the matter couplings.

However, as we discuss further in App.~\ref{app:vortex}, dark vortex formation can deplete the DM abundance, invalidating many vector DM production mechanisms~\cite{East:2022rsi}. For our benchmark UV completion, this somewhat lowers the accessible couplings, though it may be ameliorated by additional model building. Dark vortex formation does not occur if the axial vector has a St\"uckelberg mass, though this leads to additional potential bounds from swampland conjectures~\cite{Reece:2018zvv}. However, we are not aware of a field-theoretic UV completion of the axial-photon coupling in this case. 

One could also investigate string-theoretic realizations of axial vector DM, which can arise from massive two-form fields. Ref.~\cite{Capanelli:2023uwv} gave a concrete example in a flux compactification, but it is not clear whether the mass can be naturally ultralight; furthermore, the desired two-form field is projected out in the orientifold.\footnote{We thank Liam McAllister for pointing this out.} As for interactions, Ref.~\cite{Anastasopoulos:2006cz} discussed how generalized Chern--Simons terms arise in orientifold models, and Ref.~\cite{Coudarchet:2025dfd} argued that dipole couplings arise generically in such models, albeit for massless vectors. It would also be interesting to understand dark vortex formation, or its absence, in the dual two-form picture.

\paragraph{Conclusion.} We have investigated the fourth and final minimal ultralight DM candidate, and found a wide variety of effects, providing new paths to discovering physics beyond the Standard Model. Our results are summarized in Figs.~\ref{fig:axial_electron},~\ref{fig:axial_neutron}, and~\ref{fig:axial_photon}.

The most straightforward effects are due to the pattern of velocity suppression: for vector DM fields the spatial components dominate, while the spatial gradients of scalar DM fields are suppressed. As a result, signals involving spin torques become much stronger, while the effects of background magnetic fields, used almost universally to probe axion DM, become much weaker. ``Heterodyne'' experiments involving an excited cavity mode become the strongest laboratory probe of the axial-photon coupling, since they contain background electric fields. A more subtle difference is that the polarization rotation signal of axial vector DM builds up coherently in an optical cavity, in contrast to the near-cancellation that occurs for axion DM. This renders polarimetry one of the strongest potential probes at lower DM mass. 

There are also enhanced processes involving relativistic longitudinal modes. Processes involving longitudinal axial vectors are enhanced by $E/m$, leading to astrophysical bounds and helioscope signatures like an axion, scaling with $gE/m = E/f_\gamma$. However, processes involving longitudinal photons are enhanced by $1/m_\gamma$ and dominate over the axion-like signatures for sufficiently high $m$. The strongest resulting signature is CMB spectral distortion, from the enhanced DM-induced conversion of transverse photons to longitudinal photons.

Finally, since the axial-photon coupling breaks electromagnetic gauge invariance, it leads to potential signatures from background vector potentials. In our UV completion, these depend in detail on the dynamics of photon vortices, providing rich avenues for further study.

\acknowledgments

We thank Asher Berlin, Naomi Gendler, Seth Koren, Liam McAllister, Giacomo Marocco, Arthur Platschorre, Benjamin Safdi, Harikrishnan Ramani, Raman Sundrum, and Natalia Toro for discussions. We acknowledge the use of GPT-6 Astra for proofreading the final draft. KZ was supported by the Office of High Energy Physics of the U.S. Department of Energy under contract DE-AC02-05CH11231. AH is supported by NSF grant PHY-2514660 and the Maryland Center for Fundamental Physics. Research at Perimeter Institute is supported in part by the Government of Canada through the Department of Innovation, Science and Economic Development and by the Province of Ontario through the Ministry of Colleges and Universities.

% \newpage
\appendix

%!TEX root = main.tex

\section{Model for the Axial-Photon Coupling}
\label{app:axial_photon}

Here we give an example UV completion for the axial-photon coupling. In App.~\ref{app:photon_model} we show how it can arise by integrating out chiral fermions with nontrivial $U(1)_D U(1)_{\mathrm{EM}}^2$ anomaly cancellation; we also outline general constraints associated with perturbativity and tuning. In App.~\ref{app:vortex} we show that a photon mass $m_\gamma = 10^{-15} \, \mathrm{eV}$ is allowed in the presence of vortex formation. Allowing vortex formation places stringent constraints on the model parameters. In App.~\ref{app:benchmarks} we show three benchmarks in parameter space: ``heavy Higgs'' and ``light Higgs'' (both for $m_\gamma = 10^{-15} \, \mathrm{eV}$), and a point without vortex formation, where $m_\gamma = 10^{-26} \, \mathrm{eV}$. 

\subsection{UV Completion and Generic Constraints}
\label{app:photon_model}

\paragraph{Charge Assignments.} We consider a dark Higgs $\Phi$ which gives mass to the axial vector $V^\mu$ and four Weyl fermions $\psi$, and a visible Higgs $\phi$ which gives mass to the photon $A^\mu$ and to four Weyl fermions $\chi$.
\begin{table}
\begin{center}
\begin{tabular}{c|cc}
& $U(1)_{\mathrm{EM}}$ & $U(1)_D$ \\
\hline
$\psi_1$ & $Q_f$ & $1/2$ \\
$\psi_2$ & $-Q_f$ & $1/2$ \\
$\psi_3$ & $-Q_f+Q_\phi$ & $-1/2$ \\
$\psi_4$ & $Q_f-Q_\phi$ & $-1/2$ \\
$\Phi$ & 0 & $1$ 
\end{tabular} \qquad
\begin{tabular}{c|cc}
& $U(1)_{\mathrm{EM}}$ & $U(1)_D$ \\
\hline
$\chi_1$ & $Q_f$ & $-1/2$ \\
$\chi_2$ & $-Q_f$ & $-1/2$ \\
$\chi_3$ & $-Q_f+Q_\phi$ & $1/2$ \\
$\chi_4$ & $Q_f-Q_\phi$ & $1/2$ \\
$\phi$ & $Q_\phi$ & $0$ 
\end{tabular}
\caption{Charge assignments in a simple UV completion, where the $\psi_i$ and $\chi_i$ are left-handed Weyl fermions. The axial-photon coupling arises due to nontrivial $U(1)_D U(1)_{\mathrm{EM}}^2$ anomaly cancellation.}
\label{tab:charges}
\end{center}
\end{table}
The charges are listed in Table~\ref{tab:charges}, and the Lagrangian is
\begin{multline} \label{eq:model_lag}
\mathcal{L} \supset - \frac{\lambda_\phi}{2} (|\phi|^2 - v_\phi^2)^2 - \frac{\lambda_\Phi}{2} (|\Phi|^2 - v_\Phi^2)^2 \\ + \left( y_\Phi (\psi_1 \psi_2 \Phi^\dagger + \psi_3 \psi_4 \Phi ) + y_\phi ( \chi_1 \chi_3 \phi^\dagger + \chi_2 \chi_4 \phi ) + \text{h.c.} \right),
\end{multline}
where the electromagnetic and dark gauge couplings are $e$ and $e_D$. We also reserve the ability to have $N_f$ copies of each fermion, related by an $SO(N_f)$ flavor symmetry. 

It is straightforward to check that all gauge anomalies cancel in this model. In addition, integrating out all the fermions generates no leading kinetic mixing, as the sum of $q_{\mathrm{EM}} q_D$ cancels among pairs of fermions (e.g.~$\psi_1$ with $\psi_2$). One could regard this as a result of a charge conjugation symmetry which flips the sign of $A^\mu$ while exchanging these pairs of fermions. The specific form of these couplings can also be enforced by global $U(1)_A$ symmetries. 

This set of fermions is far from unique, and not even minimal; we simply choose it because it is convenient to reason about. Different sets of charge assignments are given in Refs.~\cite{Antoniadis:2006wp,Antoniadis:2007sp,Antoniadis:2009ze}, and Ref.~\cite{Costa:2020dph} gives relevant examples with only $6$ fermions. 

\paragraph{Inducing the Axial-Photon Coupling.} From the calculations in Refs.~\cite{DHoker:1984izu,DHoker:1984mif}, it follows that integrating out all of the fermions yields an axial-photon coupling
\begin{equation}
g = \frac{e^2 e_D}{8 \pi^2} \, \mathcal{C}, \qquad \mathcal{C} = 2 N_f (2 Q_f Q_\phi - Q_\phi^2).
\end{equation}
In lieu of evaluating the triangle diagrams directly, we will give a simple self-consistency argument, in the spirit of Ref.~\cite{Preskill:1990fr}, why this must be the case. 

First, suppose that only $\Phi$ obtains a vev, so that the fermions $\psi$ become heavy. Defining $\Phi = v_\Phi e^{i \theta_\Phi}$, integrating out the fermions $\psi$ must induce the term 
\begin{equation} \label{eq:no_psi_lag}
\mathcal{L} \supset \frac{e^2 \, \mathcal{C}}{8 \pi^2} \, \frac14 \theta_\Phi F^{\mu\nu} \tilde{F}_{\mu\nu} = - \frac{g}{4} \, \epsilon^{\mu\nu\rho\sigma} \left(\frac{\del_\mu \theta_\Phi}{e_D} \right) A_\nu F_{\rho\sigma}.
\end{equation}
If $U(1)_D$ were not gauged, this would be the familiar axion-photon coupling; instead the would-be Goldstone $\theta_\Phi$ is eaten by $V_\mu$. Alternatively, suppose only $\phi$ obtains a vev, so that the fermions $\chi$ become heavy. Defining $\phi = v_\phi e^{i \theta_\phi}$, integrating out the fermions $\chi$ must induce 
\begin{equation} \label{eq:no_chi_lag}
\mathcal{L} \supset - \frac{e e_D \, \mathcal{C}}{8 \pi^2 Q_\phi} \, \frac14 \theta_\phi F_{\mu\nu}^V \tilde{F}^{\mu\nu} = - \frac{g}{4} \, \epsilon^{\mu\nu\rho\sigma} V_\mu \left(\frac{\del_\nu \theta_\phi}{e Q_\phi} \right) F_{\rho\sigma}.
\end{equation}
We are interested in the situation where both $\Phi$ and $\phi$ obtain a vev, and all fermions are integrated out. Since the full theory is anomaly free, the final Lagrangian must be gauge invariant, regardless of how one regulates the individual triangle diagrams. Furthermore, the Lagrangian must contain both the terms in Eq.~\eqref{eq:no_psi_lag} and Eq.~\eqref{eq:no_chi_lag}. This is only possible if we have the term 
\begin{equation} \label{eq:full_axial_photon}
\mathcal{L} \supset - \frac{g}{4} \, \epsilon^{\mu\nu\rho\sigma} \left(V_\mu + \frac{\del_\mu \theta_\Phi}{e_D} \right) \left(A_\nu + \frac{\del_\nu \theta_\phi}{e Q_\phi} \right) F_{\rho\sigma}
\end{equation}
which, in unitary gauge, is precisely the axial-photon coupling. 

\paragraph{Perturbativity Constraints.} The argument above shows that the scale $f_\gamma$ in $g = m/f_\gamma$ is not necessarily set by a single high scale in the UV completion. However, it is related to the scale at which the infrared theory \textit{must} break down. 

In particular, Eq.~\eqref{eq:full_axial_photon} shows that the axial-photon coupling is effectively a dimension-$5$ operator, producing amplitudes that grow with energy. This is ultimately due to enhanced processes involving longitudinal modes, as we discussed in Sec.~\ref{sec:axial_photon}. To ensure perturbativity at the scale $\Lambda$, we must have 
\begin{equation} \label{eq:axial_photon_lambda}
\Lambda \lesssim 4 \pi f_\gamma \min(1, m_\gamma/m) = \frac{4\pi}{g} \min(m, m_\gamma)
\end{equation}
due to longitudinal vectors and photons, respectively. This holds independently of the parameters of the UV completion; demanding the theory is perturbative up to $\Lambda \sim \mathrm{TeV}$ yields the constraint shown in Fig.~\ref{fig:axial_photon}, which is several orders of magnitude weaker than astrophysical bounds. The condition Eq.~\eqref{eq:axial_photon_lambda} is equivalent to demanding that the vector and photon masses are not tuned at one loop, $\delta m^2 \sim g^2 \Lambda^2 / (4 \pi)^2 \lesssim m^2$ and $\delta m_\gamma^2 \sim g^2 \Lambda^2 / (4 \pi)^2 \lesssim m_\gamma^2$. 

There is also a perturbativity bound on the fermion Yukawas, 
\begin{equation} \label{eq:yukawa_pert_bound}
N_f \max(y_\phi^2, y_\Phi^2) \lesssim 4 \pi.
\end{equation}
In addition, at two loops the axial-photon coupling induces a $g-2$ shift for charged fermions. However, this is much weaker than astrophysical bounds, and too weak to appear in Fig.~\ref{fig:axial_photon}.

\paragraph{Tuning Constraints.} For reference, the Higgs, photon, vector, and fermion masses are related to the parameters in Eq.~\eqref{eq:model_lag} by
\begin{subequations}
\begin{align}
m_\phi &= \sqrt{\lambda_\phi} \, v_\phi, \ \qquad m_\gamma = e Q_\phi v_\phi, \quad \qquad \ \ \, m_\chi \sim y_\phi v_\phi, \\
m_\Phi &= \sqrt{\lambda_\Phi} \, v_\Phi, \qquad m_V = e_D v_\Phi = m, \qquad m_\psi \sim y_\Phi v_\Phi.
\end{align}
\end{subequations}
To avoid a naive tuning of the Higgs mass, we must have 
\begin{subequations}
\begin{align}
m_\phi^2 &\gtrsim \frac{\Lambda^2}{(4 \pi)^2} \, \max(e^2 Q_\phi^2, N_f y_\phi^2, \lambda_\phi) \\
m_\Phi^2 &\gtrsim \frac{\Lambda^2}{(4 \pi)^2} \, \max(e_D^2, N_f y_\Phi^2, \lambda_\Phi) 
\end{align}
\end{subequations}
where $\Lambda$ is the UV cutoff. We will have extremely low $Q_\phi$ and $e_D$ to make the photon and vector ultralight, so the main constraint comes from the Yukawas. The Higgs quartics are also renormalized by a fermion box diagram. To avoid tuning of the quartics, we need 
\begin{equation}
\lambda_\phi \gtrsim \frac{N_f y_\phi^4}{(4\pi)^2}, \qquad \lambda_\Phi \gtrsim \frac{N_f y_\Phi^4}{(4\pi)^2}.
\end{equation}

\subsection{Photon Mass Bounds and Vortex Formation}
\label{app:vortex}

So far, the model parameters are relatively unconstrained. We are free to take high Higgs masses; the corresponding small values of $Q_\phi$ and $e_D$ render $g$ extremely small, but as indicated in Fig.~\ref{fig:axial_photon}, experiments can probe down to $g \sim 10^{-40}$. The main difficulty is associated with the photon mass $m_\gamma$. As we now discuss, bounds on $m_\gamma$ are much weaker when vortex formation occurs, but allowing vortex formation implies additional constraints on $m_\phi$ and $v_\phi$.

\paragraph{Criteria for Vortex Formation.} The impact of vortex formation on photon mass bounds was first discussed in Ref.~\cite{Adelberger:2003qx}. Here we review these results, using the more accurate criteria discussed in Ref.~\cite{East:2022rsi}, which are standard results for superconductors. 

We will always be in the regime $m_\gamma \ll m_\phi$, corresponding to a type II superconductor. A vortex string has a core of radius $\sim 1/m_\phi$ within which the Higgs $\phi$ goes to zero, allowing $\theta_\phi$ to have nontrivial winding. Since the observed vector potential is $A_\mu + \del_\mu \theta_\phi / (e Q_\phi)$, this can discharge large vector potentials, suppressing signals of a photon mass. 

Within a magnetic field $B$, uniform over scales $\sim 1/m_\gamma$, there are three relevant scales: 
\begin{equation}
B \gtrsim \begin{cases} B_{c1} \sim m_\gamma v_\phi & \text{lower critical field, vortices energetically favorable}, \\ B_{\mathrm{sh}} \sim m_\phi v_\phi & \text{superheating field, vortices spontaneously form}, \\ B_{c2} \sim m_\phi^2 v_\phi / m_\gamma & \text{upper critical field, restore } \phi = 0, \end{cases}
\end{equation}
where $B_{c1} \ll B_{\mathrm{sh}} \ll B_{c2}$. When vortices are present, the energy is minimized for a surface density $\sim e Q_\phi B$ of vortices, corresponding to a vortex spacing
\begin{equation}
d \sim \sqrt{\frac{1}{e Q_\phi B}} \sim \begin{cases} 1/m_\gamma & B = B_{c1}, \\ 1/m_\phi & B = B_{c2}, \end{cases}
\end{equation}
i.e.~the symmetry is restored when the vortices overlap. As for the superheating field\footnote{Ref.~\cite{Adelberger:2003qx} assumes vortex formation occurs spontaneously in a vector potential $A \gtrsim B_{\mathrm{sh}} / m_\gamma$, which is strictly weaker than the criterion we use. Ref.~\cite{photon_paper} will discuss refinements of our criteria involving plasma effects, as well as the possibility of even higher $m_\gamma$, but these do not affect the validity of our benchmarks here.}, it is roughly when the magnetic field energy density exceeds the Higgs vacuum energy, $B^2 \gtrsim m_\phi^2 v_\phi^2$.

The effects of a photon mass are equivalent to an effective current $\v{J}_{\mathrm{eff}} = - m_\gamma^2 \v{A}$, and in the absence of vortices, a magnetic field $B$ extending over length $L$ yields a vector potential $A \sim B L$. Following Ref.~\cite{PhysRevLett.80.1826}, one can estimate
\begin{equation} \label{eq:vector_benchmarks}
A \sim \begin{cases} B_{\mathrm{mag}} L_{\mathrm{mag}} \sim (10 \, \mathrm{T}) (1 \, \mathrm{m}) & \text{large laboratory magnet,} \\ B_\oplus R_\oplus \sim (3 \times 10^{-5} \, \mathrm{T}) (6 \times 10^6 \, \mathrm{m}) \sim 200 \, \mathrm{T} \, \mathrm{m} & \text{Earth's field,} \\ B_{\mathrm{gal}} L_{\mathrm{gal}} \sim (10^{-10} \, \mathrm{T}) (600 \, \mathrm{pc}) \sim 2 \times 10^9 \, \mathrm{T} \, \mathrm{m} & \text{galactic field,} \end{cases}
\end{equation}
where $600 \, \mathrm{pc}$ is an estimate for the scale of reversals of direction of the galactic magnetic field. It is also useful to note that $1 \, \mathrm{GeV} \simeq 1 \, \mathrm{T} \, \mathrm{m}$.

The large Galactic vector potential is responsible for the strongest constraints on the photon mass. However, in the presence of vortex formation, it is discharged to 
\begin{equation} \label{eq:A_res_bound}
A_{\mathrm{res}} \sim B d \sim \sqrt{\frac{B}{e Q_\phi}}.
\end{equation}
Note that when vortices are energetically favorable at all, $B \gtrsim B_{c1}$, we automatically have $A_{\mathrm{res}} \lesssim B/m_\gamma$, which is already enough to invalidate many photon mass bounds, as we will discuss below. If vortex formation is spontaneous, $B \gtrsim B_{\mathrm{sh}}$, we typically have $A_{\mathrm{res}} \ll B/m_\gamma$. 

\paragraph{Photon Mass Bounds.} The photon mass $m_\gamma$ has been probed by a variety of experiments, reviewed in Refs.~\cite{Tu_2005,Goldhaber:2008xy}. We first consider experiments insensitive to vortex formation. The best laboratory bounds are from a $1971$ Cavendish test of Coulomb's law~\cite{Williams:1971ms}, which found $m_\gamma \leq 10^{-14} \, \mathrm{eV}$, and from production and detection of longitudinal photons in DarkSRF, which found $m_\gamma \leq 1.6 \times 10^{-15} \, \mathrm{eV}$~\cite{Kalia:2025afc}. One can also probe the photon mass using the dispersion of light; the effect is enhanced for low-frequency light traveling over long distances. Ref.~\cite{Malta:2022zdb} considered Schumann resonances, and claimed $m_\gamma \lesssim 3 \times 10^{-14} \, \mathrm{eV}$. Several works considered radio waves from pulsars, with the strongest recent claim being $m_\gamma \lesssim 2 \times 10^{-15} \, \mathrm{eV}$~\cite{Wang:2023fnn}, but these tests depend on systematics, such as dispersion due to the interstellar medium. Based on these results, we take $m_\gamma = 10^{-15} \, \mathrm{eV}$ as an allowed benchmark value. 

The strongest proposed constraints depend on the galactic vector potential. One simple and very old estimate~\cite{yamaguchi1959composite,chibisov1976astrophysical} is to demand that the vector potential energy does not exceed the magnetic field energy, $m_\gamma^2 A^2 \leq B_{\mathrm{gal}}^2$, to avoid changing galactic plasma dynamics (though see Ref.~\cite{Goldhaber:2008xy} for counterarguments). Without vortex formation, we have $A \sim B_{\mathrm{gal}} L_{\mathrm{gal}}$, which implies $m_\gamma \lesssim 1/L_{\mathrm{gal}} \sim 10^{-26} \, \mathrm{eV}$. However, in the presence of a vortex lattice at equilibrium, Eq.~\eqref{eq:A_res_bound} implies $m_\gamma^2 A_{\mathrm{res}}^2 \leq B_{\mathrm{gal}}^2$, invalidating the bound. Similarly, torsion balance experiments using the galactic vector potential~\cite{PhysRevLett.80.1826,PhysRevLett.90.081801} would constrain $m_\gamma \lesssim 10^{-18} \, \mathrm{eV}$ without vortex formation, but with vortex formation they permit $m_\gamma = 10^{-15} \, \mathrm{eV}$. 

Other constraints depend on the dynamics of the solar wind~\cite{Ryutov:1997zz,Ryutov:2007zz,PhysRevLett.103.201803} and Jupiter's magnetic field profile~\cite{Yan:2023kdg}. For $m_\gamma = 10^{-15} \, \mathrm{eV}$, these all involve magnetic fields at least as strong as $B_{\mathrm{gal}}$, extending over a length $\gtrsim 1/m_\gamma$, so they are also weakened by vortex formation. 

\paragraph{Dark Vortex Formation.} Though we do not investigate DM production mechanisms in detail in this work, we note that dark vortex formation can deplete the DM abundance, invalidating many vector DM production mechanisms~\cite{East:2022rsi}. This applies to both ordinary vector DM and axial vector DM, as long as the DM mass arises from the Higgs mechanism. If the DM is produced at a Hubble rate $H \sim m$, then dark vortex formation is avoided if~\cite{Cyncynates:2023zwj}
\begin{equation}
e_D \lesssim 10^{-14} \, \lambda_\Phi^{1/4} \left( \frac{m}{10^{-6} \, \mathrm{eV}} \right)^{5/8}.
\end{equation}
For comparison, $e_D \sim 10^{-13}$ for the heavy Higgs benchmark in App.~\ref{app:benchmarks}. In that case one could simply take a somewhat lower $e_D$ and a somewhat lower axial-photon coupling, and it would be interesting to explore the full parameter space in more detail. For instance, one could ask if the nonminimal dark photon models in Refs.~\cite{Cyncynates:2023zwj,Cyncynates:2024yxm} are compatible with an observably large axial-photon coupling. 

\subsection{Three Benchmark Points}
\label{app:benchmarks}

Now that we have laid out all the constraints, we consider three benchmark points.

\paragraph{Heavy Higgs Benchmark.} Suppose vortices have formed by some mechanism, and discharge the vector potential of the galaxy. In this case, we need only demand the vortex phase is energetically favorable, $B_{\mathrm{gal}} \gtrsim B_{c1}$, which allows a relatively high value of $v_\phi$. This permits both Higgs bosons to be relatively heavy. For simplicity, we can take $\lambda_\phi \sim \lambda_\Phi \sim 1$ and
\begin{equation}
m_\phi \sim m_\Phi \sim v_\phi \sim v_\Phi \sim 10^7 \, \mathrm{eV}
\end{equation}
which implies the small couplings 
\begin{equation}
e Q_\phi = \frac{m_\gamma}{v_\phi} \sim 10^{-22}, \qquad e_D = \frac{m}{v_\Phi} \sim 10^{-13} \, \frac{m}{10^{-6} \, \mathrm{eV}}.
\end{equation}
For simplicity we take Yukawa couplings $y_\phi \sim y_\Phi \sim 1$, which implies $m_\chi \sim m_\psi \sim 10^7 \, \mathrm{eV}$, so that the relevant millicharge constraints are from supernova bounds~\cite{Chang:2018rso},  
\begin{equation}
N_f \max(Q_f^2, Q_\phi^2) \lesssim 10^{-16}.
\end{equation}
The simplest case is to take $N_f = 1$ and $Q_f \sim 10^{-8}$, which gives a coupling
\begin{equation}
g \simeq \frac{e}{2 \pi^2} \, N_f Q_f (e Q_\phi) e_D \sim 10^{-45} \, \frac{m}{10^{-6} \, \mathrm{eV}}.
\end{equation}
This is near what can be probed by near-term experiments, albeit several orders of magnitude smaller. The Higgs quartics are untuned, and the Higgs masses are natural up to $\Lambda \sim 10^8 \, \mathrm{eV}$.

It would also be possible to further increase $g$. For instance, we could take $Q_f \sim 10^{-5}$, where the supernova bound is inapplicable due to trapping. We could also increase $N_f$ while keeping $N_f Q_f^2$ constant, at the cost of making the Higgs masses less natural. 

We will not consider vortex formation mechanisms in detail, but one possibility is that vortices were formed earlier in the universe; in this case one would have to decrease $B_{c1}$ by several orders of magnitude, so that it is smaller than the primordial magnetic field. Vortices could also form spontaneously around compact objects such as magnetars. (Our point has $B_{\mathrm{sh}} \sim 10^{12} \, \mathrm{T}$, so one would have to decrease $B_{\mathrm{sh}}$ by a few orders of magnitude.) 

As for the vector potential in the lab, note that the vortices in the galaxy have very small Higgs cores ($1/m_\phi \sim 10^{-14} \, \mathrm{m}$), but are widely spaced ($d_{\mathrm{gal}} \sim 10^5 \, \mathrm{km}$). Over the lifespan of the Earth, galactic vortices may have accumulated to discharge the Earth's vector potential. If this process reaches equilibrium, they will be separated by $d_\oplus \sim 10^2 \, \mathrm{km}$, yielding a residual $A_{\mathrm{res}} \sim B_\oplus d_\oplus \sim 10 \, \mathrm{T} \, \mathrm{m}$. The vector potential of a laboratory magnet may also be suppressed by accumulating Earth-bound vortices near it, though it is not clear if this would happen on a realistic timescale, since the size of a magnet is much smaller than $d_\oplus$. 

\paragraph{Light Higgs Benchmark.} We can also impose the more stringent criterion $B_{\mathrm{gal}} \gtrsim B_{\mathrm{sh}}$, so that galactic vortex formation is inevitable. This forces $m_\phi$ to be much lower, so that we cannot avoid a naively unnatural Higgs mass; instead we simply allow the mass to be tuned. 

Concretely, we may take $m_\phi \sim 10^{-10} \, \mathrm{eV}$ and $v_\phi \sim 10^3 \, \mathrm{eV}$, which implies $e Q_\phi \sim 10^{-18}$ and $B_{\mathrm{sh}} \sim 5 \times 10^{-10} \, \mathrm{T}$. It also implies $B_{c2} \sim 5 \times 10^{-5} \, \mathrm{T}$, so that the symmetry is not restored in the laboratory by the Earth's magnetic field; however, the Earth will contain vortices of radius $1/m_\phi \sim \mathrm{km}$ separated by $\sim \mathrm{few} \, \mathrm{km}$. A laboratory magnet would be too small to create a vortex by itself, but its vector potential could be suppressed if a vortex sits over the laboratory.

As for the fermions, we may choose them to be relatively heavy or light. In the heavy case, we take $y_\phi \sim 1$ so that $m_\chi \sim 10^3 \, \mathrm{eV}$. The strongest millicharge bound is from red giants~\cite{Fung:2023euv},
\begin{equation}
N_f \max(Q_f^2, Q_\phi^2) \lesssim 10^{-28},
\end{equation}
and for simplicity we take $N_f = 1$ and $Q_f \sim 10^{-14}$. The dark Higgs is free to be heavier, and for concreteness we can take $m_\Phi \sim v_\Phi \sim 10^3 \, \mathrm{eV}$ along with $y_\Phi \sim 1$. This gives a coupling
\begin{equation}
g \simeq \frac{e}{2 \pi^2} \, N_f Q_f (e Q_\phi) e_D \sim 10^{-43} \, \frac{m}{10^{-6} \, \mathrm{eV}}
\end{equation}
comparable to the heavy Higgs case; the smaller $Q_f$ compensates for the larger $Q_\phi$ and $e_D$. One could increase $g$ by several orders of magnitude by increasing $N_f$ or decreasing $v_\Phi$.

The price of taking heavy fermions here is that the quartic $\lambda_\phi$ is also tuned. To avoid this problem one could take $y_\phi \sim 10^{-6}$ so that $m_\chi \sim 10^{-3} \, \mathrm{eV}$. Note that the fermion mass is high enough for the axial-photon EFT to be valid in most laboratory DM experiments, but low enough to require reconsidering many astrophysical bounds, along the lines of Refs.~\cite{Masso:2006gc,Brzeminski:2026xly}.

\paragraph{Benchmark Without Vortex Formation.} Finally, we could suppose vortices do not form, and simply accept a much lower photon mass, $m_\gamma = 10^{-26} \, \mathrm{eV}$. 

\begin{figure} 
\centering
\includegraphics[width=\textwidth]{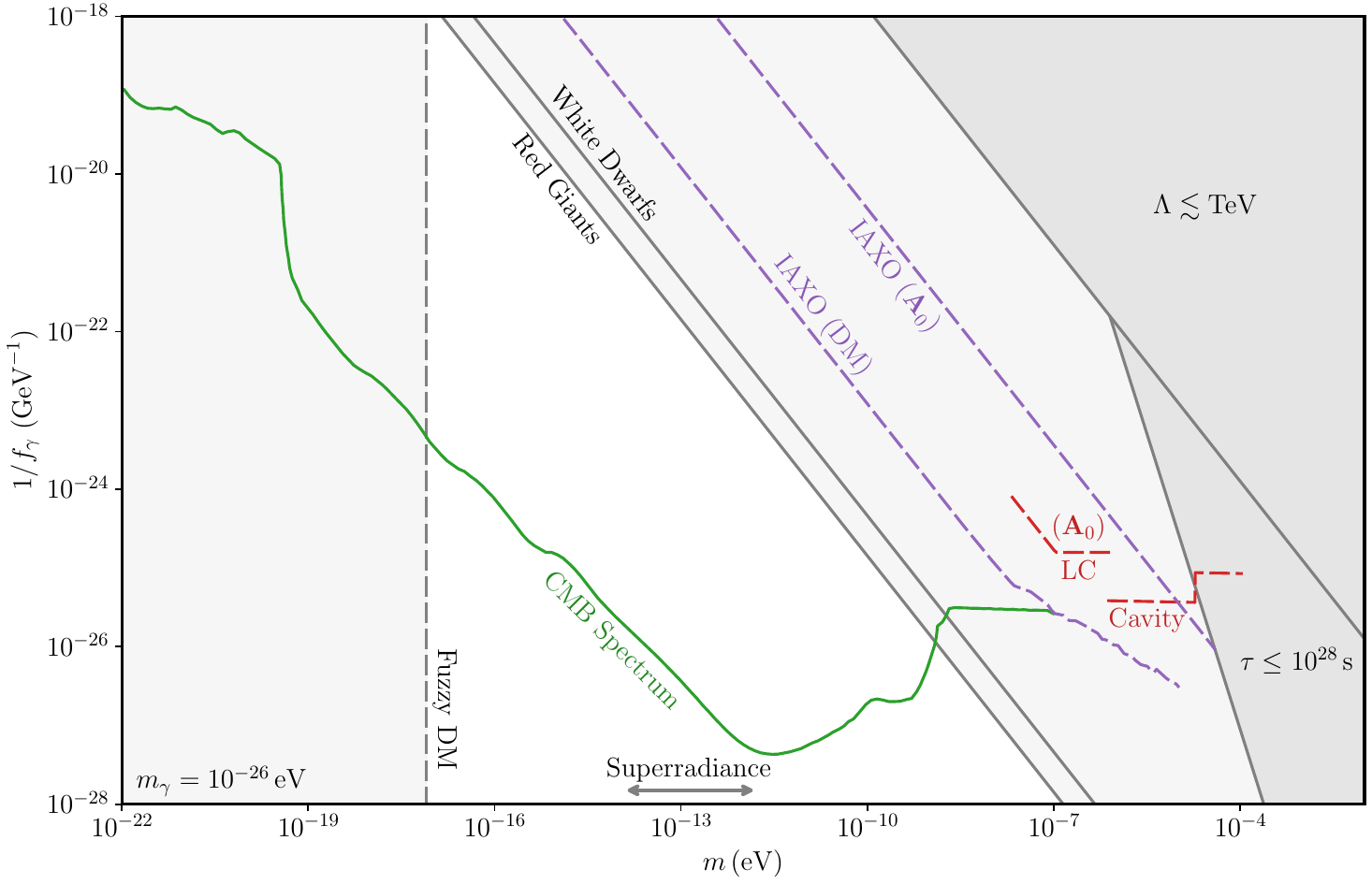}
\caption{Current constraints and projected future sensitivity to the axial-photon coupling, for a photon mass $m_\gamma = 10^{-26} \, \mathrm{eV}$ assuming no vortex formation. The color scheme is the same as in Fig.~\ref{fig:axial_photon}.}
\label{fig:axial_photon_low}
\end{figure}

In Fig.~\ref{fig:axial_photon_low} we show the analogue of Fig.~\ref{fig:axial_photon} for this alternative benchmark. The main difference is that signatures enhanced by $1/m_\gamma$ become much stronger, so that the CMB spectrum becomes the strongest probe by far. Some laboratory probes are also enhanced, to a lesser extent, by the large vector potential sourced by the galaxy (see Eq.~\eqref{eq:vector_benchmarks}). We have not worked out second-order effects in detail, but since they scale with an additional factor of $\sqrt{\rho_\DM} / (f_\gamma m_\gamma)$, they may become dominant for laboratory haloscopes and polarimeters. 

A very wide range of parameters are possible. For example, we could take $m_\phi \sim m_\Phi \sim v_\phi \sim v_\Phi \sim m_\chi \sim m_\psi \sim 10^8 \, \mathrm{eV}$ and $N_f = 1$, which is untuned up to $\Lambda \sim \mathrm{GeV}$. For these masses, millicharge constraints are much weaker, and we can take $Q_f \sim 10^{-3}$. The coupling is
\begin{equation}
g \sim 10^{-53} \, \frac{m}{10^{-6} \, \mathrm{eV}},
\end{equation}
and for reference, at this photon mass the strongest astrophysical bounds reach $g \sim 10^{-44}$. Alternatively, a substantially higher coupling could be achieved for very light fermions, $m_\chi \sim m_\psi \sim 10^{-3} \, \mathrm{eV}$, again at the cost of tuning. 

%!TEX root = main.tex

\section{Massive Photons in Medium}
\label{app:propagation}

When a massive photon propagates through a conducting medium, there are four physical polarizations: two transverse plasmons $\gamma_T^p$, the longitudinal plasmon $\gamma_L^p$, and the longitudinal photon $\gamma_L^0$. In App.~\ref{app:photon_dof} we show that $\gamma_L^0$ has potentials enhanced by $1/m_\gamma$ and field strengths suppressed by $m_\gamma$, and remains largely distinct from the longitudinal plasmon. In App.~\ref{app:photon_prop} we show how axial vector DM rotates photons among the transverse and longitudinal polarizations. Finally, in App.~\ref{app:plasmons} we consider channels for plasmon decay. 

\subsection{Transverse and Longitudinal Modes}
\label{app:photon_dof}

We consider photons propagating through a cold, collisionless, nonrelativistic plasma, as relevant for a number of astrophysical and cosmological probes.

\begin{figure}[t]
\centering
\includegraphics[scale=1.0]{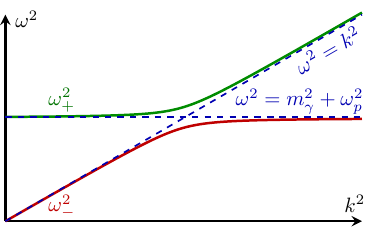}
\caption{Longitudinal modes for a massive photon in medium. There is a narrow avoided crossing near $k = \omega_p$, whose width is controlled by the very small photon mass $m_\gamma$. Below the avoided crossing, the $\omega_-$ and $\omega_+$ branches correspond to the longitudinal photon $\gamma_L^0$ and longitudinal plasmon $\gamma_L^p$ respectively, while above the avoided crossing they switch roles.}
\label{fig:dispersion}
\end{figure}

\paragraph{Mode Frequencies.} The equation of motion is 
\begin{equation}
(\del^2 + m_\gamma^2) A^\mu = J^\mu 
\end{equation}
where the current in the plasma is $\v{J} = \sigma \v{E}$, and in frequency space $\sigma = i \omega_p^2 / \omega$ where $\omega_p$ is the plasma frequency. Now consider plane waves proportional to $e^{i (k z - \omega t)}$. Then we have $J^x = - \omega_p^2 A^x$ and $J^y = - \omega_p^2 A^y$, which implies 
\begin{equation} \label{eq:transverse_disp}
\omega^2 = k^2 + m_\gamma^2 + \omega_p^2
\end{equation}
for the transverse modes. As for the longitudinal modes, note that conservation of $J^\mu$ implies $\del_\mu A^\mu = 0$, so that $\omega A^0 = k A^z$. Then the longitudinal electric field is 
\begin{equation} \label{eq:ez_expr}
E^z = - i k A^0 + i \omega A^z = \frac{i}{\omega} (\omega^2 - k^2) A^z. 
\end{equation}
Substituting this into the $z$-component of the equation of motion yields 
\begin{equation} \label{eq:m_gamma_expr}
(\omega^2 - k^2) (1 - \omega_p^2 / \omega^2) = m_\gamma^2
\end{equation}
which has solutions 
\begin{subequations}
\begin{align}
\omega_\pm^2 &= \frac12 \left( k^2 + m_\gamma^2 + \omega_p^2 \pm \Delta \right), \\
\Delta &= \sqrt{\left(k^2+m_\gamma^2+\omega_p^2\right)^2 -4k^2\omega_p^2} \,.
\end{align}
\end{subequations}
As shown in Fig.~\ref{fig:dispersion}, away from the avoided crossing, there is always a longitudinal photon mode with $\omega \simeq k$, and a longitudinal plasmon mode with $\omega \simeq \omega_p$. 

\paragraph{Mode Polarizations.} To compute matrix elements, we also need to find canonically normalized polarization vectors for each mode. In general, this does not correspond to imposing $\epsilon^\mu \epsilon_\mu = -1$. Instead, the proper criterion is that if one sets $A^\mu \to \epsilon^\mu e^{i (kz - \omega t)}$, then the total energy density becomes $T^{00} \to \omega^2$. Moreover, in addition to the field energy in Eq.~\eqref{eq:vacuum_energy_density}, the medium carries electron kinetic energy density $(\omega_p^2 / \omega^2) E^2/2$.

It is straightforward to show that the transverse modes still have $\epsilon^\mu = (0, 1, 0, 0)$ and $(0, 0, 1, 0)$. The longitudinal polarization vectors are more involved; one must use $\omega \epsilon^0 = k \epsilon^z$, then apply Eq.~\eqref{eq:m_gamma_expr} and the identity $\omega_\pm^4 - k^2 \omega_p^2 = \pm \omega_\pm^2 \Delta$, to show that
\begin{equation}
\epsilon^\mu_\pm = \sqrt{\frac{|\omega_\pm^2 - \omega_p^2|}{\Delta}} \, (k/m_\gamma, 0, 0, \omega_\pm/m_\gamma).
\end{equation}
Despite appearances, this polarization vector is not always enhanced by $1/m_\gamma$. For instance, in the limit $k \gg \omega_p$, relevant for most thermal plasmons in many astrophysical systems, 
\begin{subequations}
\begin{align}
\epsilon^\mu_+ &\simeq (k/m_\gamma, 0, 0, \omega_+/m_\gamma), \\
\epsilon^\mu_- &\simeq (\omega_p/k, 0, 0, \omega_p^2 / k^2).
\end{align}
\end{subequations}
In general, the longitudinal photon $\gamma_L^0$ has a polarization vector enhanced by $1/m_\gamma$, and an electric field suppressed by $m_\gamma$, while the longitudinal plasmon $\gamma_L^p$ has neither. This confirms the statements made in Sec.~\ref{sec:axial_photon}.

\paragraph{Landau--Zener Transitions.} As a longitudinal photon propagates out of a star, the environmental value of $\omega_p$ decreases. In principle, it can transition to a longitudinal plasmon when $k$ passes $\omega_p$, but the probability for this to occur is highly suppressed. 

To see this, note that in terms of the variables $\mathcal{A} = m_\gamma A_z$ and $\mathcal{E} = - i E_z$, Eq.~\eqref{eq:ez_expr} and the $z$-component of the equation of motion yield
\begin{equation}
\omega^2 \begin{pmatrix} \mathcal{A} \\ \mathcal{E} \end{pmatrix} = \begin{pmatrix} k^2 & m_\gamma \omega \\ m_\gamma \omega & \omega_p^2 \end{pmatrix} \begin{pmatrix} \mathcal{A} \\ \mathcal{E} \end{pmatrix}.
\end{equation}
Far from the avoided crossing, which has width $\Delta k \sim m_\gamma$, the longitudinal plasmon and photon correspond to $\mathcal{E}$ and $\mathcal{A}$ respectively. 

As we will discuss in App.~\ref{app:photon_prop}, the above equation corresponds to a first-order differential equation for the amplitudes, 
\begin{equation}
i \del_t \begin{pmatrix} \bar{\mathcal{A}} \\ \bar{\mathcal{E}} \end{pmatrix} \simeq \frac12 \begin{pmatrix} k & m_\gamma \omega/k \\ m_\gamma \omega/k & \omega_p^2/k \end{pmatrix} \begin{pmatrix} \bar{\mathcal{A}} \\ \bar{\mathcal{E}} \end{pmatrix}.
\end{equation}
Near the avoided crossing, $\omega_p^2 / k \simeq 2 \omega_p - k$, and the transition probability is given by the Landau--Zener formula, 
\begin{equation}
P(\gamma_L^0 \to \gamma_L^p) \sim 1 - \exp\left( -\frac{\pi}{2} \frac{m_\gamma^2}{|\del \omega_p / \del t|} \right) \sim \frac{m_\gamma^2}{\omega_p} \bigg|\frac{\del \log \omega_p}{\del t}\bigg|^{-1}
\end{equation}
with the derivatives evaluated at the avoided crossing. At our benchmark $m_\gamma = 10^{-15} \, \mathrm{eV}$, this is numerically very small in any astrophysical setting. The effect of the avoided crossing is negligible, and we can think of the longitudinal photon and plasmon as separate modes. 

\subsection{Photon Propagation in Axial Dark Matter}
\label{app:photon_prop}

Now we consider a plane-wave photon of energy $\omega$ propagating through axial vector DM, where $\omega \gg m, m_\gamma, \omega_p$. We will find a transverse polarization rotation, similar to axion DM, and mixing with the longitudinal photon, which is enhanced by powers of $1/m_\gamma$. 

\paragraph{Defining the Longitudinal Amplitude.} The equation of motion is
\begin{equation}
(\del^2 + m_\gamma^2) A^\nu = J^\nu + J_{\mathrm{eff}}^\nu + \frac{1}{m_\gamma^2} \, \del^\nu (\del_\mu J^\mu_{\mathrm{eff}}). 
\end{equation}
However, solving it directly is subtle, because the vector potential $A^\mu$ contains large contributions from the longitudinal photon, which mostly cancel in the field strength. To make the power counting manifest, it is convenient to introduce an auxiliary variable $\alpha$ by
\begin{equation} \label{eq:scalar_part_def}
A^\mu \equiv \tilde{A}^\mu + \frac{\del^\mu \alpha}{m_\gamma} = \left(\tilde{A}^0 + \dot{\alpha}/m_\gamma, \tilde{\v{A}} - \nabla \alpha / m_\gamma \right)
\end{equation}
and then impose the ``massive Lorenz'' condition $\del_\mu \tilde{A}^\mu = m_\gamma \alpha$, which implies
\begin{subequations} \label{eq:reduced_eoms}
\begin{align} \label{eq:reduced_alpha_eom}
(\del^2 + m_\gamma^2) \alpha &= \frac{1}{m_\gamma} \, \del_\mu J_{\mathrm{eff}}^\mu, \\
(\del^2 + m_\gamma^2) \tilde{A}^\nu &= J^\nu + \frac{g}{4} \, \epsilon^{\mu\nu\rho\sigma} (2 V_\mu F_{\rho\sigma} + (\tilde{A}_\sigma + \del_\sigma \alpha / m_\gamma) F^V_{\rho \mu}). \label{eq:reduced_A_eom}
\end{align}
\end{subequations}
Above, only $\tilde{A}^\mu$ contributes to the field strength. The longitudinal photon is dominantly encoded in $\alpha$, and the smaller $\tilde{A}^0$ and $\tilde{A}^z$ yield its longitudinal field.

Here $\alpha$ is like a St\"uckelberg scalar: it introduces a redundancy $\delta \tilde{A}^\mu = - \del^\mu (\delta \alpha) / m_\gamma$, related to the (broken) electromagnetic gauge symmetry. It is not needed, but it will make the $1/m_\gamma$ enhancement of conversion between transverse and longitudinal photons manifest.

\paragraph{Equations of Motion.} We now evaluate the equations of motion Eqs.~\eqref{eq:reduced_eoms} explicitly. For simplicity, we neglect DM velocity effects, so that all quantities depend only on $t$ and $z$, and we discard $\tilde{A}^0$, which is fixed by the massive Lorenz condition. This yields
\begin{subequations}
\begin{align}
(\del^2 + m_\gamma^2) \alpha &\simeq - \frac{g}{2 m_\gamma} \, \v{B} \cdot \dot{\v{V}} \label{eq:alpha_eom} \\
(\del^2 + m_\gamma^2) \tilde{A}^z &\simeq J^z - g \left(\v{E} \times \v{V} + \frac12 \dot{\v{V}} \times \tilde{\v{A}} \right) \cdot \hat{\v{z}} \label{A_tilde_z_eom} \\
(\del^2 + m_\gamma^2) \tilde{A}^i &\simeq J^i - g \left(\v{E} \times \v{V} + \frac12 \dot{\v{V}} \times \tilde{\v{A}} - \frac{\del_z \alpha}{2 m_\gamma} \dot{\v{V}} \times \hat{\v{z}} \right)^i \label{A_tilde_eom}
\end{align}
\end{subequations}
where the last line describes transverse photons, with $i = \{x, y\}$. 

We now identify the dominant source terms. First, if we start with a transverse photon, we can ignore the effect of $\dot{\v{V}} \times \tilde{\v{A}}$, as it is suppressed by $m/\omega$ relative to $\v{E} \times \v{V}$. Next, transverse-longitudinal conversion is only competitive when it is enhanced, $m \gg m_\gamma$, since the longitudinal photon is almost undetectable. In this limit, comparing the source terms above yields $\tilde{A}^z \sim (m_\gamma / m) \alpha$, so the longitudinal photon is almost entirely encoded in $\alpha$. Moreover, given a longitudinal photon amplitude, the final source term in Eq.~\eqref{A_tilde_eom} dominates.

Keeping only the most relevant terms, the equations of motion simplify to
\begin{subequations} \label{eq:simplified_eom}
\begin{align}
(\del^2 + m_\gamma^2) \alpha &\simeq - \frac{g}{2 m_\gamma} \, \v{B} \cdot \dot{\v{V}}, \label{eq:alpha_eom_2} \\
(\del^2 + m_\gamma^2) \tilde{A}^i &\simeq J^i - g \left(V_z \, \v{E} \times \hat{\v{z}} - \frac{\del_z \alpha}{2 m_\gamma} \dot{\v{V}} \times \hat{\v{z}} \right)^i. \label{A_tilde_eom_2}
\end{align}
\end{subequations}
To solve for the evolution of the amplitudes, we follow the procedure in Ref.~\cite{PhysRevD.37.1237}. For $f \in \{\tilde{A}^x, \tilde{A}^y, \alpha\}$, we define $f = \bar{f} e^{i (kz - \omega t)}$ where $\bar{f}(z)$ is a slowly varying amplitude, and $k$ is determined by the transverse dispersion relation Eq.~\eqref{eq:transverse_disp}. We then have
\begin{equation}
(\del^2 + m_\gamma^2) f = - (\omega^2 - m_\gamma^2 + \del_z^2) f \simeq - e^{i (kz - \omega t)} (\omega_p^2 \bar{f} + 2 i k \del_z \bar{f}).
\end{equation}
To define the amplitudes of the currents on the right-hand sides of Eqs.~\eqref{eq:simplified_eom}, we neglect the time variation of the DM, as its effect is suppressed by $m/\omega$. Then Eqs.~\eqref{eq:simplified_eom} become
\begin{equation} \label{eq:amplitude_evolution}
\del_z \begin{pmatrix} \bar{A}^x \\ \bar{A}^y \\ \bar{\alpha} \end{pmatrix} \simeq \frac{g}{2} \begin{pmatrix} 0 & V^z - V^0 & - \dot{V}^y / 2 m_\gamma \\ V^0 - V^z & 0 & \dot{V}^x / 2 m_\gamma \\ \dot{V}^y / 2 m_\gamma & - \dot{V}^x / 2 m_\gamma & 0 \end{pmatrix} \begin{pmatrix} \bar{A}^x \\ \bar{A}^y \\ \bar{\alpha} \end{pmatrix} + \frac{i \omega_p^2}{2 \omega} \begin{pmatrix} 0 \\ 0 \\ \bar{\alpha} \end{pmatrix}
\end{equation}
where we used $\omega \simeq k$ and restored the dependence on $V^0$, which is velocity suppressed. The final term appears because the longitudinal photon has $\omega^2 \simeq k^2 + m_\gamma^2$, without the additional $\omega_p^2$ for the transverse modes. This result is the starting point for Sec.~\ref{sec:propagation}.

\paragraph{Enhanced Second-Order Effects.} At second order, polarization rotation effects need not average to zero over a full DM oscillation, because three-dimensional rotations do not commute. For instance, for a circularly polarized vector which has $V^x$ positive, then $V^y$ positive, then $V^x$ negative, and then $V^y$ negative, the net effect of a full DM oscillation is to rotate the photon's polarization about the $z$-axis. 

To see this in detail, define the vector amplitude $\v{A} = (\bar{A}^x, \bar{A}^y, \bar{\alpha})$ and the rotation matrices $(J_i)_{jk} = - \epsilon_{ijk}$. Neglecting $\omega_p$ for simplicity, Eq.~\eqref{eq:amplitude_evolution} is equivalent to 
\begin{equation} \label{eq:A_eom}
\del_z \v{A} = -\frac{g}{4} \left(\frac{\dot{V}^x}{m_\gamma} J_x + \frac{\dot{V}^y}{m_\gamma} J_y + 2 V^z J_z \right) \v{A} = - \frac{g}{4} M(t) \v{A},
\end{equation}
where we dropped the velocity-suppressed $V^0$ term, and the vector field is evaluated at $(t, z) = (t, t)$. The solution after propagating a distance $\tau$ is the path-ordered exponential
\begin{align}
\v{A}(\tau) &= \mathcal{P} \exp \left(- \frac{g}{4} \int_0^\tau dt\, M(t) \right) \v{A}(0) \\
&= \bigg( 1 - \frac{g}{4} \int_0^\tau dt\, M(t) + \frac{g^2}{16} \int_0^\tau dt \int_0^t dt' \, M(t) M(t') \bigg) \v{A}(0) + \mathcal{O}(g^3).
\end{align}
Within the second-order term, terms proportional to anticommutators of the $J_i$ can be rewritten as a product of two independent integrals from $0$ to $\tau$ which each average to zero. The contributions which grow with $\tau$ come only from commutators. The term proportional to $[J_x, J_y] = J_z$ is the largest, as it is enhanced by $1/m_\gamma^2$, and the most important, since it causes an observable transverse polarization rotation. Focusing on it, we have
\begin{equation}
I_2 = \int_0^\tau dt \int_0^t dt' \, M(t) M(t') \supset \frac{J_z}{2 m_\gamma^2} \int_0^\tau dt \int_0^t dt' \, \dot{V}^x(t)\dot{V}^y(t') - \dot{V}^y(t)\dot{V}^x(t').
\end{equation}
Neglecting spatial gradients of the DM, we may perform the $t'$ integral to find 
\begin{equation}
I_2 \supset \frac{J_z}{2 m_\gamma^2} \int_0^\tau dt \, \dot{V}^x(t) (V^y(t) - V^y(0)) - \dot{V}^y(t) (V^x(t) - V^x(0)).
\end{equation}
The $\dot{V}^x(t) V^y(0)$ and $\dot{V}^y(t) V^x(0)$ terms average to zero. Discarding them, we have
\begin{equation}
I_2 \supset \frac{J_z}{2 m_\gamma^2} \int_0^\tau dt \, \dot{V}^x(t) V^y(t) - \dot{V}^y(t) V^x(t) \simeq \frac{J_z m}{2 m_\gamma^2} \int_0^\tau dt \, V^x_0 V^y_0 \, \sin \phi,
\end{equation}
where $V^x_0$ and $V^y_0$ are the oscillation amplitudes of $V^x$ and $V^y$, and $\phi$ is their phase difference. Then for a transverse photon traveling towards $+\hat{\v{z}}$, the net polarization rotation is
\begin{equation} \label{eq:total_rotation}
\Delta \theta \simeq -\frac{g}{2} \int_0^\tau dt \, V^z(t) + \frac{g^2 m}{32 m_\gamma^2} \int_0^\tau dt \, V^x_0(t) V^y_0(t) \sin \phi(t).
\end{equation}
The second-order effect depends on the vector's ellipticity; it vanishes for a linearly polarized vector, but is generically nonzero for a randomly polarized vector. 

\paragraph{Photon Conversion on Astrophysical Scales.} Axial vector DM can resonantly convert to transverse photons when the plasma frequency matches the DM frequency, $\omega_p \simeq m$. We have not considered these signals in detail, because on cosmological scales they are suppressed by plasma effects~\cite{Hook:2025pbn}, which may also be relevant on astrophysical scales. 

Still, if these plasma effects do not suppress the result, then we have the scaling
\begin{equation} \label{eq:axion_dp_resonant}
P(\mathrm{DM} \to \gamma) \sim \frac{1}{m v_r} \bigg| \frac{\del \log \omega_p^2}{\del r} \bigg|^{-1} \times \begin{cases} g_{a\gamma\gamma}^2 B_\perp^2 & \text{axion\ DM, background}\ \v{B}, \\ v_\DM^2 B_\perp^2 / f_\gamma^2 & \text{background}\ \v{B}, \\ E^2 / f_\gamma^2 & \text{background} \ \v{E}, \\ m^2 A^2 / f_\gamma^2 & \text{background}\ \v{A}. \end{cases}
\end{equation}
where the axial vector results are inferred from the transverse part of the effective current in Eq.~\eqref{eq:jbar_full_expr}. Since the effect of background $\v{B}$ is velocity suppressed, the axion DM results of Refs.~\cite{An:2023wij,Todarello:2023ptf,Beadle:2024jlr} become much weaker than other astrophysical bounds when recast to axial vector DM. As for background $\v{E}$, in many astrophysical systems it is much smaller than $\v{B}$, but the Earth's typical electric field $E_\oplus \sim 100 \, \mathrm{V}/\mathrm{m}$ obeys $E_\oplus / (v_\DM B_\oplus) \sim 10$. Thus, there could be an enhanced electric analogue of the geomagnetic signal in Ref.~\cite{Arza:2021ekq}. 

The background $\v{A}$ effect can dominate if the length $L$ of the background magnetic field satisfies $m L \gtrsim v_\DM$. This is a potentially interesting signature, but it depends on how much the vector potential is suppressed by vortices, as discussed in App.~\ref{app:vortex}. `'

Finally, the reverse process can also occur. For instance, in a background magnetic field, photons convert to longitudinal vectors similarly to an axion, so one can directly recast axion constraints from polarization induced around magnetic white dwarfs~\cite{Dessert:2022yqq,Benabou:2025jcv}, and ``patchy screening'' of the CMB~\cite{Mondino:2024rif,Goldstein:2024mfp}. 

\subsection{Plasmon Decay}
\label{app:plasmons}

In Sec.~\ref{sec:cooling}, we considered astrophysical bounds associated with on-shell plasmon decay. Here we show that the processes considered there are indeed the dominant ones. 

We suppose a plasmon ($\gamma_T^p$ or $\gamma_L^p$) decays into an axial vector ($V_T$ or $V_L$) and a plasmon/photon ($\gamma_T^p$ or $\gamma_L^p$ or $\gamma_L^0$). Many combinations are kinematically forbidden in the simple medium model of App.~\ref{app:photon_dof}; more generally, they would have a suppressed phase space. In addition, only processes with a final $V_L$ or $\gamma_L^0$ are relevant, as their matrix elements can be enhanced by $\omega_0/m$ or $\omega_0/m_\gamma$, respectively, where $\omega_0$ is a typical stellar energy scale. 

This leaves three processes of interest: 
\begin{subequations}
\begin{align}
\gamma_T^p(q, \epsilon^{(i)}) &\to V_L(p, \epsilon^{(V)}) + \gamma_L^p(k, \epsilon^{(f)}), \\
\gamma_T^p/\gamma_L^p(q, \epsilon^{(i)}) &\to V_L(p, \epsilon^{(V)}) + \gamma_L^0(k, \epsilon^{(f)}), \\
\gamma_T^p/\gamma_L^p(q, \epsilon^{(i)}) &\to V_T(p, \epsilon^{(V)}) + \gamma_L^0(k, \epsilon^{(f)}).
\end{align}
\end{subequations}
The first process is the same as for an axion, recast via $g_{a\gamma\gamma} \lra g/m$, and it is known to be a subdominant part of the usual Primakoff emission rate~\cite{Raffelt:1987np}. The second process is naively enhanced by $1/(m_\gamma m)$, but is actually much smaller. To see this, note that if one uses the longitudinal polarizations $\epsilon^{(f)}_\mu \simeq k_\mu / m_\gamma$ and $\epsilon_\mu^{(V)} \simeq p_\mu / m$, then the matrix element vanishes, 
\begin{equation}
\mathcal{M} \sim g \epsilon^{\mu\nu\rho\sigma} (q_\mu + k_\mu) \epsilon^{(i)}_\nu \epsilon^{(V)}_\rho \epsilon^{(f)}_\sigma \simeq \frac{g}{m m_\gamma} \epsilon^{\mu\nu\rho\sigma} q_\mu \epsilon^{(i)}_\nu p_\rho k_\sigma = 0.
\end{equation}
We can only get a nonzero contribution from $\epsilon^{(f)}_\mu - k_\mu/m_\gamma$ or $\epsilon_\mu^{(V)} - p_\mu / m$, which are suppressed by $m_\gamma / \omega_0$ and $m / \omega_0$ respectively. Then the matrix element can at most scale as 
\begin{equation}
\mathcal{M} \sim \frac{g \omega_0}{m m_\gamma} \min(m^2, m_\gamma^2)
\end{equation}
which renders it always subdominant to either the first or third process. (This is analogous to Eq.~\eqref{eq:double_long_mixing}, and the discussion below implies that the scale $\omega_0$ here should be $\omega_p$.)

The third process is enhanced by $1/m_\gamma$, and can also occur in the dark axion portal, recast via $1/f_a \lra g / 2 m_\gamma$, if one identifies $\gamma_L^0$ with the axion. In our notation, Ref.~\cite{Arias:2020tzl} showed that for an initial transverse plasmon, $\mathcal{M} \sim g \omega_p^2 / m_\gamma$, corresponding to a decay rate $\Gamma \sim (g^2 / m_\gamma^2) (\omega_p^4 / \omega_0)$ in the plasma frame. For $T \gtrsim \omega_p$, the emitted power per volume is
\begin{equation} \label{eq:trans_plas_decay}
\frac{P}{V} \sim \int_0^T (\omega_0^2 \, d \omega_0) \left(\frac{T}{\omega_0} \right) (\omega_0) \, \Gamma \sim \frac{g^2}{m_\gamma^2} \, \omega_p^4 T^3
\end{equation}
where $T/\omega_0$ accounts for the thermal occupancy.

The contribution from longitudinal plasmon decay is usually neglected in the dark axion portal literature. To see that it is indeed subdominant for $T \gtrsim \omega_p$, note that the decay is kinematically forbidden unless $|\v{q}| \lesssim \omega_p$, and in this region the matrix element scales as $\mathcal{M} \sim g \omega_p^2 / m_\gamma$, which corresponds to $\Gamma \sim g^2 \omega_p^3 / m_\gamma^2$. The emitted power per volume scales as 
\begin{equation}
\frac{P}{V} \sim \int_0^{\omega_p} (|\v{q}|^2 \, d|\v{q}|) \left(\frac{T}{\omega_p} \right) (\omega_p) \, \Gamma \sim \frac{g^2}{m_\gamma^2} \, \omega_p^6 T
\end{equation}
which is suppressed relative to Eq.~\eqref{eq:trans_plas_decay} by $\sim \omega_p^2 / T^2$. 

As a result, all cooling bounds relevant for the axial-photon coupling can be directly recast from axion or dark axion portal constraints, as we do in Sec.~\ref{sec:cooling}. As a rough consistency check, note that the Primakoff process gives $P/V \sim g^2 T^7 / m^2$, so it is comparable to Eq.~\eqref{eq:trans_plas_decay} when $m/m_\gamma \sim (T/\omega_p)^2$. For solar emission, this corresponds to $m \sim 10^{-13} \, \mathrm{eV}$, which is consistent within an order of magnitude to the point where the recast bounds from Ref.~\cite{Arias:2020tzl} and Ref.~\cite{Vinyoles:2015aba} intersect.

%!TEX root = main.tex

\section{Other Haloscopes with Electric Background Fields}
\label{app:electric}

In Sec.~\ref{sec:haloscopes}, we stated that the axial-photon coupling is the first example of an ultralight DM effect for which electric background fields yield resonantly enhanced sensitivity, unsuppressed by the DM velocity. Here we justify this statement, comparing to prior claims in the literature. 

First, for the standard axion-photon coupling it is well-known that a static background $\v{E}_0$ yields only velocity-suppressed effects; see Ref.~\cite{Ouellet:2018nfr} for a pedagogical discussion. Accordingly, Refs.~\cite{Gao:2022zxc,Chen:2020cbs} considered targeting the velocity-suppressed axion gradient with electric background fields. For the rest of this appendix, we set DM gradients to zero for simplicity. 

Second, dilaton DM $\phi$ coupled to photons yields an effective current $\v{J}_{\mathrm{eff}} = g_\phi (\phi \v{J}_0 - \dot{\phi} \v{E}_0)$. Refs.~\cite{Flambaum:2022zuq} considered searching for it in a cavity with a static electric background field. In this case the signal power depends on the form factor 
\begin{equation}
C_\phi = \int_{V_c} \tilde{\v{E}}_0 \cdot \tilde{\v{E}}_n.
\end{equation}
However, this form factor always vanishes. To see this, note that the cavity mode $n$ obeys
\begin{subequations}
\begin{align}
\nabla \times \tilde{\v{B}}_n &= \omega_n \tilde{\v{E}}_n + \tilde{\v{J}}_n \label{eq:curl_b_mode} \\
\nabla \times \tilde{\v{E}}_n &= \omega_n \tilde{\v{B}}_n 
\end{align}
\end{subequations}
where $\tilde{\v{J}}_n$ is confined to the cavity surface $S$, with $\tilde{\v{J}}_n \perp d\v{S}$. Since $\tilde{\v{E}}_0 \parallel d \v{S}$ for conducting surfaces, we have $\tilde{\v{E}}_0 \cdot \tilde{\v{J}}_n = 0$, so Eq.~\eqref{eq:curl_b_mode} implies
\begin{equation} \label{eq:c_phi_cancels}
C_\phi = \frac{1}{\omega_n} \int_{V_c}\tilde{\v{E}}_0 \cdot (\nabla \times \tilde{\v{B}}_n) = \frac{1}{\omega_n} \int_S (\tilde{\v{B}}_n \times \tilde{\v{E}}_0) \cdot d \v{S} + \frac{1}{\omega_n} \int_{V_c}(\nabla \times \tilde{\v{E}}_0) \cdot \tilde{\v{B}}_n
\end{equation}
where we integrated by parts in the second step. These two terms both vanish, because $\v{E}_0 \parallel d \v{S}$ and $\nabla \times \v{E}_0 = 0$, respectively. The dilaton's oscillation simply scales the curl-free electric background by a factor of $1 + g_\phi \phi$, without exciting the resonant cavity modes. The same applies to resonant LC circuit modes, if one takes $S$ to be the combined surface of all the circuit elements and the circuit's conducting shield; this rules out the proposal of Ref.~\cite{Donohue:2021jbv}.

Third, Ref.~\cite{Sokolov:2022fvs} proposed alternative axion-photon couplings. Away from Standard Model (SM) sources, they correspond to adding effective electric and magnetic currents, 
\begin{subequations}
\begin{align}
\nabla \times \v{B} - \dot{\v{E}} &= - g \dot{a} \v{B} + \sl{g} \dot{a} \v{E} = \v{J}_{\mathrm{eff}}, \\ 
\nabla \times \v{E} + \dot{\v{B}} &= - \bar{g} \dot{a} \v{E} + \sl{g} \dot{a} \v{B} = -\v{J}^m_{\mathrm{eff}}, \label{eq:modified_faraday}
\end{align}
\end{subequations}
where $g = g_{a\gamma\gamma}$ is the usual axion-photon coupling. However, depending on how one treats SM sources, the new terms are either phenomenologically unviable, or equivalent to the usual axion-photon or dilaton-photon couplings. 

To see this, suppose only $\bar{g}$ is nonzero and static SM charges are present, $\nabla \cdot \v{E} = \rho$. Taking the divergence of Eq.~\eqref{eq:modified_faraday} yields $\nabla \cdot \dot{\v{B}} = - \bar{g} \dot{a} \rho$, so that $\nabla \cdot \v{B} = - \bar{g} a \rho$. In terms of the rotated fields $(\v{E}', \v{B}') = (\v{E} - \bar{g} a \v{B}, \v{B} + \bar{g} a \v{E})$, the modified Maxwell's equations read
\begin{subequations}
\begin{align}
\nabla \cdot \v{E}' &\simeq \rho, \\ 
\nabla \cdot \v{B}' &\simeq 0, \\
\nabla \times \v{B}' - \dot{\v{E}}' &\simeq \bar{g} \dot{a} \v{B}', \\
\nabla \times \v{E}' + \dot{\v{B}}' &\simeq 0,
\end{align}
\end{subequations}
at leading order in $\bar{g}$. These are just the ordinary equations of axion electrodynamics with $g_{a\gamma\gamma} = - \bar{g}$, so $\bar{g}$ simply encodes the usual axion effect in a more complicated way. If both $g$ and $\bar{g}$ are nonzero, we can always rotate to eliminate $\bar{g}$, and the physical coupling is $g_{a\gamma\gamma} = g - \bar{g}$.\footnote{Alternatively, if one takes $\nabla \cdot \v{B} = 0$, the modified Maxwell's equations can only hold away from SM charges. We would then have $\nabla \cdot \v{B}' \simeq \bar{g} a \rho$, i.e.~the coupling $\bar{g}$ is equivalent to the usual axion-photon coupling plus a ``dual Witten effect'' which gives SM electric charges a magnetic charge. Ref.~\cite{Heidenreich:2023pbi} gave a clean derivation of this effect, invoking only fields away from charges, and found that it is inconsistent with the SM.}\footnote{The duality-motivated equations in Ref.~\cite{Visinelli:2013mzg} correspond to turning on $g = \bar{g}$. These equations are either equivalent to having no axion at all (if $\nabla \cdot \v{B} = - \bar{g} a \rho$) or unviable due to the dual Witten effect (if $\nabla \cdot \v{B} \neq - \bar{g} a \rho$). Using these equations, Ref.~\cite{Anzuini:2022bqd} showed that the dual Witten effect leads to a magnetic dynamo in neutron stars, but this does not occur in standard axion electrodynamics.}

Nonetheless, Refs.~\cite{Sokolov:2022fvs,Tobar:2022rko,Tobar:2023rga,Li:2022oel,Li:2023aow} claimed that when $\bar{g}$ (also called $g_{aBB}$) is nonzero, one can obtain a large signal in a haloscope with static background $\v{E}_0$. We can show this is not the case, even without rotating fields. The electric and magnetic wave equations are 
\begin{subequations}
\begin{align}
\nabla \times \nabla \times \v{E} + \ddot{\v{E}} &= -\dot{\v{J}}_{\mathrm{eff}} - \nabla \times \v{J}^m_{\mathrm{eff}}, \\
\nabla \times \nabla \times \v{B} + \ddot{\v{B}} &= -\dot{\v{J}}^m_{\mathrm{eff}} + \nabla \times \v{J}_{\mathrm{eff}}.
\end{align}
\end{subequations}
The usual cavity haloscope signal power is derived by integrating the electric wave equation against a cavity mode $\tilde{\v{E}}_n$. For magnetic source currents, it is more convenient to integrate the magnetic wave equation against $\tilde{\v{B}}_n$, which yields a signal power depending on the form factor
\begin{align}
\bar{C} &= \int_{V_c} \tilde{\v{E}}_0 \cdot \tilde{\v{B}}_n \\
&= \frac{1}{\omega_n} \int_{V_c}\tilde{\v{E}}_0 \cdot (\nabla \times \tilde{\v{E}}_n) = \frac{1}{\omega_n} \int_S (\tilde{\v{E}}_n \times \tilde{\v{E}}_0) \cdot d \v{S} + \frac{1}{\omega_n} \int_{V_c}\tilde{\v{E}}_n \cdot (\nabla \times \tilde{\v{E}}_0) = 0
\end{align}
where both terms vanish for the same reasons as in Eq.~\eqref{eq:c_phi_cancels}.

Similarly, the coupling $\sl{g}$ (also called $g_{aAB}$) is equivalent to the dilaton-photon coupling. This is easiest to see in covariant notation, where including SM sources self-consistently yields 
\begin{subequations}
\begin{align}
\del_\mu ((1 + \sl{g}a) F^{\mu\nu}) &= J^\nu, \\
\del_\mu ((1 - \sl{g}a) \tilde{F}^{\mu\nu}) &= 0.
\end{align}
\end{subequations}
The scaled fields ${F'}^{\mu\nu} = (1 - \sl{g} a) F^{\mu\nu}$ obey the usual Bianchi identity and $\del_\mu ((1 + 2 \sl{g}a) {F'}^{\mu\nu}) \simeq J^\nu$, which after identifying $2 \sl{g} a \to - g_\phi \phi$ yields the usual dilaton-photon coupling at leading order.

For completeness, we show that one can reach the same conclusion working with the unscaled fields. Here the magnetic current yields the form factor
\begin{align}
\sl{C} &= \int_{V_c} \tilde{\v{B}}_0 \cdot \tilde{\v{B}}_n \\
&= \frac{1}{\omega_n} \int_{V_c}\tilde{\v{B}}_0 \cdot (\nabla \times \tilde{\v{E}}_n) = \frac{1}{\omega_n} \int_S (\tilde{\v{E}}_n \times \tilde{\v{B}}_0) \cdot d \v{S} + \frac{1}{\omega_n} \int_{V_c}(\nabla \times \tilde{\v{B}}_0) \cdot \tilde{\v{E}}_n
\end{align}
where the surface term vanishes because $\tilde{\v{E}}_n \parallel d\v{S}$. The remaining term is proportional to $\int_{V_c} \v{J}_0 \cdot \tilde{\v{E}}_n$, which is precisely what one expects from the usual form of the dilaton coupling, since $\v{J}_{\mathrm{eff}} \supset g_\phi \phi \v{J}_0$. However, in a typical cavity haloscope where the magnet is outside the cavity, $\v{J}_0$ vanishes inside the cavity, so that $\sl{C}$ (claimed to be nonzero in Ref.~\cite{McAllister:2022ibe}) vanishes. 

The vanishing form factors $C_\phi$, $\bar{C}$, and $\sl{C}$ were all identified in at least one of the works cited above, but these works then either assumed them to be $\mathcal{O}(1)$, took an unphysical choice of $\v{E}_0$, or claimed they were small but nonzero, possibly due to numeric error. For heterodyne detection, we substitute $(\tilde{\v{E}}_0, \tilde{\v{B}}_0) \to (\tilde{\v{E}}_\ell^*, \tilde{\v{B}}_\ell^*)$, and the dilaton form factors $C_\phi$ and $\sl{C}$ both vanish by the orthogonality of cavity modes (Ref.~\cite{Dai:2024dkr} claimed a nonzero value in an extremely wide cavity, possibly due to numeric error), while the alternative axion form factor $\bar{C}$ is equal in magnitude to $(\omega_\ell / \omega_n) C_B$ and ultimately leads to the same physical result. 

%!TEX root = main.tex

\section{Derivation of Two-Form Dualities}
\label{app:dualities}

Here we derive the dualities in Eqs.~\eqref{eq:massive_duality_a} and~\eqref{eq:massive_duality_b} between a massive two-form and a vector. To warm up, we will consider the duality between a massless two-form and a scalar. 

\paragraph{Massless Two-Form Duality.} In this case, we take the parent Lagrangian
\begin{equation}
\mathcal{L}[H, a] = \frac{1}{12} H_{\mu\nu\rho} H^{\mu\nu\rho} + \frac16 \epsilon^{\mu\nu\rho\sigma} H_{\mu\nu\rho} \del_\sigma a + \frac16 H_{\mu\nu\rho} J^{\mu\nu\rho}
\end{equation}
where $a$ and the three-form $H$ both lack a kinetic term, and are thus nondynamical, and $J$ is an external current. Integrating the second term by parts shows that $a$ is just a Lagrange multiplier, and integrating it out enforces the Bianchi identity $dH = 0$, which implies that in Minkowski space, we can write $H = dB$. This yields the massless two-form action $\mathcal{L}[B]$. 

Alternatively, we can integrate out $H$. Its equation of motion is algebraic,
\begin{equation}
H_{\mu\nu\rho} + \epsilon_{\mu\nu\rho\sigma} \del^\sigma a + J_{\mu\nu\rho} = 0.
\end{equation}
Since it is nondynamical, we can integrate it out by plugging this into the action, yielding
\begin{equation}
\mathcal{L}[a] = -\frac{1}{12} (\epsilon_{\mu\nu\rho\sigma} \del^\sigma a + J_{\mu\nu\rho})^2 = \frac12 (\del_\mu a)^2 - \frac16 \epsilon_{\mu\nu\rho\sigma} J^{\mu\nu\rho} \del^\sigma a - \frac{1}{12} J_{\mu\nu\rho} J^{\mu\nu\rho}
\end{equation}
which involves a standard derivatively coupled axion. Defining $J_\sigma = \epsilon_{\mu\nu\rho\sigma} J^{\mu\nu\rho}/6$ for convenience, which implies $J_{\mu\nu\rho} = - \epsilon_{\mu\nu\rho\sigma} J^\sigma$, we have thus shown the duality
\begin{equation} \label{eq:massless_duality_1}
\mathcal{L}[B] = \frac{1}{12} H_{\mu\nu\rho} H^{\mu\nu\rho} - \frac16 \epsilon^{\mu\nu\rho\sigma} H_{\mu\nu\rho} J_\sigma \quad \lra \quad \mathcal{L}[a] = \frac12 (\del_\mu a - J_\mu)^2.
\end{equation}
On the axion side of the duality, we also have a contact term $J^\mu J_\mu / 2$. However, we could just as well have subtracted this term from the original parent Lagrangian. This shifts it to the other side of the duality, giving
\begin{equation} \label{eq:massless_duality_2}
\mathcal{L}[B] = \frac{1}{12} (H_{\mu\nu\rho} + J_{\mu\nu\rho})^2 \quad \lra \quad \mathcal{L}[a] = \frac12 (\del_\mu a)^2 - (\del_\mu a) J^\mu.
\end{equation}
Note that in both cases the current couples directly to $H_{\mu\nu\rho}$, not to $B_{\mu\nu}$, as required by the massless two-form field's gauge symmetry. 

For example, the axion-fermion coupling corresponds to $J^\mu = - g_{a\psi} \bar{\psi} \gamma^\mu \gamma^5 \psi$, while the axion-photon coupling corresponds to $J^\mu = g_{a\gamma\gamma} K^\mu / 4$. In the latter case, Eq.~\eqref{eq:massless_duality_2} implies that on the two-form side, we have 
\begin{equation} \label{eq:two_photon_massless}
\mathcal{L}[B] = \frac{1}{12} \left(H_{\mu\nu\rho} - \frac32 g_{a\gamma\gamma} A_{[\mu} F_{\nu\rho]} \right)^2 
\end{equation}
where brackets denote the antisymmetric part, normalized with a factor of $1/6$. This satisfies electromagnetic gauge invariance, provided we simultaneously transform $\delta A_\mu = \del_\mu \alpha$ and $\delta B_{\mu\nu} = \alpha g_{a\gamma\gamma} F_{\mu\nu}/2$. Finally, if we took $J^\mu = m_\gamma A^\mu$ in~\eqref{eq:massless_duality_1}, we would have a photon with St\"uckelberg mass $m_\gamma$, where gauge symmetry requires the simultaneous transformation $\delta A_\mu = \del_\mu \alpha$ and $\delta a = m_\gamma \alpha$. On the two-form side this becomes the topological ``BF'' term,
\begin{equation} \label{eq:Stuckelberg_a}
\mathcal{L}[B] = \frac{1}{12} H_{\mu\nu\rho} H^{\mu\nu\rho} + \frac{m_\gamma}{4} \epsilon^{\mu\nu\rho\sigma} B_{\mu\nu} F_{\rho\sigma}.
\end{equation}
The fact that this is really a photon mass term was discussed pedagogically in Ref.~\cite{Reece:2023czb}.

\paragraph{Massive Two-Form Duality.} Now we turn to the massive case. First, to derive Eq.~\eqref{eq:massive_duality_a}, we use the parent Lagrangian 
\begin{equation}
\mathcal{L}[V, B] = \frac12 m^2 V_\mu V^\mu - \frac{m}{2} \epsilon^{\mu\nu\rho\sigma} V_\mu \del_\nu B_{\rho\sigma} - \frac14 m^2 B_{\mu\nu} B^{\mu\nu} + m V_\mu J^\mu
\end{equation}
After integrating by parts, $B_{\mu\nu}$ has the algebraic equation of motion
\begin{equation}
m B^{\mu\nu} = - \epsilon^{\mu\nu\rho\sigma} \del_\rho V_\sigma.
\end{equation}
Plugging this equation of motion back into the action directly yields the right-hand side of Eq.~\eqref{eq:massive_duality_a}. Alternatively, the equation of motion for $V_\mu$ is
\begin{equation}
V^\mu = \frac{1}{6m} \epsilon^{\mu\nu\rho\sigma} H_{\nu\rho\sigma} - \frac{1}{m} J^\mu, 
\end{equation}
and plugging this in gives, after some algebra,
\begin{align}
\mathcal{L}[B] &= \frac{1}{12} H_{\mu\nu\rho} H^{\mu\nu\rho} - \frac14 m^2 B_{\mu\nu} B^{\mu\nu} + \frac{1}{6} \epsilon^{\mu\nu\rho\sigma} J_\mu H_{\nu\rho\sigma} - \frac{1}{2} J^\mu J_\mu \\
&= \frac{1}{12} (H_{\mu\nu\rho} + J_{\mu\nu\rho})^2 - \frac14 m^2 B_{\mu\nu} B^{\mu\nu}
\end{align}
which is the left-hand side of Eq.~\eqref{eq:massive_duality_a}.

Similarly, to derive Eq.~\eqref{eq:massive_duality_b}, we use the parent Lagrangian 
\begin{equation}
\mathcal{L}[V, B] = \frac12 m^2 V_\mu V^\mu - \frac{m}{2} \epsilon^{\mu\nu\rho\sigma} V_\mu \del_\nu B_{\rho\sigma} - \frac14 m^2 B_{\mu\nu} B^{\mu\nu} + \frac{m}{2} B_{\mu\nu} J^{\mu\nu}.
\end{equation}
The equation of motion of $V_\mu$ is
\begin{equation}
m V^\mu = \frac{1}{6} \epsilon^{\mu\nu\rho\sigma} H_{\nu\rho\sigma}.
\end{equation}
Plugging this in directly gives the left-hand side of Eq.~\eqref{eq:massive_duality_b}. Alternatively, we can integrate the second term by parts, so that $B_{\mu\nu}$ has the equation of motion
\begin{equation}
m B^{\mu\nu} = J^{\mu\nu} - \epsilon^{\mu\nu\rho\sigma} \del_\rho V_\sigma.
\end{equation}
Plugging this back in gives, after some algebra, the right-hand side of Eq.~\eqref{eq:massive_duality_b}. This second duality was also reviewed in Ref.~\cite{Plantier:2025hcm}, which called it the ``massive'' duality. 

\paragraph{Applying Both Dualities.} As an example, we consider starting with an arbitrary vector interaction $m V_\mu J^\mu$, applying Eq.~\eqref{eq:massive_duality_a} to write it as a two-form interaction, then applying Eq.~\eqref{eq:massive_duality_b} to return to a vector. After the first step, we can write the Lagrangian as 
\begin{equation}
\mathcal{L}[B] = \frac{1}{12} H^2 - \frac14 m^2 B^2 + \frac{m}{2} B_{\mu\nu} \bar{J}^{\mu\nu} - \frac12 J_\mu J^\mu
\end{equation}
where we defined $\bar{J}^{\mu\nu} = \epsilon^{\mu\nu\rho\sigma} \del_\rho J_\sigma/m$. Subsequently applying Eq.~\eqref{eq:massive_duality_b} gives 
\begin{align}
\mathcal{L}[V] &= - \frac14 F_V^2 + \frac12 m^2 V^2 - \frac{1}{2} \epsilon^{\mu\nu\rho\sigma} V_\mu \del_\nu \bar{J}_{\rho\sigma} + \frac{1}{4} \bar{J}_{\mu\nu} \bar{J}^{\mu\nu} - \frac12 J_\mu J^\mu \\
&= - \frac14 F_V^2 + \frac12 m^2 V^2 + V_\mu \, \frac{\del^\mu (\del^\nu J_\nu) - \del^2 J^\mu}{m} + \frac{J^\mu \del^2 J_\mu}{2m^2} + \frac{(\del^\mu J_\mu)^2}{2m^2} - \frac12 J_\mu J^\mu
\end{align}
where we dropped total derivatives in the second step. The original interaction has been rewritten as a set of higher-derivative and contact interactions. Though it looks very different, it is equivalent, because it can be reached from the original Lagrangian $\mathcal{L}[V] = - F_V^2/4 + m^2 V^2/2 + m V_\mu J^\mu$ by the field redefinition $V^\mu \to V^\mu - J^\mu / m$. 

\bibliographystyle{utphys3}
{\scriptsize
\bibliography{refs}}

\end{document}